\documentclass[a4paper,12pt]{fullarticle}
\usepackage[british]{babel}
\usepackage{csquotes}
\usepackage{sciencestuff}
\usepackage[algoruled,algosection,vlined,shortend,linesnumbered]{algorithm2e}

\usepackage[dvipsnames]{xcolor}
\usepackage{esint}
\usepackage{booktabs}
\usepackage{array}
\usepackage[table]{xcolor}
\usepackage{graphicx}
\usepackage{xcolor}
\usepackage{nicematrix}
\usepackage{booktabs}
\usepackage{listings}
\usepackage{mdframed}
\usepackage{tcolorbox}

\newtcolorbox{corollary}{
    colback=teal!4!white,
    colframe=teal!75!black, 
    title=\textbf{Corollary},
    arc=2pt,
    boxrule=0.5pt,
    left=6pt, right=6pt, top=5pt, bottom=5pt
}

\newtcolorbox{proposition}{
    colback=NavyBlue!5!white,
    colframe=NavyBlue!75!black,
    title=\textbf{Proposition},
    arc=3pt,
    boxrule=0.5pt,
    left=5pt, right=5pt, top=5pt, bottom=5pt}

\newtcolorbox{claim}{
    colback=BrickRed!5!white,
    colframe=BrickRed!80!black,
    title=\textbf{Claim},
    arc=3pt,
    boxrule=0.5pt,
    left=5pt, right=5pt, top=5pt, bottom=5pt}
\newtcolorbox{theorem}{
    colback=Maroon!5!white,
    colframe=Maroon!75!black,
    title=\textbf{Theorem},
    arc=3pt,
    boxrule=0.5pt,
    left=5pt, right=5pt, top=5pt, bottom=5pt}

\newtcolorbox{example}{
    colback=ForestGreen!5!white,
    colframe=ForestGreen!75!black,
    title=\textbf{Example},
    arc=3pt,
    boxrule=0.5pt,
    left=5pt, right=5pt, top=5pt, bottom=5pt}

    \newtcolorbox{definition}{
    colback=SlateGray!5!white,
    colframe=SlateGray!75!black,
    title=\textbf{Definition},
    arc=3pt,
    boxrule=0.5pt,
    left=5pt, right=5pt, top=5pt, bottom=5pt}
\title{%
    Functional Renormalization for Random Matrix Theory:\\ \relax
    {\Large The relational background field method}}

\author[1]{Vincent Lahoche\emailfoot{vincent.lahoche@cea.fr}}
\author[2]{Dine Ousmane Samary\emailfoot{dine.ousmanesamary@uac.bj}}

\affil[1]{%
  Université Paris-Saclay, CEA, 
  \protect \\
  Palaiseau, F-91120, France
}
\affil[2]{%
	Faculté des Sciences et Techniques (ICMPA-UNESCO Chair)
	\protect \\
	Université d'Abomey-Calavi, 072 BP 50, Benin
}

\date{}

\hypersetup{%
  pdftitle={Functional Renormalization for Random Matrix Theory I},
  pdfkeywords={%
  functional renormalization group,
  theoretical physics,
  data science,
  signal analysis,
  signal detection,
    random matrix theory
    },
  pdfsubject={signal detection}
}

\newtheorem{remark}{Remark}

\begin{document}

\maketitle

\begin{abstract}
The construction of a reliable renormalization group (RG) flow for discrete gravity models that preserves their underlying symmetry group, typically $U(N)$ or $O(N)$, remains an open problem. For random matrix models, which are the focus of this paper, this symmetry is intrinsically tied to the interactions encoding the random geometry of two-dimensional quantum Euclidean spacetime.
We develop a novel approach based on the introduction of a partial matrix-valued intermediate field. In the large-$N$ limit, measure concentration strongly suppresses fluctuations of its singular values, allowing it to play the role of a self-consistent background field. This provides the basis for a relational RG in which the notion of scale is dynamically induced by the effective Gaussian measure in the basis where the intermediate field is diagonal.
Our construction preserves the symmetry of the original model and admits a well-defined continuum limit. We show that the resulting infrared theory is described by a three-dimensional non-local Euclidean field theory with a non-trivial Wilson–Fisher-like fixed point and a single relevant direction. Once the constraint of vanishing anomalous dimension imposed by the Ward identities is taken into account, the vertex expansion resums into a single functional equation and the associated critical exponent takes the exact value $\theta=1/2$, that of the large-$N$ spherical universality class. We discuss its physical interpretation in relation to the confining and non-confining sectors of the matrix model, and compare it qualitatively with standard double-scaling expectations. We finally discuss extensions of the background-field approach to other discrete gravity models, such as random tensor models, and to different symmetry groups, as well as connections with more formal RG frameworks and information geometry.
\end{abstract}

\highlights{%
By exploiting the concentration (self-averaging) of large random matrices, we develop a novel renormalization group framework that preserves unitary symmetry by construction. The flow is organized around the effective spectrum of the intermediate Hubbard-Stratonovich field, providing a natural and symmetry-compatible notion of scale.
}


\clearpage

{\small\tableofcontents}

\clearpage

\section{Introduction, motivation and related works}\label{sec1}

\begin{figure}
\begin{center}
\includegraphics[scale=0.6]{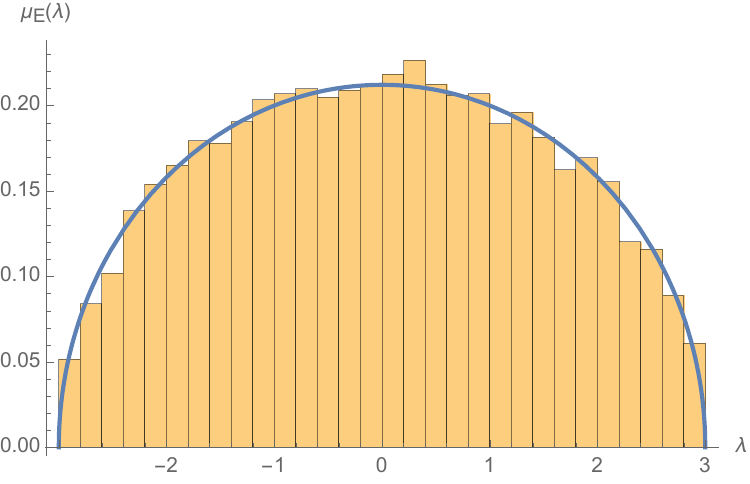}
\end{center}
\caption{Illustration of the concentration mechanism: the empirical eigenvalue distribution (yellow histogram) is close to the Wigner semicircle for sufficiently large $N$ (here, $N=10^4$). }\label{fig1}
\end{figure}

\paragraph{General background.} Random matrix theory originated in Wigner’s seminal work on nuclear physics \cite{wigner1967random}. Wigner proposed that the complicated yet deterministic Hamiltonian of a large atomic nucleus could be modeled by a random matrix. For a Gaussian Hermitian matrix, this idea leads to the famous semicircle law: as $N\to\infty$, the empirical spectral distribution converges to a deterministic, symmetric limiting density (Fig. \ref{fig1}). More precisely, for a Hermitian matrix whose entries have variance $\sigma^2/N$, Wigner’s law states that the limiting spectral density is given by \cite{potters2020first}:

\begin{theorem}\label{th0}
Let $M$ be a Hermitian random matrix with independent, centered Gaussian entries of variance $\sigma^2/N$. Let $\{\lambda_i\}$ be its $N$ eigenvalues, $i=1,2,\cdots, N$. Then, as $N\to \infty$, the empirical eigenvalue distribution $\mu_{\text{E}}(\lambda):=\frac{1}{N}\sum_{i=1}^N\delta(\lambda-\lambda_i)$ converges weakly to the semicircle law $\mu(\lambda)$:

\begin{equation}
\mu_{\text{E}}(\lambda) \to \mu(\lambda) := \left\{
    \begin{array}{ll}
        \frac{\sqrt{4\sigma^2-\lambda^2}}{2\pi \sigma^2} & \mbox{if}\,  -2\sigma \leq \lambda \leq 2\sigma \\
        0 & \mbox{if} \quad  \lambda^2 > 4\sigma^2\,.
    \end{array}
\right.
\end{equation}
\end{theorem}

In the late 1980s, and particularly during the 1990s, a connection emerged between random matrix theory and two-dimensional random geometry, which is itself closely related to the long-standing problem of quantum gravity
\cite{di19952d,eynard2015random}.

\begin{figure}
\begin{center}
\includegraphics[scale=0.5]{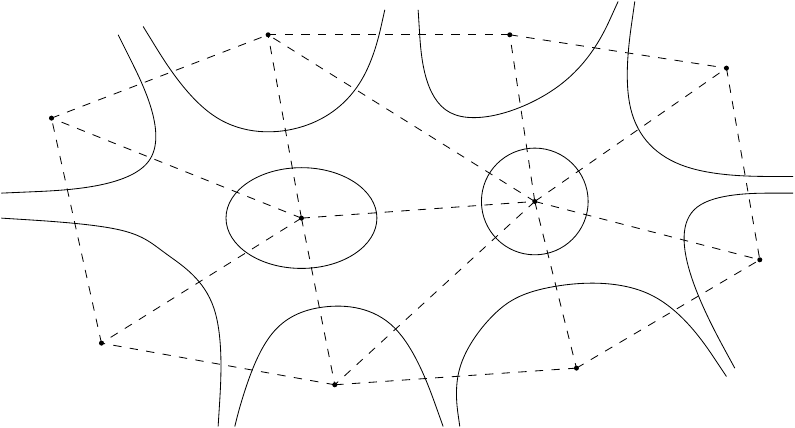}\quad \includegraphics[scale=0.6]{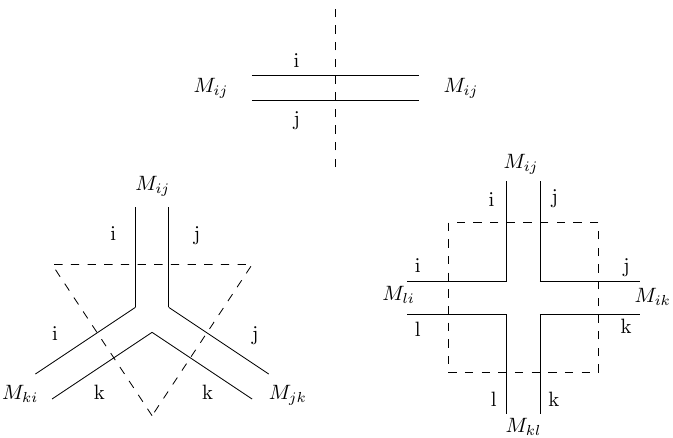}
\end{center}
\caption{Right: Illustration of the duality between the Feynman graphs of a Hermitian random matrix model and the triangulation of a surface. The correspondence is illustrated for cubic $\Tr M^3$ and quartic $\Tr M^4$ interactions, the duality between the line and the propagator $\delta_{ij}\delta_{kl}$ being illustrated on the left. Left: a typical Feynman diagram, and its dual triangulation (in dashed lines).}\label{fig2}
\end{figure}

Historically, this construction emerged in two stages. First, it was recognized that the path integral over geometries in Feynman’s approach to the quantization of gravity could be formulated as a sum over randomly triangulated two-dimensional surfaces, with the underlying discretization corresponding more generally to Regge calculus \cite{barrett2018tullio}. Later, in the 1980s, it was realized that these triangulated surfaces could be associated, through a duality, with the Feynman graphs of random matrix models, particularly Hermitian matrix models (Fig. \ref{fig2}). Random matrix theory thus became one of the first concrete frameworks for discrete quantum gravity. Furthermore, since the worldsheet of a string is a two-dimensional surface, the theory of random surfaces encoded by random matrices naturally connects to string theory. This connection was also extensively explored during the 1990s \cite{brezin1990exactly}.

One of the major achievements of the theory was the identification of its continuum limit with Liouville quantum gravity \cite{seiberg1990notes}, a development that culminated in the establishment of the famous KPZ relations \cite{duplantier2011liouville}. Other approaches to discrete quantum gravity subsequently sought to extend the success of random matrix models to higher dimensions. Among the most prominent examples in the literature are colored random tensor models \cite{gurau2024quantum,gurau2017random}, group field theories (GFTs) \cite{boulatov1992model,freidel2005group}, and their extensions inspired by colored tensor models, known as tensorial group field theories (TGFTs) \cite{carrozza2014tensorial}. The latter are also closely related to spin foams and loop quantum gravity \cite{rovelli2004quantum}. In particular, the Feynman diagrams of GFTs encode simplicial decompositions of spacetime, while their Feynman amplitudes provide the corresponding spin-foam amplitudes. Finally, other approaches to discrete quantum gravity should also be mentioned, notably dynamical triangulations \cite{loll2020quantum}.

For discrete quantum gravity models inspired by random matrix theory, such as tensor models and, in particular, group field theories (GFTs), a major open problem is to understand their phase structure and the possible emergence of a continuum limit. In GFTs, condensate-based approaches have indeed been developed to reconstruct semiclassical solutions of gravity. When applied to cosmology, these approaches can describe Friedmann–Robertson–Walker-type geometries while resolving the Big Bang singularity\footnote{Similar singularity-resolution mechanisms also arise in other approaches to quantum gravity; see, for example, \cite{Bojowald:2025ocr}.} \cite{Oriti:2024elx}. These condensates, constructed from coherent states, provide a (semi-)classical limit, much like coherent photon states in quantum electrodynamics, which reproduce the classical behavior of the electromagnetic field. However, the precise microscopic mechanism underlying the emergence of such condensates remains only partly understood. In particular, it is not yet established whether a renormalization-group flow can naturally drive the microscopic dynamics of spacetime quanta toward a condensed phase through a phase transition. The existence of fixed points, corresponding to non-trivial resummations of spin-foam amplitudes, is therefore a central objective.
Although the study of phase transitions through the renormalization group is well established, with critical phenomena providing a particularly successful example \cite{Wegner:1972ih,Wilson:1972zzb,Zinn-Justin:2019jix}, its application to GFTs remains challenging. Their distinctive non-locality, which encodes how geometric quanta are glued together through the Feynman rules, complicates the direct application of standard renormalization methods and raises questions about the reliability of conventional approximations \cite{Lahoche:2020pjo}.

In the case of random matrices, which share the non-local character of GFTs, a continuous phase transition associated with Liouville quantum gravity has been identified in the \textit{double-scaling limit} \cite{duplantier2010liouvillequantumgravitykpz}. Moreover, various renormalization-group approaches have shown that this transition is governed by a specific fixed point with a single relevant direction. Random matrix theory shares several important features with GFTs and therefore provides an excellent testing ground for benchmarking renormalization-group approximation schemes by comparing their predictions with analytical results. In particular, as in tensor models, matrix models possess an exact $U(N)$ symmetry arising from the trace structure of their interactions. This symmetry presents an additional challenge for renormalization-group approaches, since the methods developed following the pioneering work of Brézin and Zinn-Justin \cite{brezin1992renormalization} generally require an explicit breaking of this symmetry. In this article, we introduce a new renormalization-group approach for matrix models, based on a background-field-type approximation, which has the advantage of formally preserving the $U(N)$ symmetry of the models.

As this work lies primarily at the interface of two fields, quantum gravity and the functional renormalization group, we devote part of the remainder of this introduction to the continuum limit of random matrices and to the so-called double-scaling limit (see the next paragraph). This discussion is intended to help non-expert readers understand the motivation and significance of the present work in its broader context. Conversely, a brief introduction to the nonperturbative renormalization group is provided at the beginning of Section \ref{FRG_sec}.

\paragraph{Continuum limit and double scaling for random matrices.} Let us begin with a brief review of the continuum limit of random matrices, with particular emphasis on the double-scaling limit. Readers already familiar with these concepts may skip this paragraph, which is primarily intended to recall the basic notions needed for the discussion that follows.

In two dimensions, the partition function of 2D Einstein gravity without matter is written as:
\begin{equation}
Z=\sum_{g\geq 0} \int [\dd h] e^{-\beta A +\gamma (2-2 g)}\label{Zdiscrete}
\end{equation}
where $g$ denotes the genus of the surface, $[\dd h]$ is the integration measure over equivalent metrics, $A$ denotes the area of the surface, and $\chi:=2-2g \equiv \frac{1}{4\pi} \int \sqrt{h} R$ is the Euler characteristic, related to the volume integral of the scalar curvature by the Gauss-Bonnet theorem. The classical Einstein-Hilbert action $S_{\text{EH}}[\mathcal{M}]:=\beta A -\gamma \chi(\mathcal{M})$ depends on a given topological manifold $\mathcal{M}$, and the integral extends over all such manifolds. In a discrete approach, each manifold can be triangulated, and quantization amounts to summing over all possible triangulations ${\Delta}$:
\begin{equation}
Z=\sum_{\Delta}\, e^{-S_{\text{EH}}[\Delta]}\,.
\end{equation}
The parameters $\beta$ and $\gamma$ are fixed by the cosmological constant $\Lambda$ and the gravitational constant $G$ ($\beta$ depends on both, $\gamma$ only on $G$):
\begin{equation}
\gamma:=\frac{4\pi}{G}\,, \qquad \beta:=\frac{\Lambda}{G}\,.
\end{equation}
For a fixed triangulation $\Delta$, the Euler characteristic is explicitly given by $\chi=V(\Delta)-E(\Delta)+T(\Delta)$, where $V$, $E$, and $T$ denote the number of vertices, edges, and triangles, respectively (i.e., the 0-simplices, 1-simplices, and 2-simplices of the simplicial decomposition of the manifold).

Generally, a random matrix model (here, Hermitian) is defined by a polynomial, and in this article, we will focus on even-degree polynomials. We therefore consider a polynomial of degree $2K$:
\begin{equation}
\mathcal{V}(M):=\sum_{l=1}^K \, \frac{a_{2 l}}{2l} \, M^{2 l}\,,
\end{equation}
where $l$ will hereafter denote the \textit{degree} of the interaction, and a partition function $Z$, constructed as a path integral over the configuration space assumed to be Haar-distributed (the measure is assumed to be Haar-invariant \cite{di19952d}):
\begin{equation}
Z:=\int \dd M\, \exp \left(-N \Tr \mathcal{V}(M)\right)\,.\label{partfunc}
\end{equation}
By definition, we will say that these $U(N)$-invariant interactions consisting of a single trace are \textit{tracial} and connected. Conversely, an interaction composed of two or more traces (but still $U(N)$-invariant) will be called tracial and disconnected.
With the conventional choice $a_2=1$, the free propagator is written as $C_{ij,kl}:=\frac{1}{N}\delta_{il}\delta_{jk}$. Perturbatively, the partition function takes the usual form of a sum of amplitudes $\mathcal{A}_G$ indexed by graphs $G$, and for a quartic model ($K=2$), we have for instance:
\begin{equation}
Z=\sum_{G}\, a_4^{v(G)}\, \mathcal{A}_G\,,
\end{equation}
where $v(G)$ is the number of quartic vertices in the graph $G$. The Feynman graphs of matrix models are ribbon graphs, thus presenting a richer structure than that of ordinary Feynman graphs. In particular, in addition to vertices and edges, they include faces:
\begin{definition}\label{defface}
A face is a closed loop of Kronecker deltas, each generating a factor of $N$.
\end{definition}
Thus:
\begin{equation}
\mathcal{A}_G \sim N^{v(G)-\ell(G)+f(G)}\,.
\end{equation}
The duality described in Fig. \ref{fig2} implies the correspondences $G \leftrightarrow \Delta$, $v(G) \leftrightarrow T(\Delta)$, $ \ell(G) \leftrightarrow E(\Delta)$, and $f(G) \leftrightarrow V(\Delta)$, so that the expansion in powers of $N$ is naturally organized according to the topological invariant $g$ (the genus) of the dual decomposition, $\mathcal{A}_G \sim N^{2-2g}$, and:
\begin{equation}
Z=\sum_{g}\, \, N^{2-2g} \mathcal{Z}_g\,. 
\end{equation}
In the naive $N\to \infty$ limit, planar maps dominate ($g=0$), which are topologically equivalent to 2D (Brownian) spheres, and $Z\approx N^2 \mathcal{Z}_0$. Furthermore, $\mathcal{Z}_0(a_4)$ exhibits the following critical behavior:
\begin{equation}
\mathcal{Z}_0(a_4) \sim \vert a_4-a_4^* \vert^{2-\gamma}\,,
\end{equation}
where the \textit{entropic exponent} is $\gamma=-1/2$ for this model \cite{di19952d} and $a_4^*=-1/12$. The expansion in powers of $a_4$ becomes singular at $a_4=a_4^*$, and the planar free energy $f_0(a_4):= \ln\mathcal{Z}_0(a_4) $ is no longer analytic at this point. The interpretation of this singularity becomes evident when calculating the average area of the surface $A(\Delta)$ corresponding to the discrete manifold $\Delta$, assuming that each $2$-simplex has an elementary area $a$. We have\footnote{The coupling appears to the power of the number of vertices, i.e., the number of triangles. The derivative of the free energy $f_0$ with respect to $a_4$ therefore yields the sum $\sum_G v(G) a_4^{v(G)} \mathcal{A}_G/\left(\sum_G  a_4^{v(G)}\mathcal{A}_G\right)$, where the sums here run over connected graphs. The probability $p(v)$ of realizing $v$ vertices is $p(v):=\sum_{G\vert v}a_4^{v(G)} \mathcal{A}_G/\left(\sum_G  a_4^{v(G)}\mathcal{A}_G\right)$.}:
\begin{equation}
\langle A(\Delta) \rangle = a  \langle T(\Delta) \rangle = a  a_4 \frac{\dd f_0}{\dd a_4} \sim \frac{a}{\vert a_4-a_4^* \vert}\,.
\end{equation}
An approach to the continuum limit therefore consists in taking $a\to 0$ simultaneously with $a_4\to a_4^*$, so as to keep the average area fixed. The transition at the critical point thus corresponds to an infinitely refined triangulation, consistent with the intuition of a continuum. In reality, however, this limit is better understood as a Brownian sphere than as a smooth continuous geometry. The resulting geometry is fractal rather than smooth, corresponding to a semiclassical rather than strictly classical regime \cite{duplantier2010rigorous}.
The so-called \textit{double scaling} limit allows one to go beyond the planar sector. It is based on the observation that, at the critical point:
\begin{equation}
\mathcal{Z}_g \sim \vert a_4-a_4^* \vert^{\frac{(2-\gamma)(2-2g)}{2}}\,,
\end{equation}
This limit consists in letting $N\to \infty$ and $a_4\to a_4^*$ simultaneously, while keeping the product $N \vert a_4-a_4^* \vert^{\frac{2-\gamma}{2}}$ constant. Thus, near the critical point, we have the following behavior:
\begin{equation}
Z \sim \sum_g A_g \left(N\vert a_4-a_4^* \vert^{\frac{2-\gamma}{2}}\right)^{2-2g}\,.
\end{equation}
In this limit, all topologies contribute to the continuum limit, as suggested by the discretized partition function \eqref{Zdiscrete}. It is worth emphasizing that, in the present construction, which includes gravity but no matter, the exponent $\gamma=-1/2$ is essential. Within the renormalization-group approach to the continuum limit, this exponent can be interpreted as a critical exponent associated with an interacting fixed point, whose existence reflects a non-trivial resummation of Feynman graphs.
Let us also note that the correspondence with \eqref{Zdiscrete} is established for
\begin{equation}
a_4 \leftrightarrow e^{-\Lambda/G}\,,\qquad N \leftrightarrow e^{4\pi/G}\,.
\end{equation}
\begin{remark}
Let us recall that the continuum limit of matrix models with vanishing central charge ($c=0$, pure gravity), as well as that of the so-called minimal models with $c\leq 1$, corresponds to two-dimensional quantum gravity, described by Liouville theory \cite{garban2012quantum,miller2018liouville,duplantier2010rigorous,duplantier2010liouvillequantumgravitykpz}. For higher central charges ($c>1$), however, the geometry degenerates into a branched-polymer phase: the Hausdorff dimension drops to $2$, while the spectral dimension becomes $4/3$, as characteristic of random trees. These models therefore no longer describe two-dimensional quantum gravity in the usual sense.
\end{remark}

\paragraph{Renormalization group approach and open issues.} In their seminal work \cite{brezin1992renormalization}, Zinn-Justin and Brézin proposed an original approach to double scaling by interpreting the constraint $N \vert a_4-a_4^* \vert^{\frac{2-\gamma}{2}}=\mathrm{const.}$ as a scaling law typical of those encountered in statistical physics in the vicinity of an infrared (IR) fixed point of the renormalization group. Indeed, it is expected that a shift $N\to N+\delta N$ can be exactly compensated by a shift in the quartic coupling, in precisely the same way that typical scaling behavior manifests near a fixed point, while preserving the continuum limit at large scales. In other words, Zinn-Justin and Brézin proposed interpreting the scale change $N\to N+\delta N$ (with $\delta N<0$ and $\vert \delta N \vert /N\ll 1$) from a Wilsonian perspective as the result of a partial integration over the "high-energy" components $N+\delta N<i,j\leq N$ of the $N\times N$ Hermitian matrix $M_{N\times N}$ (whose entries are assumed to be random). More precisely, the partial integration relies on partitioning the $N\times N$ matrix $M_{N\times N}$ into four blocks: an $(N+\delta N) \times (N+\delta N)$ matrix $M_{(N+\delta N) \times (N+\delta N)}$, a rectangular $(N+\delta N) \times \delta N$ matrix $X$, and a square $\delta N \times \delta N$ matrix $Y$:
\begin{equation}
M_{N\times N} := \begin{bNiceArray}{c|c}[margin]
\CodeBefore
  \cellcolor{gray!8}{1-1}
  \cellcolor{gray!20}{1-2, 2-1, 2-2}
\Body
  M_{N+\delta N\times N+ \delta N} & X \\
  \hline
  X^\dagger & Y
\end{bNiceArray}\,.
\end{equation}
Thus, partial integration over $X$, $X^\dagger$, and $Y$ effectively transforms the path integral by defining an effective theory describing a random matrix of size $(N+\delta N) \times (N+\delta N)$. If we denote by $\mathcal{V}_N$ the potential for a matrix of size $N$, the effective potential $\mathcal{V}_{N+\delta N}^\prime$ for $M_{(N+\delta N) \times (N+\delta N)}$ is defined by:
\begin{equation}
\exp \left(-(N+\delta N) \mathcal{V}_{N+\delta N}^\prime \right) \propto \int \dd X \dd X^\dagger \dd Y \exp \left(-N \mathcal{V}_{N}\right)\,.\label{partial}
\end{equation}
By restricting ourselves to a quartic truncation, this hypothesis, according to the authors of \cite{brezin1992renormalization}, translates into a Callan–Symanzik (CS)-type equation for the partition function, which for large $N$ is written as:
\begin{equation}
\left[N \frac{\partial }{\partial N}+\beta(a_4)\frac{\partial}{\partial a_4}+\delta\right] Z(N,a_4) \approx 0\,,
\end{equation}
where $\beta(a_4) := N \, \dd a_4/\dd N$. Assuming the existence of a stable fixed point, i.e., such that $\beta(a_4^*) = 0$ and $\beta^\prime(a_4^*) < 0$ (required for $\theta:=-\beta^\prime(a_4^*)>0$; indeed \eqref{oneloopRG} gives $\beta^\prime(-1/6)=-1$), this equation is equivalent to assuming the following scaling law:
\begin{equation}
Z(N,a_4) = N^{-\delta} f(N^{-\beta^\prime(a_4^*)} \vert a_4-a_4^* \vert)\,.
\end{equation}
The correspondence with the double scaling limit is thus established provided that the critical exponent $\theta := -\beta^\prime(a_4^*)$ is related to the entropic exponent $\gamma$ by:
\begin{equation}
\boxed{2-\gamma=\frac{2}{\theta}\,,}
\end{equation}
with the exponent $\delta$ fixed, in the planar regime, at the value $\delta=-2$. By explicitly calculating the integral \eqref{partial} in the $N\to \infty$ limit, one finds, at one-loop order:
\begin{equation}
\beta(a_4)=a_4+6a_4^2\,,\label{oneloopRG}
\end{equation}
which admits a fixed point at $a_4^*=-1/6$, with critical exponent $\theta=1$, corresponding to $\gamma=0$, to be compared with the exact value $\gamma=-1/2$. A substantial methodological improvement was achieved in the 2010s through non-perturbative techniques based on the Wetterich equation \cite{delamotte2012introduction}. These methods rely on a modification of the propagator “à la Kontsevich”, which explicitly breaks the $U(N)$ invariance of the model at the kinetic level, while preserving it at the level of the interactions:
\begin{equation}
\mathcal{V}(M):=\frac{1}{2} \sum_{i,j} M_{ij} R_N(i,j) \bar{M}_{i,j} +\sum_{l=1}^K \, \frac{a_{2 l}}{2l} \, M^{2 l}\,,
\end{equation}
where the function $R_N(i,j)$ is typically:
\begin{equation}
R_N(i,j):=Z_N \left(\frac{2N}{i+j}-1\right) \Theta \left(1-\frac{i+j}{2N}\right)\,,\label{regul}
\end{equation}
$\Theta(x)$ is the standard Heaviside step function, and $N$ plays the role of an IR cut-off on the matrix size, while $Z_N$ corresponds to the wavefunction renormalization. The regulator function $R_N$ is constructed so as to interpolate between two regimes:
\begin{itemize}
    \item An ultraviolet (UV) regime, $N \to \Lambda$ (for some large cut-off $\Lambda$ on the size of the matrix), for which $R_{N\to \Lambda} \sim \Lambda$ and quantum fluctuations are frozen.
    \item An IR regime, for which $R_N$ formally vanishes when $N \ll 1$ and all fluctuations are integrated out.
    \item The IR fluctuations relative to the scale $2N$, i.e., such that $i+j \leq 2N$, acquire a relatively large effective mass (of the order of $2N$) and therefore decouple from the flow. Conversely, for the UV modes ($i+j > 2N$), the regulator vanishes, and they are simply integrated into the effective action.
\end{itemize} 
By construction, the regulator \eqref{regul} incorporates these limits, as well as the Zinn-Justin–Brézin prescription of partially integrating over a specific ordering of the matrix rows and columns. It is precisely this ordering, encoded in $R_N$, that explicitly breaks the $U(N)$ invariance of the initial model. In principle, constructing a renormalization group requires defining a preferred order in which fluctuations are integrated out, usually dictated by the spectrum of the free covariance matrix. For a matrix model strictly respecting $U(N)$ invariance, the free covariance is
$
C_{ij,kl}=N^{-1}\delta_{il}\delta_{jk},
$
which makes no distinction between fluctuations around the Gaussian point. Strict $U(N)$ invariance therefore does not provide an unambiguous notion of scale through a preferred ordering of fluctuations, and hence does not directly define UV and IR regimes.
Breaking the strict $U(N)$ invariance makes it possible to introduce such a notion of scale and to formulate an ordinary Wilsonian renormalization group. In the present context, this construction is further supported by the canonical power counting provided by the $1/N$ topological expansion of random matrices. In \cite{eichhorn2019status,eichhorn2013continuum,eichhorn2014towards,eichhorn2019towards,eichhorn2018flowing,geloun2018functional,castro2026quantitativecharacterizationgravitationaluniversality}, the authors explored various approximation schemes and models, both matrix and tensorial, based on this type of construction. Their results consistently suggest qualitative, and in some cases quantitative, agreement with theoretical expectations. In particular, they find evidence for an interacting fixed point with a relevant direction numerically compatible with the critical exponent $\theta=4/5$ predicted by double scaling.
However, this symmetry breaking is not necessarily benign. The reliability of the resulting flow in exploring regions of phase space beyond the reach of analytical results therefore remains, at least in part, an open question.

A first difficulty, pointed out in \cite{gurau2024quantum}, stems from the fact that the IR limit is, so to speak, “cut off” at zero index. Consequently, the existence of formal fixed points is, strictly speaking, meaningful only in the deep UV regime, $N\to\infty$. This limitation is particularly significant because, by construction, all canonical dimensions of the couplings, as determined by power counting, are negative (see the sign of the linear term in \eqref{oneloopRG}), with their magnitude increasing with the degree of the interaction.
The absence of a stable renormalizable surface introduces a further difficulty, since the existence of such a surface underlies many standard truncation schemes. This also raises questions about the reliability of the UV fixed points, as any finite truncation necessarily neglects potentially relevant contributions. Finally, constructing the flow in the mesoscopic regime, $1\ll N\ll\Lambda$, presents the additional technical challenge of incorporating non-planar contributions.

More seriously, this symmetry breaking is not necessarily without consequences. Indeed, although the regulator formally vanishes in the IR limit ($N\ll 1$), the restoration of the symmetry is not guaranteed. In practice, approximation schemes can induce a residual dependence on the regulator in the IR, potentially affecting the reliability of the resulting flow \cite{pawlowski2017physics}.

The anomalous dimension $\eta := N \, \dd \ln (Z_N)/\dd N$ is relatively large at the fixed point identified in these studies, raising questions about the convergence of the derivative expansion \cite{balog2019convergence}. In the present context, this issue manifests itself through the proliferation of interactions whose couplings depend explicitly on the matrix indices, such as
$
b_4(i,j,k,l) \sim i^a j^b k^c l^d
$
for a quartic interaction. Such interactions are naturally generated by the Ward identities modified by the presence of the regulator and are required for their consistency. Moreover, power counting assigns them canonical dimensions that are just as relevant as those of the connected-trace interactions of the initial model.
The fact that the flow inevitably leaves the trace sector has been investigated in \cite{Lahoche:2019ocf,Lahoche:2020pjo}. There, the authors showed that modifying the regulator to include a corrective term in the Ward identities, designed to cancel the first-order derivative interactions, improves the convergence of the flow. Nevertheless, the proliferation of index-dependent interactions is itself unavoidable, as it is intrinsically linked to the explicit breaking of $U(N)$ symmetry that is essential to the construction of the flow.

In this work, we propose a different approach. We introduce an auxiliary matrix field through a Hubbard–Stratonovich transformation \cite{Hubbard:1959ub}, which leaves the initial quartic model unchanged and preserves its effective $U(N)$ invariance. We then exploit concentration of measure in the large-$N$ limit to define a renormalization group on the space of singular values of the intermediate field.
This approach has the advantage of preserving $U(N)$ symmetry explicitly while addressing several of the difficulties associated with the previous construction. In particular, because the flow is no longer tied to a coarse-graining of the matrix size, the large-$N$ limit decouples from the construction of the flow equations. This removes the need to question the validity of the planar approximation at each stage of the flow. Furthermore, the method provides a natural definition of a genuine continuum (IR) limit.
We note that related approaches have recently been considered in several papers:

\begin{enumerate}
\item In \cite{Achitouv:2025ghy,Achitouv:2024kqh,Lahoche:2024huc,Natta:2024nke,Lahoche:2024qqq,Lahoche:2024puq,Lahoche:2024hox,Lahoche:2024gal,Lahoche_2024PFP}, the authors study glassy systems (classical and quantum) by constructing a coarse-graining on the disorder spectrum in the quenched limit.
\item In \cite{finotello2026datafieldtheorytheory,finotello2026fieldtheorydataanomaly2,Finotello_2026}, the authors construct an RG from a coarse-graining on the spectrum of the empirical data correlation matrix in a weak signal-to-noise regime.
\end{enumerate}

The use of a background field to define fluctuations, and subsequently construct a renormalization group, is not new. It is, in particular, a common strategy in non-perturbative approaches to the renormalization group of gravity \cite{reuter1998nonperturbative}, where the background metric provides a notion of scale through the spectrum of the Laplacian.
In our construction, however, the role of the background field is more fundamental: it fully determines the notion of scale in a relational manner. The fluctuations of the matrix field are defined relative to a background field that is itself a particular configuration, or vacuum, of the initial theory. This strategy therefore follows a relational tradition \cite{rovelli2004quantum}, in which spacetime is understood as emerging from relations between fields. A field configuration can, for instance, serve as a clock relative to certain physical processes, as illustrated by the cosmological background, whose state defines cosmological time \cite{rovelli1993statistical}.

\begin{remark}
In cases where symmetry is broken by the initial model rather than by the regulator, as is the case for TGFTs\footnote{In addition to a non-trivial propagator, these theories often include a geometric constraint ensuring the local flatness of the triangulated spaces generated by the perturbative theory. This constraint manifests as a projector in momentum space, which affects the power counting.}, the previous objections largely fall away, and a reliable renormalization group can be constructed. See \cite{carrozza2014tensorial,carrozza2014renormalization,carrozza2016flowing,carrozza2014renormalization,lahoche2015renormalization} for general work on this subject, \cite{benedetti2015functional,benedetti2016functional,geloun2015functional,geloun2016functional,geloun2018functional,marchetti2023mean,carrozza2017renormalizable,lahoche2017functional,carrozza2017asymptotic,lahoche2017renormalization,pithis2020phase} for non-perturbative applications, and \cite{lahoche2020renormalization,lahoche2019ward,lahoche2021no,lahoche2018nonperturbative,lahoche2019ward,lahoche2020pedagogical,lahoche2019ward,Lahoche:2025bmp} for studies explicitly investigating the impact of symmetry breaking by the kinetic term.
\end{remark}

\paragraph{Outline.} The paper is organized  as follows. In Section \ref{sec_rel}, we construct the relational background-field method by exploiting concentration of measure. We also discuss the notion of dimension, which is unusual in that it depends on the scale.
In Section \ref{FRG_sec}, we consider the Wetterich formalism and construct an approximate solution using a vertex expansion in the deep IR, where the theory resembles a three-dimensional non-local Euclidean theory. We demonstrate the existence of a Wilson–Fisher-type fixed point with a stable relevant direction, and discuss its critical exponent and physical interpretation.
In Section \ref{Sec_Ward}, we show how the Ward identities associated with gauge fixing allow the hierarchy of flow equations to close, in the deep IR, within the connected-trace sector. This result confirms the existence and stability of the fixed point.
Section \ref{info} is devoted to a discussion of the construction of the flow from the perspective of information geometry. Finally, the conclusion addresses certain limitations and perspectives of the method. Two appendices, \ref{App1} and \ref{App2}, summarize basic results useful for the reader, concerning respectively the spectrum of $U(N)$ random matrices and the addition of two random matrices.


\section{Relational background and the infrared limit}\label{sec_rel}

This first section lays the foundations of the formalism, supported by Appendices \ref{App1}, \ref{App2}, and \ref{App3}. The main result demonstrates how a random matrix model for a given Hermitian matrix $M$ can be decomposed into two fields, $M \to (\sigma, M)$, where the field $\sigma$ emerges in the large $N$ limit as a background upon which the remaining degrees of freedom of $M$ can be partially integrated out.

\subsection{Partial Hubbard-Stratonovich and large N limit}\label{sectionfree}

We consider a model of Hermitian random matrices whose partition function is given by \eqref{partfunc}. As a first step, we will focus on a quartic potential involving a single trace:
\begin{equation}
\mathcal{V}(M)=\frac{1}{2}\, M^2+\frac{a_4}{4}\, M^4\,.\label{Quarticini}
\end{equation}
Usually, the intermediate field formalism breaks the quartic interaction $M^4$ into a three-body interaction $\sigma M^2$, where $\sigma$ is a Gaussian intermediate matrix field with variance $1/N$. Here, we normalize this variance to $1$ by adding and subtracting a reference quartic interaction, so that the new effective potential including the intermediate field is written as:
\begin{equation}
\tilde{\mathcal{V}}(M,\sigma)=\frac{1}{2}\, \sigma^2+\frac{1}{2}\, M^2+\beta \sigma M^2+\frac{a_4+\alpha}{4}\, M^4\,.\label{quarticmodifies}
\end{equation}
The parameters $\alpha$ and $\beta$ are chosen so that partial integration over the intermediate field $\sigma$ yields exactly the original potential \eqref{Quarticini}. For a linearly interacting field, the result of the Gaussian integration amounts to substituting into \eqref{quarticmodifies} the value of $\sigma$ at the saddle point:
\begin{equation}
\frac{\partial \tilde{\mathcal{V}}}{\partial \sigma}= \sigma+\beta M^2\,,
\end{equation}
so that by substituting into \eqref{quarticmodifies}, we find:
\begin{equation}
\tilde{\mathcal{V}}(M,\sigma)\Big\vert_{\text{saddle}}=\frac{1}{2}\, M^2+\frac{a_4}{4}\, M^4+\left(\frac{\alpha}{4}-\frac{\beta^2}{2}\right)\,M^4\,,
\end{equation}
so that we recover the original potential under the condition $\alpha=2\beta^2$. By fixing $\alpha$ in such a way, the total partition function of the model can be written, setting $u:=a_4+2\beta^2$:
\begin{equation}
Z_N\propto \int \dd \sigma \dd M \, e^{-N \Tr \left(\frac{1}{2}\,\sigma^2+\frac{1}{2}(1+2\beta \sigma)\, M^2+\frac{u}{4}\,M^4\right)}\,.\label{defZN}
\end{equation}
The integration measures $\dd M$ and $\dd \sigma$ are here Lebesgue measures, and the boundary conditions of the integration domain are defined such that, for any asymptotically convex function $f[\sigma]$, the integral of a total derivative vanishes:
\begin{equation}
\int \dd \sigma \, \frac{\partial }{\partial \sigma_{ij}} e^{-\Tr f[\sigma]}=0\,.
\end{equation}
Explicitly, $\dd \sigma= \prod_i \dd \sigma_{ii}\times \prod_{i<j} \dd \Re \sigma_{ij} \dd \Im \sigma_{ij}$. Any Hermitian matrix can be decomposed in the form $\sigma = \mathbf{U} \Lambda \mathbf{U}^\dagger$, where $\mathbf{U}$ is a unitary matrix and $\Lambda$ is a diagonal matrix $\Lambda:= \mathrm{diag} (\lambda_1,\lambda_2,\cdots, \lambda_N)$. The change of variable $(\sigma) \to (\Lambda, \mathbf{U})$ (see Appendix \ref{App1}) introduces a Jacobian equal to the square of the Vandermonde determinant $\Delta(\Lambda)=\prod_{i<j} \, \vert \lambda_i-\lambda_j \vert$. The partition function can therefore also be written as:
\begin{equation}
Z_N=\int \dd \mathbf{U} \int \dd \Lambda  \int \dd M \, e^{-N \sum_{\lambda} \mathcal{L}_{\lambda}[\Lambda,M]}\,,\label{Zeigen}
\end{equation}
where $\dd \Lambda := \prod_{i=1}^N \dd \lambda_i$, $M_{\lambda\lambda^\prime}:=\sum_{i,j}\, \bar{u}_i^{(\lambda)} u_j^{(\lambda^\prime)} M_{ij}$, where $u_i^{(\lambda)}$ denotes the eigenvector associated with the eigenvalue  $\lambda$, $\dd \mathbf{U}$ is the Haar measure on unitary matrices ($\int \dd \mathbf{U}=1$), and:
\begin{equation}
\mathcal{L}_{\lambda}[\Lambda,M]:= \frac{1}{2}\,\lambda^2-\frac{1}{N}\sum_{\lambda^\prime \neq \lambda} \ln \vert \lambda-\lambda^\prime \vert+\frac{1}{2} (1+2\beta\lambda)\, (M^2)_{\lambda\lambda} +\frac{u}{4}\,(M^4)_{\lambda\lambda}\,.
\end{equation}
The unitary invariance of the integral linked to the cyclicity of traces in \eqref{Zeigen} and to the invariance of the measure $\dd (\mathbf{U} M \mathbf{U}^\dagger)=\dd M$ decouples the integral over $\dd \mathbf{U}$, and:
\begin{equation}
Z_N= \int \dd \Lambda  \int \dd M \, e^{-N \sum_{\lambda} \mathcal{L}_\lambda[\Lambda,M]}\,.\label{Zeigen2}
\end{equation}
The most probable configuration of eigenvalues is determined by the saddle point of the Lagrangian $\mathcal{L}_\lambda$. Following the method recalled in Appendix \ref{App1}, we will use the density formalism, and we therefore introduce a field $\varphi(x)$, continuous in the limit $N\to \infty$, defined by:
\begin{equation}
\varphi(x):=\frac{1}{N}\sum_{i=1}^N\, \delta(x-\lambda_i)\,.
\end{equation}
As recalled in the appendix, the change of variable $\{\Lambda\} \to \{\varphi(x) \}$ introduces an entropy term $e^{S[\varphi]}\sim \exp \left(-N \int \dd x \varphi(x) \ln \varphi(x)\right)$, and, in the continuous limit, the integral approaches an ordinary path integral:
\begin{equation}
Z_N\propto \int [\dd \varphi(x)] \, e^{-N^2\mathcal{L}_1[\varphi]} \times \int [\dd \psi(x,y)]\, e^{-\mathcal{L}_2[\varphi,\psi]}\,,\label{partitionfield}
\end{equation}
where the integration runs over the nearly continuous fields $\varphi(x)$ and $\psi(x,y):=N^{3/2} M_{xy}$ in a slightly abusive notation evoking a genuinely continuous field, and
\begin{align}
&\mathcal{L}_1[\varphi]:=\int \dd x \,\varphi(x) \frac{x^2}{2} - \fint \dd x \dd y \, \varphi(x) \varphi(y) \ln \vert x-y \vert + \frac{1}{N} \, \int \dd x \, \varphi(x) \ln \varphi(x)\,,\\
&\mathcal{L}_2[\varphi,\psi]:=\frac{1}{2}\int \dd x\, \varphi(x) (1+2 \beta x) \psi\star\psi(x,x)+\frac{u}{4N}\,\int \dd x \varphi(x) (\psi\star \psi \star \psi \star \psi)(x,x)\,,\label{equationL2}
\end{align}
where $\fint$ is the Cauchy principal value, and where we have introduced the product $\star$ by:
\begin{equation}\psi\star \phi (x,y):=\int \dd z\varphi(z)\, \psi(x,z) \phi(z,y)\,.
\end{equation}
At large $N$, the measures concentrate around the saddle point of the integral. In this limit, the eigenvalues exhibit very small fluctuations, with global fluctuations suppressed by powers of $1/N^2$, due to the balance between the confinement induced by the effective potential and the Coulomb repulsion arising from the integration measure. The spectrum therefore self-averages, with its limiting eigenvalue distribution determined by the zeros of a family of orthogonal polynomials \cite{potters2020first,di19952d,mehta2004random}.
Moreover, $\sigma$ is the free sum of a Wigner matrix $A$ and the random matrix $-\beta M^2$, both of which converge, in the large-$N$ limit, to deterministic spectral laws.


The rigidity of the asymptotic eigenvalue spectrum is further reflected in the characteristic polynomial. At the saddle point, one has
$
\mathbb{E}\left[\prod_{i=1}^N(z-\lambda_i)\right]
\simeq
\prod_{i=1}^N(z-\lambda_i^*),
$
where the $\lambda_i^*$ denote the most probable eigenvalues. More precisely, the equality becomes exact only in the strict large-$N$ limit, where the eigenvalue fluctuations become negligible. The family of orthogonal polynomials $p_n(x)$ whose zeros determine the asymptotic eigenvalue distribution of $\sigma$ satisfies the orthogonality relation:
\begin{equation}
\int \dd x\, p_n(x) p_m(x) e^{-N \mathcal{V}_{\text{eff}}(x)}=Z_n \delta_{mn}\,,\label{orthogonal}
\end{equation}
for a normalization factor $Z_n$ of no importance, and where $\mathcal{V}_{\text{eff}}(x)$ denotes the effective potential for an eigenvalue of $\sigma$, once the matrix $M$ has been integrated out. One of the major results of this approach is that the characteristic polynomial $Q(z):=\prod_{i=1}^N (z-\lambda_i)$ coincides, up to a factor, with the polynomial $p_N(z)$, whose zeros determine the position of the matrix eigenvalues.

\begin{remark}
Let us denote by $\varphi_0(x)$ the limiting saddle-point distribution (assuming it exists, see the following section) in the limit $N\to \infty$. The effective theory described by the Lagrangian $\mathcal{L}_2[\varphi_0,M]$ establishes a correspondence between an initial quartic matrix theory, say $\mathcal{M}_4$, and a non-local quartic field theory, denoted by $\mathcal{F}_4^\star(\beta)$, according to a certain state-dependent star product, such that:

\begin{enumerate}
\item The effective distribution $\varphi_0(x)$ is not independent of the eigenvalue distribution of $M^2$ (see the discussion below).
\item The initial theory admits a confinement of the eigenvalues of the matrix $M$ for $a_4\in [-1/12,+\infty)$. Below the critical value $a_{4c}=-1/12$, the potential well no longer confines an eigenvalue gas. Note that the Gaussian point $a_4=0$ of $\mathcal{M}_4$ is mapped onto an interacting model of $\mathcal{F}_4^\star(\beta)$, $u=2\beta^2$. Conversely, $u=0$ corresponds to a non-confining potential $a_4=-2\beta^2$, and specifically for the choice $\beta=1$, $a_4=-2<a_{4c}$, the spectrum of $M$ at large $N$ is no longer bounded (see Appendix \ref{App1}).
\end{enumerate}
\end{remark}

The relational background field method consists of partially integrating out the modes indexed by the eigenvalues of the effective free propagator $C:=N^{-1}(\mathbb{I}\otimes \mathbb{I}+\beta\Lambda\otimes \mathbb{I}+\beta\mathbb{I}\otimes \Lambda)^{-1}$ of the theory $\mathcal{F}_4^\star(\beta)$. Here, the parameter $\beta$ acts as a slider that controls how much of the original theory's interaction structure is transferred to the intermediate field. In particular, the choice $\alpha=1/12\to \beta=1/\sqrt{24}$ maps the critical point $a_4=-1/12$ to the Gaussian point $u=0$. At this value, the field $\psi(x,y)$ becomes free, while the entire non-linear interaction structure is transferred to the background field $\varphi_0(x)$. Once the residual Gaussian field is integrated out, the resulting theory resembles the Penner model \cite{penner1988perturbative}. From the perspective of the strategy developed in this article, however, this choice is of limited interest for the generic flow: because $\psi(x,y)$ no longer interacts with itself, it generates no nontrivial renormalization-group flow. It will nevertheless prove useful in Section \ref{secCritical} to analyze the critical behavior. More precisely, the flow reduces to purely dimensional scaling.

Let us state the analogy once more: the parallel with the background field method in Einstein gravity is now clear. In this approach, the metric (the gravitational field) $g_{\mu\nu}=g_{\mu\nu}^{(0)}+\eta_{\mu\nu}$ is split into two parts: a background field $g_{\mu\nu}^{(0)}$, which determines the geometric objects (connection and curvature), as well as the spectrum of the Laplacian $\Delta:=\sqrt{\vert g_0 \vert^{-1}} \partial_\mu (\sqrt{\vert g_0 \vert} g_0^{\mu\nu} \partial_\nu)$, on which the coarse-graining of the fluctuating field $\eta_{\mu\nu}$ is based. The splitting is partly arbitrary (much like the choice of $\beta$), but it resolves a central obstacle to the quantization of general relativity: the need to preserve diffeomorphism invariance, since gauge symmetry is intimately linked to the dynamics of spacetime itself. The background field method notably allows for gauge fixing while maintaining manifest geometric invariance throughout the loop calculations, with the consistency of the construction ultimately ensured by the Ward identities. The relational background field $\sigma$ plays a role analogous to that of the background metric $g_{\mu\nu}^{(0)}$: it is a non-trivial configuration of the gravitational field that provides the geometric support for the fluctuations of the field $\psi(x,y)$. In summary, we have the following heuristic correspondence:
\begin{align}
g_{\mu\nu}^{(0)} &\longleftrightarrow \sigma\,,\\
\eta_{\mu\nu} &\longleftrightarrow  \psi\,.
\end{align}
Let us also clarify the meaning of the continuum limit in the present context. All quantities, such as Feynman amplitudes, as well as the effective loop functions that we will compute later, are first evaluated within the discrete theory. The transition to the continuum limit is performed only at the very end of the calculation, most notably by replacing the effective distribution $\varphi(x)$ with its saddle-point solution. We will illustrate this procedure explicitly in the context of perturbation theory in Section \ref{secscaling}.

Let us mention here that the propagator $C:=N^{-1}(\mathbb{I}\otimes \mathbb{I}+\beta\Lambda\otimes \mathbb{I}+\beta\mathbb{I}\otimes \Lambda)^{-1}$ of the $\mathcal{F}_4^\star$ theory\footnote{Note that $\sum_{i,j} (1+2\beta\lambda_i) M_{ij}M_{ji}=\sum_{ij}(1+\beta\lambda_i+\beta\lambda_j) M_{ij}M_{ji}$.}, is written in its discrete version (for the rescaled field $\psi = N^{3/2}M$) as:
\begin{equation}
\langle \psi(x,y) \psi(x^\prime,y^\prime) \rangle_{0,\text{dis}} = N^2\frac{\delta_{x y^\prime}\delta_{x^\prime y}}{1+\beta x+\beta y}\,,
\end{equation}
where $\langle X \rangle_{0,\text{dis}}$ is the average over the free distribution, without interactions. 

\subsection{Free addition and IR limit}

The relational background field method relies on three hypotheses:
\begin{enumerate}
\item The fluctuations of the spectrum $\varphi(x)$ tend to zero in the limit $N\to \infty$ and the background field converges to a deterministic field $\varphi_0(x)$.
\item The deterministic spectrum $\varphi_0(x)$ is confined and vanishes outside a region $[x_-,x_+]$ (the edges defining the IR ($x_-$) and UV ($x_+$) regions, respectively).
\item At least in the IR regime (around $x_-$), the background field essentially decouples from the field $\psi(x,y)$, and follows a universal power law $\varphi_0(x) \sim (x-x_{-})^\theta$, with $\theta=1/2$.
\end{enumerate}
In this section, we will essentially prove points 2) and 3), assuming point 1) to be true, the latter following from the general theory of random matrices, as mentioned in the previous section (but, point 2), indeed, implies point 1)). More precisely, we will show the following result:

\begin{claim}
In the regime where the potential still confines the eigenvalue gas ($a_4\geq -1/12$), there always exists a finite interval $\beta\in (0, \beta_c)$, $\beta_c >0$ such that $\varphi_0(x) \sim (x-x_{-})^{1/2}$ around the smallest eigenvalue $x_-$.
\end{claim}

The proof we propose fundamentally relies on Voiculescu's free probability theory \cite{voiculescu1992free,potters2020first}, which generalizes the notion of independence in ordinary probability theory to the non-commutative setting. We recall some of its basic concepts here. Readers unfamiliar with free probability theory may instead approach the essential results of this section through the more elementary method reviewed in Appendix \ref{App2}.
\begin{definition}\label{definitonfree}
Two random variables $A$ and $B$ are free if, for a certain positive and linear functional $\tau(.)$ and any family of polynomials $p_1(A),\cdots, p_n(A)$ and $q_1(B),\cdots, q_n(B)$ such that $\tau(p_k(A))=0$ and $\tau(q_k(B))=0$, then, for any alternating polynomial, the following condition holds:
\begin{equation}
\tau (p_1(A) q_1(B) p_2(A) q_2(B)\cdots p_l(A) q_l(B))=0\,.
\end{equation}
\end{definition}Note that the function $\tau(.)$ defines the moments of the distribution, $m_k(A):=\tau(A^k)$, whose generating functional is the Stieltjes transform of $A$, and for sufficiently large $\vert z \vert$:
\begin{equation}
g_A(z)=\sum_{k=0}^\infty \, \frac{m_k(A)}{z^{k+1}}\,.\label{gAdef}
\end{equation}
An alternative definition of "freeness" is that the mixed free cumulants vanish, implying in particular that for two free random variables $A$ and $B$, with $n$-point cumulants $\kappa_n(A)$ and $\kappa_n(B)$, $\kappa_n(A+B)=\kappa_n(A)+\kappa_n(B)$, $\forall\, n$. The generating functional of the cumulants $R_{A}:=\sum_{n=1}^\infty\, \kappa_n(A) g^{n-1}$ (the $R$-transform of $A$) thus follows a simple addition law in the case of free variables:
\begin{equation}
R_{A+B}(g)=R_A(g)+R_B(g)\,.\label{Rprop}
\end{equation}
Note that free cumulants are not related to moments by the same relations as in ordinary probability. Thus, for a centered random variable $A$, the free cumulants satisfy $\tau(A^4)=\kappa_4(A)+2 \kappa_2^2(A)$ and not $\tau(A^4)=\kappa_4(A)+3 \kappa_2^2(A)$, as would be the case for an ordinary random variable. Random matrices generally constitute an example of free variables in the limit $N\to \infty$. Let us give an elementary example here:

\begin{example}
Let $A$ and $B$ be two Hermitian Gaussian random matrices, with variance $1/N$ and zero mean. In the case of random matrices, $\tau(.):=\frac{1}{N} \mathbb{E} \Tr(.)$ \cite{potters2020first}, and by definition $\tau(A)=\tau(B)=0$. According to Definition \ref{definitonfree}, the polynomials here are thus the linear functions $p_1(A)=A$ and $q_1(B)=B$. According to Definition \ref{definitonfree}, for $A$ and $B$ to be free, we must have $\tau(ABAB)=0$ for example. It is easy to see that this is indeed the case in the limit $N\to \infty$ by applying Wick's theorem. We have $\mathbb{E}(A_{ij} A_{kl})=N^{-1} \delta_{il} \delta_{jk}$, so that the average $\mathbb{E} \Tr(ABAB)=N^{-2} \sum_i \delta_{ii}=N^{-1}$ (the contraction creates only one cycle). In the limit $N\to \infty$, we therefore have $\Tr(ABAB)\to 0$. Conversely, $\tau(A^2 B^2)$ creates three independent cycles, compensating for the power $N^{-3}$ coming from the definition of $\tau$ and the two free propagators involved in the Wick contraction. More generally, $\tau(A^2 B^2)=\tau(A^2)\tau(B^2)$. Note that $\tau(A^4)=2(\tau(A^2))^2$, $2$ and not $3$, because while there are indeed three Wick contractions, only two are planar and generate three cycles, the third term being "suppressed" in the limit $N\to \infty$.
\end{example}
Note also that, for matrices, the usual definition of the Stieltjes transform is:
\begin{equation}
g_A(z):=\frac{1}{N} \mathbb{E} \Tr (z \mathbb{I}-A)^{-1}\,.
\end{equation}
In light of these definitions, it is easy to see that in the limit $N\to \infty$, the intermediate field $\sigma$ is the free sum
\begin{equation}
\sigma=A-\beta M^2\,,
\end{equation}
where $A$ is a Hermitian Gaussian Wigner matrix of size $N\times N$ and variance $1/N$. This property effectively decouples the two matrices by construction. Indeed, if we consider a random matrix $A$ with distribution $Q(A):=e^{-\frac{N}{2}\Tr A^2}/Z_A$, the latter is independent of the random matrix $M$, with distribution $P(M):=e^{-N\Tr \mathcal{V}(M)}/Z_N$. The matrices $A$ and $M$ are therefore free in the limit $N\to \infty$, just like\footnote{By construction $\mathbb{E}[\Tr A]=0$; however, $\tau(B)=N^{-1} \Tr \mathbb{E}(B)=:N^{-1}\Tr \bar{B}\neq 0$. On the other hand, we do have $\tau (B-\tau(B) \mathbb{I})=0$, and it is easy to see that, for example, $\lim_{N\to \infty}\tau (A(B-\tau(B)\mathbb{I}) A (B-\tau(B)\mathbb{I}))=0$, where $\mathbb{I}$ is the identity matrix.} $A$ and $B:=M^2$. The freeness of the two matrices follows from the same diagrammatic argument given in the previous example\footnote{A more advanced argument relies on the fact that the matrices $A$ and $M$ differ by a random unitary transformation uniformly distributed according to the Haar measure on the group. The moments of these unitary transformations, the Weingarten functions, cancel out terms involving alternating sequences of polynomials whose $\tau$-average is zero.}: For $\tau(AMAM)$, and assuming $\tau(A)=\tau(M)=0$, Wick's theorem yields $\tau(AMAM)=\mathbb{E} [(N^{-1} \Tr M)^2]\to 0$. Once this observation is made, we can \textit{define} $\sigma$ as the free sum $\sigma=A-\beta M^2$, and deduce the distribution $P^\prime(\sigma)$ from $Q(A)$ and $P(M)$.
\begin{align}
\nonumber P^\prime(\sigma)&=\int \dd M \dd A\, Q(A) P(M) \delta (\sigma-A+\beta M^2) = \int \dd M \, Q(\sigma+ \beta M^2) P(M)\\
&\propto \int \dd M\, e^{-\frac{N}{2}\Tr \sigma^2-N \beta \Tr \sigma M^2-\frac{N \beta^2}{2} \Tr M^4-N\Tr \mathcal{V}(M)}\,,
\end{align}
the normalization being precisely the partition function \eqref{defZN}, provided that $\mathcal{V}(M)=M^2/2+(u-2\beta^2)M^4/4$. Defining $\mathfrak{z}_A(g)$ as the inverse function of $g_A(z)$: $g_A(\mathfrak{z}_A(g))=g$, it is possible to show that $R_A(g)=\mathfrak{z}_A(g)-g^{-1}$. The relation \eqref{Rprop} can also be written as $\mathfrak{z}_{A+B}(g)=\mathfrak{z}_A(g)+\mathfrak{z}_B(g)-g^{-1}$, which is to say (see Appendix \ref{App2}):
\begin{equation}
g_{A+B}(z)=g_{A}(z-R_B(g_{A+B}(z)))\,.
\end{equation}
Applying this relation to the case $\sigma=A-\beta M^2$, we have $R_\sigma(g)=R_A(g)-\beta R_{M^2}(-\beta g)$, having used the relation $R_{a A}(g)=a R_A(a g)$, and thus:
\begin{equation}
g_{\sigma}(z)=g_{-\beta M^2}(z-R_A(g_{\sigma}(z)))\,.\label{subordination}
\end{equation}
The composition of this equation with the inverse function $\mathfrak{z}_\sigma(g)$ leads to the equation:
\begin{equation}
g=g_{-\beta M^2}(\mathfrak{z}_\sigma(g)-R_A(g))\,.
\end{equation}
Let us now note that for a Wigner matrix of variance $1/N$, $R_A(g)=g$ (see Appendix \ref{App2}). Differentiating both sides with respect to $g$, we have:
\begin{equation}
1=g_{-\beta M^2}^\prime(\mathfrak{z}_\sigma(g)-g) \left(\mathfrak{z}_\sigma^\prime(g)-1\right)\,.\label{equationzprime}
\end{equation}
Assuming that its support is compact along the real axis, the spectrum of the matrix $\sigma$ forms, in the limit $N\to \infty$, a branch cut of the complex function $g_\sigma(z)$. At a spectral edge $z_*$, the Stieltjes transform generally remains finite, $g_\sigma(z_*)=g_*$, while its derivative becomes singular, $\vert g_\sigma^\prime(z\to z_*) \vert \to\infty$. Since $\mathfrak{z}_\sigma(g)$ is the inverse function of $g_\sigma(z)$, this divergence implies $\mathfrak{z}_\sigma^\prime(g=g_*)=0$. Applying this criticality condition to the previous equation then gives:
\begin{equation}
g_{-\beta M^2}^\prime(z_*-g_*)=-1\,.\label{equationboundary}
\end{equation}

\begin{example}
For a hermitian Wigner matrix with variance $1/N$, the Stieltjes transform is given by:
\begin{equation}
g(z)=\frac{z-\sqrt{z^2-4}}{2}\,.
\end{equation}
The spectral density $\mu(x)$ is determined by the Sokhotski–Plemelj formula $\lim_{\epsilon\to 0} \Im \, g(x-i \epsilon)=\pi \mu(x)$. Its position coincides with the cut $[-2,2]$ along the real axis, where the square root is imaginary. Furthermore, $g^{\prime}(z) \sim \frac{1}{\sqrt{z^2-4}}$, and $\vert g^{\prime}(z\to \pm 2) \vert \to \infty$.
\end{example}

Inside the support of the spectrum, the Stieltjes transform acquires an imaginary part, fixing the definition of the eigenvalue distribution $\lim_{\epsilon\to 0} \Im \, g_{-\beta M^2}(x-i \epsilon)=\pi\, \mu_{-\beta M^2}(x)$. Equation \eqref{equationboundary} can therefore only have a solution outside the support of the spectrum $\mu_{-\beta M^2}(x)$. Outside this spectral support, the Stieltjes transform is explicitly written as:
\begin{equation}
g_{-\beta M^2}(z)=\int \dd x \,\frac{\mu_{-\beta M^2}(x)}{z-x}\,,
\end{equation}
and then:
\begin{equation}
g_{-\beta M^2}^\prime(z)=-\int \dd x \,\frac{\mu_{-\beta M^2}(x)}{(z-x)^2}\,. 
\end{equation}
It is easy to relate the distributions and Stieltjes transforms of $-\beta M^2$ to those of $M$. Given that the law of the matrix $M$ is symmetric $\mathcal{V}(M)=\mathcal{V}(-M)$, we have:
\begin{equation}
\mu_{-\beta M^2}(x)=\frac{\mu_M(\sqrt{-x/\beta})}{\sqrt{-\beta x}}\,,\label{muM2}
\end{equation}
and, because the potential of the random matrix $M$ is symmetric:
\begin{equation}
g_{-\beta M^2}(z)=-\frac{1}{\sqrt{-\beta z}} \, g_M(\sqrt{-z/\beta}\,)\,. 
\end{equation}

\begin{remark}
To derive the previous formula, we used the identity $g_M(z)=-g_M(-z)$, which holds for a random matrix with a symmetric potential. Indeed, according to the definition of the Stieltjes transform given above, equation \eqref{gAdef}, and for a symmetric potential, all odd moments vanish. Let us also note that in the formula given in the previous example, $z$ is assumed to be such that $\Re(z)>0$. In general, the solution for the Gaussian case is written as:
\begin{equation}
g(z)=\frac{z-\mathrm{sign}(\Re(z))\sqrt{z^2-4}}{2}\,,
\end{equation}
which is indeed odd.
\end{remark}

The explicit form of the distribution $\mu_M(x)$ can be calculated using standard tools from random matrix theory, and a derivation is reviewed in appendix \ref{App1}:
\begin{equation}
\mu_M(x)=\frac{\sqrt{(b-x) (b+x)} \left(a_4 \left(b^2+2 x^2\right)+2\right)}{4\pi }\,,\label{densityM4bis}
\end{equation}
with
\begin{equation}
b^2=\frac{2}{3} \frac{\sqrt{12 a_4+1}-1}{a_4}\,.
\end{equation}
\begin{figure}
\begin{center}
\includegraphics[scale=0.7]{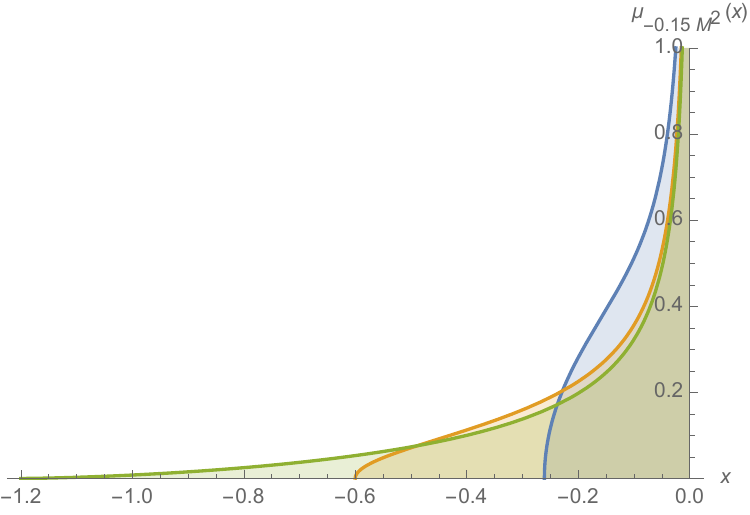}
\end{center}
\caption{Profile of the distribution $\mu_{-\beta M^2}(x)$ for $\beta=0.15$ and $a_4=1$ (blue curve), $a_4=0$ (yellow curve), $a_4=-1/12$ (green curve).}\label{spectrummoinsbetaM2}
\end{figure}
We can analyze the existence of solutions to equation \eqref{equationboundary} by distinguishing between two possibilities\footnote{Below the critical point $a_4 < -1/12$, the potential no longer confines the eigenvalues, and the free sum is no longer well-defined.}:

\begin{enumerate}
\item The matrix $M$ is not at the critical point $a_4 > -1/12$, so that according to \eqref{muM2}, $-\beta M^2$ has an edge at zero where it behaves as $1/\sqrt{-x}$ and a negative edge at $x=-\beta b^2$, where it behaves as a square root $\sqrt{\beta b^2+x}$ (Fig. \ref{spectrummoinsbetaM2}).
\item The matrix $M$ is at the critical point, the spectrum of the matrix $-\beta M^2$ thus retains its singular edge $1/\sqrt{-x}$ at zero, but exhibits an edge behaving as $(8\beta+x)^{3/2}$ at $-8\beta$ (Fig. \ref{spectrummoinsbetaM2}).
\end{enumerate}

\begin{figure}
\begin{center}
\includegraphics[scale=0.5]{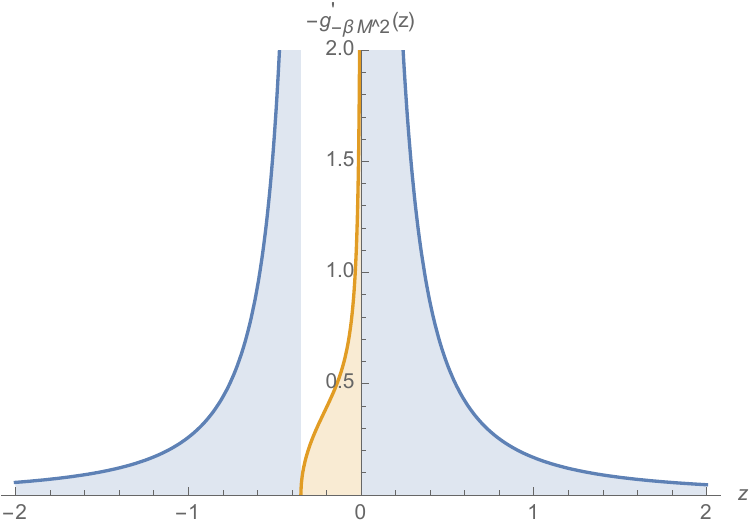}\quad \includegraphics[scale=0.5]{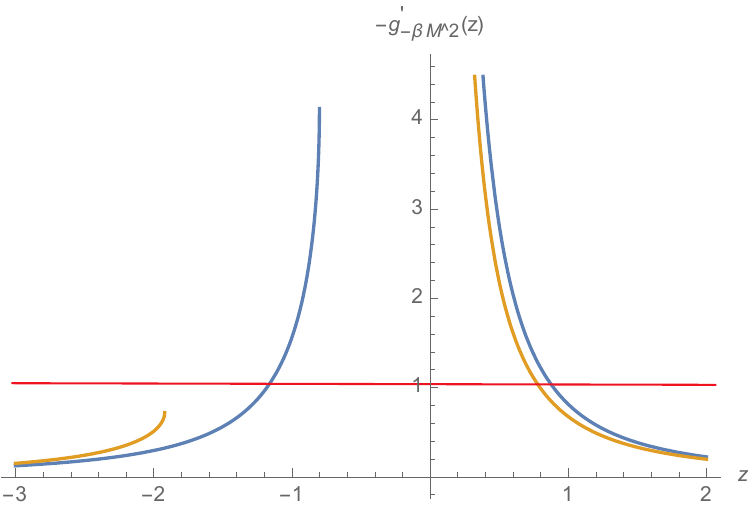}
\end{center}
\caption{On the left: In blue, typical behavior of $-g_{-\beta M^2}^\prime(z)$ for $\beta=1/5$ and $a_4=1$. In yellow, the corresponding distribution $\mu_{-\beta M^2}(x)$. The equation $-g_{-\beta M^2}^\prime(w_*)=1$ has a solution on both the negative and positive sides, beyond the boundaries of $\mu_{-\beta M^2}(x)$. On the right: Behavior of $-g_{-\beta M^2}^\prime(z)$ at the critical point $a_4=-1/12$, for $\beta=0.1$ (blue curve) and $\beta=0.24$ (yellow curve). There is no longer a solution to the equation $-g_{-\beta M^2}^\prime(w_*)=1$ on the left in the second case.}\label{figMoutcrit}
\end{figure}

In the first case, it is easy to see that the equation $g_{-\beta M^2}^\prime(w_*)=-1$ necessarily has a solution in both the positive and negative parts. Indeed, $g_{-\beta M^2}^\prime(z\to \pm \infty) = 0^-$. At the edge $z=0$, the distribution is singular, and $g_{-\beta M^2}^\prime(z\to 0^+) \to -\infty$; and at the edge $z=-\beta b^2$, because $\sqrt{\beta b^2+x}/(\beta b^2+x)^2 \sim 1/(\beta b^2+x)^{3/2}$, we again have $g_{-\beta M^2}^\prime(z\to (-\beta b^2)^-) \to -\infty$. Since the function is continuous, the intermediate value theorem implies that the function necessarily passes at least once through $-1$ on the left and on the right. Figure \ref{figMoutcrit} (left) illustrates this for the choice $\beta=1/5$ and $a_4=1$. It is easy to see that the existence of a solution implies that at the edge of the spectrum $z_*$ in question (left or right), the function $g_{\sigma}(z)$ behaves as $g_{\sigma}(z)\sim \sqrt{\vert z-z_* \vert }$. Indeed, according to \eqref{equationzprime},
\begin{equation}
\mathfrak{z}^\prime(g)=\frac{1}{g_{-\beta M^2}^\prime(\mathfrak{z}(g)-g)}+1\,,
\end{equation}
and, taking the derivative with respect to $g$:
\begin{equation}
\boxed{\mathfrak{z}^{\prime\prime}(g_*)= g_{-\beta M^2}^{\prime\prime}(z_*-g_*)\,.} \label{conditionzsecond}
\end{equation}

\begin{figure}
\begin{center}
\includegraphics[scale=0.5]{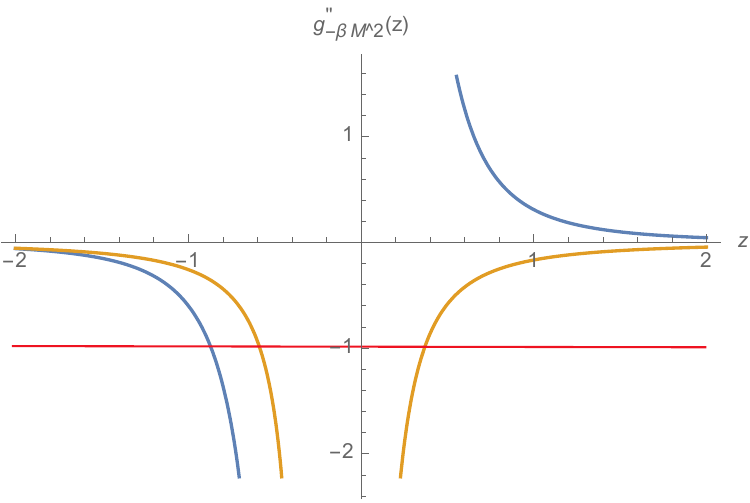}\quad \includegraphics[scale=0.5]{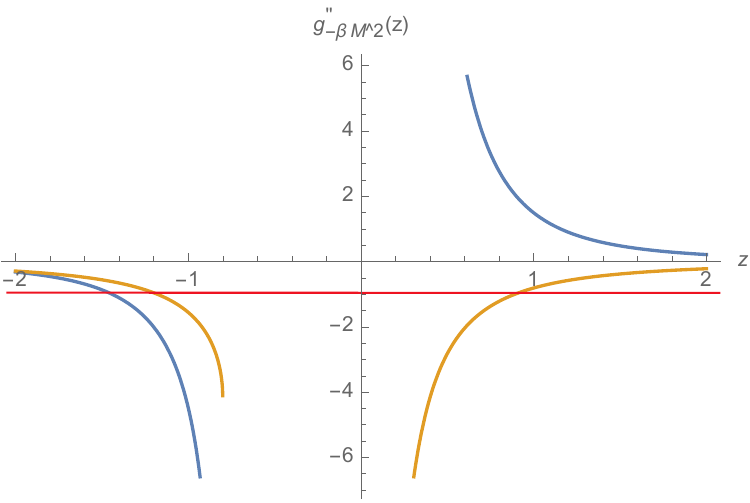}
\end{center}
\caption{On the left: Typical behavior of the second derivative $g_{-\beta M^2}^{\prime\prime}(z)$ (in blue) versus the first derivative (in yellow), outside the critical domain ($a_4=1, \beta=1/5$). On the right: Behavior of the same functions at the critical point for $\beta=0.1$.}\label{secondderiv}
\end{figure}

We have just seen that, above the critical point, there always exists a solution to the equation $g_{-\beta M^2}^\prime(w_*)=-1$, and that this solution lies outside the support of the spectrum of $-\beta M^2$. The second derivative $g_{-\beta M^2}^{\prime\prime}(w_*)$ is therefore finite at this point (Fig. \ref{secondderiv}). Consequently, in a neighborhood of $z_*\equiv x_-$, we have:
\begin{equation}
z=x_-+\frac{\dd \mathfrak{z}(g)}{\dd g}\Big\vert_{g=g_*}(g-g_*)+\frac{1}{2} \frac{\dd^2 \mathfrak{z}(g)}{\dd g^2}\Big\vert_{g=g_*}(g-g_*)^2+ \mathcal{O}((g-g_*)^3)\,.\label{equationexpansion}
\end{equation}
The second term being zero by definition, we have, taking into account that $g_{-\beta M^2}^{\prime\prime}(z_*-g_*) < 0$,
\begin{equation}
g_\sigma(z)\simeq g_\sigma(z_*)+\sqrt{\frac{2}{\vert g_{-\beta M^2}^{\prime\prime}(z_*-g_*) \vert}  (x_- -z)}\,.
\end{equation}
Taking into account the Sokhotski–Plemelj formula, we deduce that the spectral density exhibits a square-root behavior at the edge, $\varphi_0(x)\sim (x-x_-)^{1/2}$. Let us now consider the case in which $M$ is at the critical point, $a_4=-1/12$. At this precise point,
\begin{equation}
g^\prime_{-\beta M^2}(z)=-\frac{z \left(-\sqrt{\frac{8 \beta }{z}+1}\right)+4 \beta  \sqrt{\frac{8 \beta }{z}+1}+z}{24 \beta ^2 z}\,.
\end{equation}

\begin{figure}
\begin{center}
\includegraphics[scale=0.5]{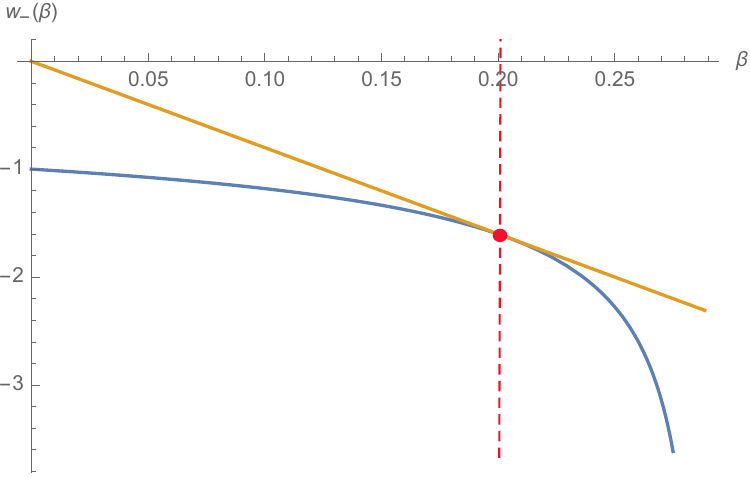}\qquad \includegraphics[scale=0.5]{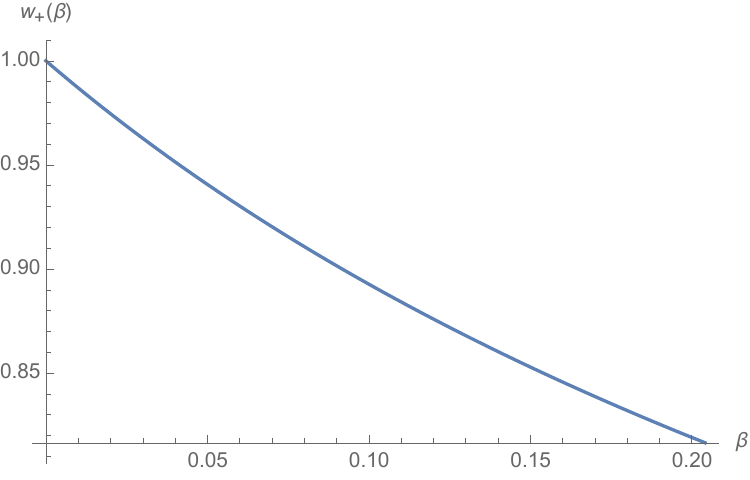}
\end{center}
\caption{On the left: behavior of $w_-(\beta)$ (blue curve) and $-8\beta$ (yellow curve). On the right: behavior of $w_+(\beta)$.}\label{solw}
\end{figure}

\begin{figure}
\begin{center}
\includegraphics[scale=0.6]{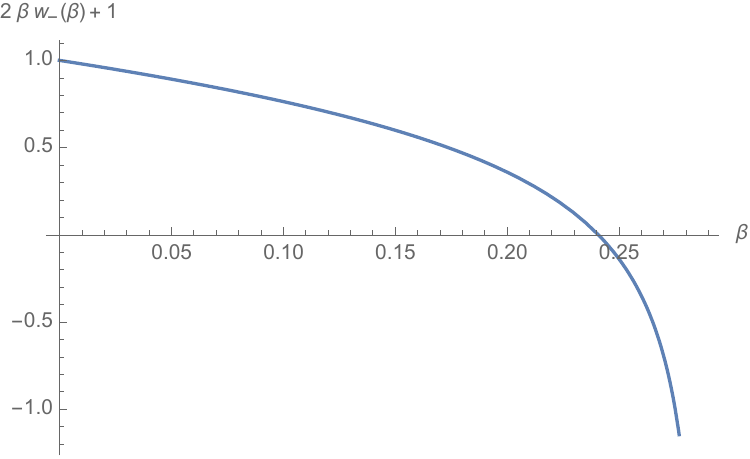}
\end{center}
\caption{Behavior of the gap $1+2\beta w_-(\beta)$ at the critical point $a_4=-1/12$.}\label{singpropa}
\end{figure}

The solution $w(\beta)$ to the equation $g^\prime_{-\beta M^2}(w)=-1$ has two branches, a positive branch $w_+(\beta)$ and a negative branch $w_-(\beta)$, as shown in Figure \ref{solw}. Note that $w_{\pm}(\beta=0)=\pm 1$: when $\beta=0$, $\sigma=A$, and its spectrum is given by a semicircle supported on the interval $[-2,2]$ (i.e. $w_\pm(0) = \pm 1$). As $\beta$ increases, the spectrum shifts to the left, until it reaches the apparent critical value $\beta_0=1/\sqrt{12}\approx 0.29$, beyond which the solution becomes imaginary. However, the true critical value of $\beta$ is reached earlier, when $w_-(\beta)=-8\beta$, that is, when the solution touches the critical edge of the spectrum of $M$. This occurs at the value (see also subsection \ref{secCritical}):
\begin{equation}
\beta_c:=\frac{1}{2\sqrt{6}}\approx 0.20\,.
\end{equation}
Note that $\beta_c=1/(2\sqrt6)=1/\sqrt{24}$ coincides exactly with the value of $\beta$ which maps the critical point $a_4=-1/12$ onto the Gaussian point $u=0$; this choice is exploited in Section \ref{secCritical}.
Below this value, the negative edge lies beyond the edge of the spectrum of $M$, so the previous argument remains valid: once again, $\varphi_0(x)\sim (x-x_-)^{1/2}$. Let us also note that, throughout the domain $(0,\beta_c)$, $1+2\beta x>0$, so that the propagator never becomes singular, as illustrated in Figure \ref{singpropa} at the critical point. Moreover, the condition $\beta =\beta_c$ implies $a_4+2\beta^2>0$ as soon as $a_4>-1/12$. Furthermore, since $a_4 =-1/12$, we have $u < 0$ at the critical point.\footnote{Note that at the critical point $\beta_c = 1/2\sqrt{6}$, the second derivative in \eqref{conditionzsecond} becomes singular because $g_{-\beta M^2}(z)$ behaves as $(8\beta+z)^{3/2}$. Consequently, the second derivative of $\mathfrak{z}$ is also singular, and the square-root behavior is lost.}

\begin{remark}
Note that the condition $\beta < \beta_c$ does more than guarantee the universal square-root behavior at the spectral edge. From a "topological" point of view, it also ensures that the system remains in a single-cut (1-cut) phase. Indeed, the equation $g_{-\beta M^2}^\prime(w_*) = -1$ implies that, as long as $\beta < \beta_c$, the spectral edge of the free sum remains separated from the spectral edge of the matrix $-\beta M^2$. This separation ensures that the edge remains regular and, in particular, supports the assumption that the measure concentrates on a unique background configuration $\varphi_0(x)$.
\end{remark}

\subsection{Renormalization, scaling and dimensions}\label{secscaling}

In this section, we focus on perturbation theory in the regime $\vert u \vert \ll 1$. This requires $a_4$ to be arbitrarily close to $-2\beta^2$, while maintaining the condition $a_4\geq -1/12$. Let us also note that, from this point onward, the symbol $\Lambda$ will be reserved for the ultraviolet cut-off.

For an ordinary field theory, the Wilsonian approach to the renormalization group consists in partially integrating out the "high" modes above a given cut-off and studying the resulting effective theory of the "low" modes, whose coupling constants are modified by the integration over the high modes. For an ordinary massive $\Phi_D^4(x)$ theory in $D$ dimensions, let us introduce an arbitrary UV cut-off $\Lambda_0$ and a lower effective cut-off $\Lambda < \Lambda_0$. The construction can then be summarized schematically as follows. We decompose the field into high modes $\Phi_>$, whose fluctuation scales (momenta in Fourier space) lie in the interval $[\Lambda, \Lambda_0]$, and low modes $\Phi_<$, corresponding to fluctuations in the complementary region $[0, \Lambda]$. By construction, $(2\pi)^{-D} \int \text{d}p\, \Phi_<(p) \Phi_>(p) = 0$, so that the Gaussian kinetic action $S_{\text{kin}}$ decomposes into a "high" part and a "low" part:
\begin{align}
\nonumber Z&=\int \, [\dd \Phi ]\, e^{-S_{\text{kin}}[\Phi_<]-S_{\text{kin}}[\Phi_>]-\frac{g}{4!}\int \dd x\, (\Phi_<(x)+\Phi_>(x))^4}\\
&=\int \, [\dd \Phi_< ]\, e^{-S_{\text{kin}}[\Phi_<]-\frac{g}{4!}\int \dd x \Phi_<^4(x) -S_{\text{eff}}[\Phi_<]}\,,
\end{align}
where the effective interaction $S_{\text{eff}}[\Phi_<]$ arises from partially integrating out the high fluctuations:
\begin{equation}
e^{-S_{\text{eff}}[\Phi_<]}:= \int \, [\dd \Phi_> ] \, e^{-S_{\text{kin}}[\Phi_>]-\frac{g}{4!} \int \dd x \, (\Phi_>^4(x)+4\Phi_>^3(x) \Phi_<(x)+4 \Phi_<^3(x) \Phi_>(x)+6 \Phi_<^2(x) \Phi_>^2(x) )} \,.
\end{equation}
We will apply this same strategy here to the spectrum defined by $\varphi_0(x)$. Let us begin by defining the notion of generalized momentum. First, by rescaling $\psi(x,y) \to \beta^{-1/2} \psi(x,y)$, the coupling constant becomes $\beta^{-2} u \to u$ and the propagator becomes $C:=N^{-1}(\beta^{-1}\mathbb{I}\otimes \mathbb{I}+\sigma\otimes \mathbb{I}+\mathbb{I}\otimes \sigma)^{-1}$. Let $K$ be a bounded linear operator between two normed spaces $K: U\to V$. The \textit{operator norm} is defined by $\Vert K \Vert_{\text{op}}:=\sup_{v\in U, \Vert v \Vert_U=1} \,\Vert K v \Vert_V$, and for an $N\times N$ Hermitian matrix, $\Vert K \Vert_{\text{op}}=\max_{1\leq i \leq N} \vert \lambda_i \vert$, i.e., the size of the spectral radius. Since the spectrum of the matrix $\sigma$ is shifted to the left (see also Appendix \ref{App3}), the norm actually corresponds to the absolute value of the smallest eigenvalue\footnote{For $N$ large enough, i.e., as long as fluctuations can be neglected, the probability that the spectral radius is given by the smallest eigenvalue tends to $1$ as $N \to \infty$.}. We then define the generalized momentum $p_\mu(\sigma)>0$ by:

\begin{equation}
\boxed{p_\mu(\sigma)=\lambda_\mu + \Vert \sigma \Vert_{\text{op}}\,,}
\end{equation}

so that the effective bare propagator $C(p_1,p_2)$ is given by:

\begin{equation}
C(p_1,p_2):=N^2 \, \frac{1}{p_1+p_2+m}\,,
\end{equation}

with:

\begin{equation}
m:=\beta^{-1}-2 \Vert \sigma \Vert_{\text{op}}\,.
\end{equation}

We define by $\rho(p):=\varphi(p-\Vert \sigma \Vert_{\text{op}})$ the empirical distribution of generalized momenta, and by $\rho_S(p):=\chi(S \leq p \leq \Lambda)\, \rho(p)$ the reduced spectrum on the interval\footnote{Here $\chi(A \leq x \leq B)$ is the characteristic function, equal to $1$ on the interval $[A,B]$ and zero elsewhere.} $[S, \Lambda]$. Here, $\Lambda$ denotes the spectral edge of $\rho(p)$:
\begin{equation}
\Lambda=\Vert \sigma \Vert_{\text{op}}+\Vert \sigma \Vert_{\text{op}}^{(+)}\,,\label{lambdadef}
\end{equation}
where $\Vert \sigma \Vert_{\text{op}}^{(+)}$ denotes the norm of the positive-definite matrix $(\sigma^{(+)})_{ij}=\sum_k U_{ik} \lambda_k U^\dagger_{kj}$ with $\lambda_k>0$. Furthermore, this $\Lambda$ defines a natural UV cut-off, since by definition $p_\mu \leq \Lambda$.

We also define the field corresponding to "UV" fluctuations (the analog of $\Phi_>$)
\begin{equation}
\Psi_S(p_1,p_2):=P_{p_1,p_2}^{(S)}\Psi(p_1,p_2)
\end{equation}
where the projector $P_{p_1,p_2}^{(S)}$ is defined as:
\begin{equation}
P_{p_1,p_2}^{(S)}:=\chi(S \leq p_1 \leq \Lambda)+\chi(S \leq p_2 \leq \Lambda)-\chi(S \leq p_1 \leq \Lambda)\chi(S \leq p_2 \leq \Lambda)\,.
\end{equation}
with the definition: $\Psi(p_1,p_2):=\psi(p_1-\Vert \sigma \Vert_{\text{op}},p_2-\Vert \sigma \Vert_{\text{op}})$ (since $p=x+\Vert\sigma\Vert_{\text{op}}$), and $\Psi_{\bar{S}}(p_1,p_2)$ being the field associated with "IR" fluctuations. Once again,
\begin{equation}
\int \dd x \, \varphi(x)\psi_S \star \psi_{\bar{S}}(x,x)= \int \dd p_1 \dd p_2 \, \rho(p_1)\rho(p_2)\, \Psi_S(p_1,p_2)\Psi_{\bar{S}}(p_2,p_1) = 0\,.
\end{equation}
We also introduce the notation $[X(\varphi) ]_{\varphi}$,
\begin{equation}
[X(\varphi) ]_{\varphi}:= \frac{\int [\dd \varphi(x)] \, X(\varphi)\, e^{-N^2\mathcal{L}_1[\varphi]}}{ \int [\dd \varphi(x)] \, e^{-N^2\mathcal{L}_1[\varphi]} }\,,
\end{equation}
for a certain functional $X(\varphi)$. The concentration of measure therefore means here that $[X(\varphi) ]_{\varphi}\to X(\varphi_0)$. Let us then consider the rewriting of the initial partition function \eqref{partitionfield}, which reads:
\begin{equation}
Z_N=: [\mathcal{Z}_N(\varphi)]_{\varphi}\,.\label{averagevarphi}
\end{equation}

\begin{figure}
\begin{center}
\includegraphics[scale=0.8]{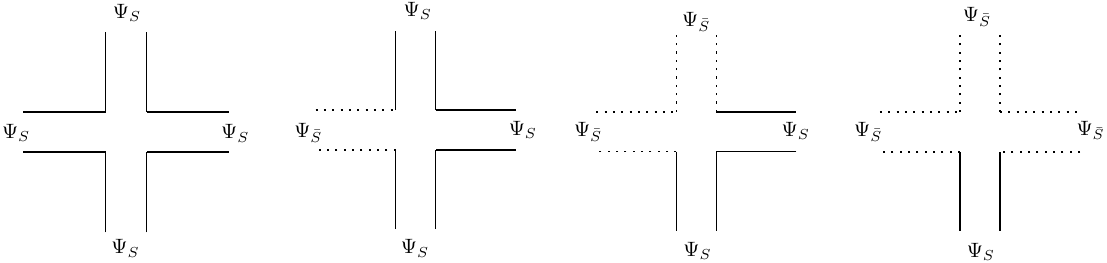}
\end{center}
\caption{Decomposition of the quartic vertex into "high" ($\Psi_S$, solid lines) and "low" ($\Psi_{\bar{S}}$, dashed lines) components.}\label{figvertexint}
\end{figure}
\begin{figure}
\begin{center}
\includegraphics[scale=0.7]{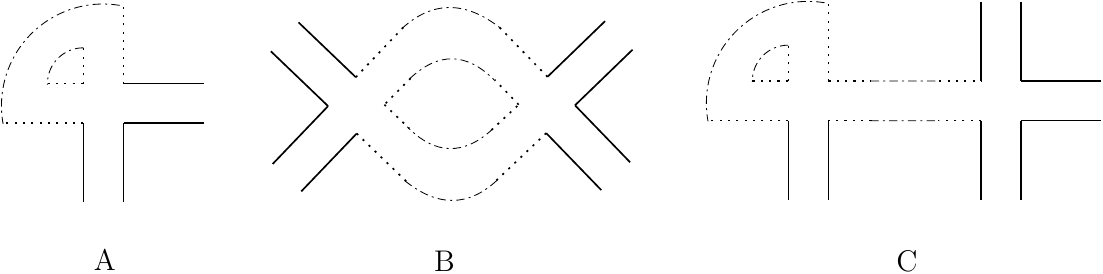}
\end{center}
\caption{One-loop corrections to the $2$-point function (A) and to the $4$-point function (B,C).}\label{oneloopvertexfunctions}
\end{figure}

By decomposing $\mathcal{Z}_N(\varphi)$, for a given realization of $\varphi(x)$, into UV and IR fluctuations,
\begin{align}
\nonumber \mathcal{Z}_N(\varphi)&= \int [\dd \Psi_S] [\dd \Psi_{\bar{S}}]\, e^{-\frac{1}{2}\int \dd p\, \rho(p)(m+2 p)( \Psi_S\star\Psi_{{S}}(p,p)+\Psi_{\bar{S}}\star\Psi_{\bar{S}}(p,p))-\frac{u}{4N}\,\int \dd p \rho(p) (\Psi\star \Psi \star \Psi \star \Psi)(p,p)}\,\\
&=\int [\dd \Psi_{\bar{S}}]\,  e^{-\frac{1}{2}\int \dd p\, \rho(p)(m+2 p) \Psi_{\bar{S}}\star\Psi_{\bar{S}}(p,p)-S_{\text{int}}[\Psi_{\bar{S}}]}\,,
\end{align}
with:
\begin{equation}
e^{-S_{\text{int}}[\Psi_{\bar{S}}]}:= \int [\dd \Psi_S]\, e^{-\frac{1}{2}\int \dd p\, \rho(p)(m+2p)\Psi_S\star\Psi_S(p,p)-\frac{u}{4N}\,\int \dd p \rho(p) (\Psi\star \Psi \star \Psi \star \Psi)(p,p)}\,.\label{Sinteff}
\end{equation}
By decomposing the field $\Psi=\Psi_S+\Psi_{\bar{S}}$, we obtain (in particular) the four vertices shown in Figure \ref{figvertexint}. One can compute the one-loop corrections by expanding \eqref{Sinteff} in powers of $u$, and using the following easily proved formula
\begin{equation}
\ln \langle e^{-U} \rangle = - \langle U \rangle + \frac{1}{2}\left( \langle U^2 \rangle - \langle U \rangle^2\right)+ \mathcal{O}(U^3)\,.
\end{equation}
The relevant one-loop diagrams contributing to the corrections of the propagator and quartic coupling of the field $\Psi_S$ are shown in Figure \ref{oneloopvertexfunctions}. We only retain planar contributions, which contain at least one face, see Definition \ref{defface}. Let us denote by $\mathcal{A}_{A,B,C}$ the corresponding Feynman amplitudes for diagrams $A$, $B$, and $C$, respectively. Let us begin with the first one:
\begin{align}
\nonumber\mathcal{A}_A&:=\frac{1}{N^5}\times N^2 \sum_{p,p_1,p_2} \, \left\{\frac{1}{p+p_1+m}+\frac{1}{p+p_2+m} \right\}\, \Psi_{\bar{S}}(p_1,p_2)\Psi_{\bar{S}}(p_2,p_1) \\\nonumber
&\to \int\, \dd p \rho(p) \dd p_1 \rho(p_1) \dd p_2 \rho(p_2)\, \left\{\frac{1}{p+p_1+m}+\frac{1}{p+p_2+m} \right\}\, \Psi_{\bar{S}}(p_1,p_2)\Psi_{\bar{S}}(p_2,p_1)\\
&=:\int \,  \dd p_1 \rho(p_1) \left(\mathcal{A}_A^{(0)}+2 p_1 \mathcal{A}_A^{(1)}+\cdots \right) \, \Psi_{\bar{S}}\star \Psi_{\bar{S}}(p_1,p_1)\,.
\end{align}
The first term gives a correction to the mass, and the second a correction to the wave-function renormalization, which arises from the fact that in diagram A of Figure \ref{oneloopvertexfunctions}, one of the external indices flows inside the loop; this is not the case for a strictly local theory. Explicitly:
\begin{align}
\mathcal{A}_A^{(0)}:=\int_S^{\Lambda}\, \dd p\, \frac{2\rho(p)}{p+m}\,,\qquad \mathcal{A}_A^{(1)}:=-\int_S^{\Lambda} \,\dd p\, \frac{\rho(p)}{(p+m)^2} \,,
\end{align}
so that the effective propagator of the field $\Psi_{\bar{S}}$ becomes:
\begin{equation}
\langle \Psi_{\bar{S}}(p_1,p_2)\Psi_{\bar{S}}(p_2^\prime, p_1^\prime)\rangle_{\textit{G}}=N^2 \, \frac{\delta_{p_1p_1^\prime}\delta_{p_2p_2^\prime}}{m_{\bar{S}}+Z_{\bar{S}}(p_1+p_2)}\,.
\end{equation}
Taking into account the different numerical factors and the coupling constant, we see that the mass and the field renormalization of $\Psi_{\bar{S}}$ are $m_{\bar{S}}:=m+\Delta m$ and $Z_{\bar{S}}=1+\delta Z_{\bar{S}}$, with:
\begin{equation}
\Delta m=u \mathcal{A}_A^{(0)}\,, \qquad \delta Z_{\bar{S}}=\,u \mathcal{A}_A^{(1)}\,.
\end{equation}

The one-loop corrections to the coupling $u$ arise from diagrams B and C. Diagram C involves a contraction of the type $\Psi_{\bar{S}} \star P^{(S)} C P^{(S)} \star \Psi_{\bar{S}}$, which vanishes by definition of the projector $P^{(S)}$. For the amplitude of diagram B, we find, keeping only the zeroth-order term in the derivative expansion of the "high" momentum loop:
\begin{equation}
\mathcal{A}_B\approx \frac{1}{N}\, \int_S^{\Lambda} \,\dd p\, \frac{\rho(p)}{(p+m)^2} \, \int \dd p_1 \rho(p_1)\, \Psi\star \Psi \star \Psi \star \Psi (p_1,p_1)\,.
\end{equation}
Taking into account the various numerical factors, we find that the variation of the coupling $u(\bar{S})=u+\Delta u (\bar{S})$, with:
\begin{equation}
\Delta u (\bar{S})= -2 u^2 \int_S^{\Lambda} \,\dd p\, \frac{\rho(p)}{(p+m)^2} \,.
\end{equation}
Since the choice of the integration interval is arbitrary, it is possible to write these variations more generally in the form:
\begin{align}
m(S^\prime)&=m(S)+u(S) \int_{S^\prime}^{S}\, \dd p\, \frac{2\rho(p)}{Z(S)p+m(S)}\,,\\
Z(S^\prime)&=Z(S)-u(S) \int_{S^\prime}^{S} \,\dd p\, \frac{\rho(p)}{(Z(S)p+m(S))^2}\,,\\
u(S^\prime)&=u(S)-2 u^2(S) \int_{S^\prime}^{S} \,\dd p\, \frac{\rho(p)}{(Z(S)p+m(S))^2}\,.
\end{align}
These equations can be translated into their infinitesimal version,
\begin{align}
S \frac{\dd m}{\dd S}&=\, -2u(S)\, \frac{S \rho(S)}{Z(S)S+m(S)}\\
S \frac{\dd Z}{\dd S}&=\,  u(S)\, \frac{S\rho(S)}{(Z(S)S+m(S))^2}\\
S \frac{\dd u}{\dd S}&=\, 2 u^2(S)\,\frac{S\rho(S)}{(Z(S)S+m(S))^2}\,.
\end{align}
It is customary to consider the flow of dimensionless renormalized parameters. However, in the present theory, there is no external dimensionful parameter with respect to which a canonical dimension can be assigned to the couplings. Nevertheless, by analogy, one may seek a natural rescaling of the parameters that removes the explicit dependence on the scale $S$ from the loop term. We therefore define the dimensionless renormalized couplings as follows:
\begin{equation}
\bar{m}(S):=Z^{-1}(S) S^{-1} m(S)\,,\qquad \bar{u}(S):=Z^{-2}(S) S^{-1} \rho(S) u(S)\,,
\end{equation}
the factors $Z(S)$ arising, as usual, from field renormalization. Expressed in terms of these couplings, the flow equations become:
\begin{align}
\beta_m &=-(1+\eta) \bar{m}(S)-\frac{2\bar{u}(S)}{1+\bar{m}(S)}\,,\\
\beta_u &=-(\dim (u)+2\eta)\bar{u}(S)+ \frac{2\bar{u}^2(S)}{(1+\bar{m}(S))^2}\,,
\end{align}
where $\beta_X:= S \frac{\dd \bar{X}}{\dd S}$, and $\eta$ is the anomalous dimension\footnote{Note that in Section \ref{FRG_sec} we adopt the opposite sign convention $\eta(k) := -k \frac{\dd}{\dd k}\ln Z(k)$, which is standard in the FRG literature.}:
\begin{equation}
\eta(S):=\frac{S}{Z(S)}\frac{\dd Z}{\dd S}(S)= \frac{\bar{u}(S)}{(1+\bar{m}(S))^2}\,,
\end{equation}
and where the \textit{canonical dimension} $\dim (u)$ is defined by:
\begin{equation}
\dim (u)=- S \frac{\dd }{\dd S} \ln (S^{-1} \rho(S))\,.
\end{equation}
The explicit dependence on $S$ is thus entirely shifted into the dimensional term, which is linear in the coupling. When $\rho(p)$ follows a power law, this canonical dimension becomes scale-independent. In particular, if the asymptotic behavior in the deep IR is given by $\rho(S)\sim S^{1/2}$, one obtains $\dim(u)=1/2$. The scale dependence of the canonical dimension implies, in particular, that no global fixed point can exist. The notion of a fixed point must therefore be replaced by that of a \textit{fixed trajectory}: a parametrized curve along which the beta functions vanish (see also \cite{Lahoche:2024gal}). These trajectories terminate in the deep IR, where the distribution approaches a power-law behavior and the canonical dimension becomes constant. The endpoints of these fixed trajectories naturally define the notion of an \textit{asymptotic fixed point}, which will be the focus of the remainder of this article.

\begin{remark}At the one-loop level, the effect of the anomalous dimension exactly cancels the one-loop vertex correction, so that, in fact, $\beta_u=-\dim(u) \bar{u}(S)+\mathcal{O}(\bar{u}^3)$. This property, well known in matrix field theories, is notably responsible for the asymptotic safety of the Grosse-Wulkenhaar model \cite{Rivasseau_2015}. Here, however, the theory is not "safe" due to the canonical dimension of the coupling, which does not vanish in the deep IR.
\end{remark}

\begin{figure}
\begin{center}
\includegraphics[scale=0.5]{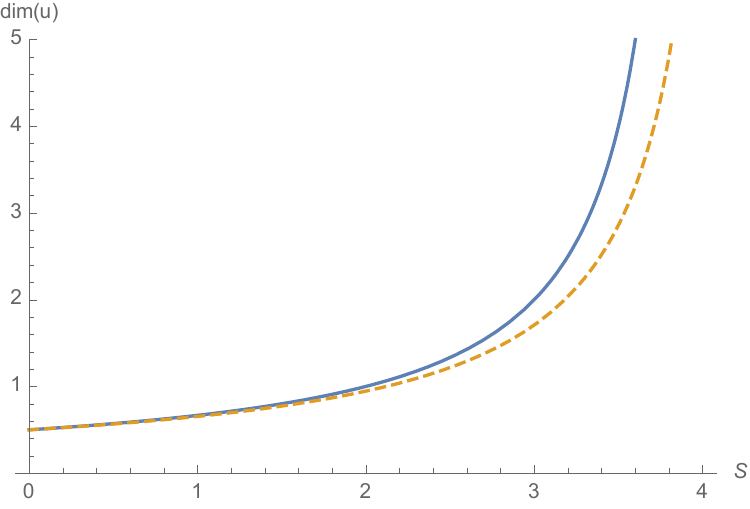}\qquad \includegraphics[scale=0.5]{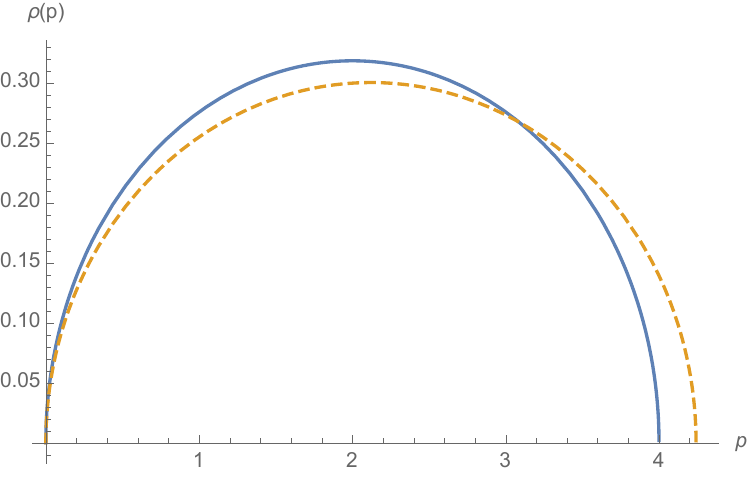}
\end{center}
\caption{Left: behavior of the canonical dimension for a pure Wigner law ($\beta=0$, in blue) and in the case where the matrix $M$ is critical ($a_4=-1/12$, $\beta=0.2$, in dashed yellow). Right: the corresponding generalized moment distributions in the limit $N\to \infty$.}\label{figdim}
\end{figure}

Figure \ref{figdim} shows the behavior of the canonical dimension for a pure Wigner distribution, and for $\beta=0.2$, with the corresponding $\rho(p)$ distributions also shown in the figures. The construction of the distributions is detailed in Appendix \ref{App3}.

\section{Functional renormalization group}\label{FRG_sec}

In the previous section, we constructed a renormalization group based on coarse-graining performed on the spectrum of the intermediate field $\sigma$. This renormalization group has the particular feature of not admitting a global fixed point, owing to the scale dependence of the canonical dimension. However, asymptotic fixed points exist, particularly in the IR limit, where, as long as $0<\beta < \beta_c$, it is guaranteed that $\rho(p) \sim \sqrt{p}$. In this limit, the theory behaves like a 3-dimensional field theory, and the search for fixed points inevitably requires non-perturbative techniques. In this section, we present a first simple method based on a vertex expansion within the framework of the Effective Average Action (EAA) method, an approach to the functional renormalization group based on the effective action of the integrated modes \cite{delamotte2012introduction,pawlowski2017physics}.

\subsection{Flow equation and truncation}\label{flowequationsection}

The EAA method, originally proposed by Wetterich \cite{wetterich1993exact}, consists in introducing an effective mass term into the action, called a \textit{regulator}, whose role is to suppress the IR modes below a scale set by a scale parameter on which this mass depends. This amounts to modifying the definition of $\mathcal{L}_2$ given by \eqref{equationL2}:
\begin{equation}
\mathcal{L}_{2,k}[\varphi,\Psi]:=\frac{1}{2}\int \dd p\, \rho(p) (m+2 p) \Psi\star\Psi(p,p)+\Delta \mathcal{L}_{2,k}+\frac{u}{4N}\,\int \dd p \rho(p) (\Psi\star \Psi \star \Psi \star \Psi)(p,p)\,,
\end{equation}
where:
\begin{equation}
\Delta \mathcal{L}_{2,k}[\Psi]:= \frac{1}{2} \int \, \dd p \dd p^\prime  \rho(p) \rho(p^\prime)\, R_{\varphi,k}(p,p^\prime) \Psi(p,p^\prime) \Psi(p^\prime,p)\,,
\end{equation}
so that the free propagator of the theory becomes:
\begin{equation}
\langle \Psi(p_1,p_2)\Psi(p_2^\prime, p_1^\prime)\rangle_{\textit{G}}=N^2 \, \frac{\delta_{p_1p_1^\prime}\delta_{p_2p_2^\prime}}{m+p_1+p_2+R_{\varphi,k}(p_1,p_2)}\,.
\end{equation}
The function $R_{\varphi,k}(p,p^\prime)=R_{\varphi,k}(p^\prime,p)$, the regulator, is assumed to have, in addition to symmetry, the following properties:
\begin{enumerate}
    \item $R_{\varphi,k\to 0}(p,p^\prime) \to 0$
    \item $R_{\varphi,k\to \Lambda}(p,p^\prime) \to \infty$
    \item $R_{\varphi,k} ( p < k,p^\prime < k) = \mathcal{O}(k) $\,.
\end{enumerate}
The last condition incorporates the idea that the IR modes relative to the scale $k\in [0,\Lambda]$ acquire a relatively large mass and therefore decouple from the long-distance physics. In what follows, we will simply denote the regulator by $R_k$ to lighten the notation, implicitly assuming that the $N \to \infty$ limit is taken at the end of the beta-function calculations and that measure concentration is operative. The first two conditions, on the other hand, concern the boundary conditions of the EAA, $\Gamma_k$, which is defined by the modified Legendre transform:
\begin{equation}
\Gamma_{N,k}[\varphi,\Pi]+\Delta \mathcal{L}_{2,k}[\Pi] :=\int \, \dd p\, \rho(p)\, \mathcal{J}\star \Pi (p,p)- \ln (\mathcal{Z}_{N,k}(\varphi,\mathcal{J}))\,,\label{Gammafunc}
\end{equation}
where, according to definition \eqref{averagevarphi}, one has, by including a source term defining $\mathcal{J}(p,p^\prime)$:
\begin{equation}
\mathcal{Z}_{N,k}(\varphi,J):= \int [\dd \Psi(p,p^\prime)]\, e^{-\mathcal{L}_{2,k}[\varphi,\Psi]+\int \, \dd p\, \rho(p) \,\mathcal{J}\star \Psi (p,p) }\,.\label{defZNk}
\end{equation}
Recall, moreover, that the cutoff $\Lambda$ was defined in the previous section, equation \eqref{lambdadef}. As for the first two conditions, they incorporate the interpolation constraints at the boundaries:
\begin{enumerate}
    \item For $k\to 0$, $\Gamma_{N,k\to 0} \to \Gamma_N$, the Legendre transform of the total free energy $\mathcal{W}_N(\varphi):=\ln \mathcal{Z}_N (\varphi)$.
    \item For $k\to \Lambda$, $\Gamma_{N,k}$ reduces to the initial microscopic action,
    \begin{equation}
        \Gamma_{N,k\to \Lambda}[\varphi,\Pi]\to \mathcal{L}_2[\varphi,\Pi]\,.
    \end{equation}
\end{enumerate}
It remains to derive the flow equations. We will present here a derivation directly in the continuum formalism, with a discrete version of the construction of the flow equation given in Appendix \ref{App4}. The proof requires specifying the definition of the functional derivative. In particular, we must impose that the chain rule is satisfied, so that, for any functional $F[J]$:
\begin{equation}
\frac{\delta F}{\delta \mathcal{J}(p_1,p_2)}= \int \dd p_3 \dd p_4\, \rho(p_3) \rho(p_4)\frac{\delta F}{\delta \mathcal{J}(p_3,p_4)} \, \frac{\delta \mathcal{J} (p_3 ,p_4)}{\delta \mathcal{J}(p_1 ,p_2)}\,,\label{chainrule}
\end{equation}
which implies the definition:
\begin{equation}
\frac{\delta \mathcal{J} (p_3, p_4)}{\delta \mathcal{J}(p_1 ,p_2)}=\frac{\delta(p_1-p_3)\delta (p_2-p_4)}{\rho(p_3) \rho(p_4)} \, \,,
\end{equation}
which corresponds here to the identity operator $\mathbb{I}(p_3,p_4; p_2,p_1)$. Let us also note that consistency with the continuum limit requires defining the value of $\delta(0)$. Since, for two sufficiently close eigenvalues $p_i$ and $p_j$ $\vert i-j \vert \approx \vert p_i-p_j \vert \, \times N \rho(p_i)$, one has the formal equivalence\footnote{Note that $f(p_i)\equiv \int \dd p_j \delta(p_i-p_j) f(p_j)$, moreover, $$\int \dd p_j \delta(p_i-p_j) f(p_j)=\frac{1}{\rho(p_i)}\int \dd p_j \delta(p_i-p_j) \rho(p_j) f(p_j) $$ Then, using the definition we propose here, and the definition of the empirical distribution $\rho(p)$, we have $$\frac{1}{\rho(p_i)}\int \dd p_j \delta(p_i-p_j) \rho(p_j) f(p_j)=\sum_i f(p_j) \delta_{ij}=f(p_i)\,.$$}:
\begin{equation}
N \rho(p_i)\delta_{ij}\equiv \delta(p_i-p_j)\,,
\end{equation}
and in particular, the coincident-point rule will be $\lim_{p_j\to p_i}\delta(p_i- p_j) \equiv N \rho(p_i)$. We thus have, according to \eqref{Gammafunc},
\begin{equation}
\frac{\delta \mathcal{W}_{N,k}}{\delta \mathcal{J}(p_2,p_1)}= \Pi(p_1 ,p_2)\,,\qquad \frac{\delta \Gamma_{N,k}}{\delta \Pi(p_1 ,p_2)}+R_k(p_1 ,p_2) \Pi(p_2, p_1) = \mathcal{J}(p_2, p_1)\,,
\end{equation}
where $\mathcal{W}_{N,k}(\varphi,\mathcal{J}):=\ln(\mathcal{Z}_{N,k}(\varphi,\mathcal{J}))$. By taking the derivative of the second relation with respect to $\mathcal{J}$, and using the chain rule \eqref{chainrule}, we have:
\begin{equation*}
\int\, \dd p_5 \dd p_6\, \rho(p_5) \rho(p_6) \frac{\delta^2}{\delta \Pi(p_5,p_6) \delta \Pi (p_1,p_2)} \left( \Gamma_{N,k}+\Delta \mathcal{L}_{2,k}\right) \frac{\delta \Pi (p_5,p_6)}{\delta \mathcal{J}(p_3,p_4)}=\frac{\delta(p_2-p_3)\delta (p_1-p_4)}{\rho(p_2)\rho(p_1)}\,,
\end{equation*}
and then:
\begin{equation}
\int\, \dd p_5 \dd p_6\, \rho(p_5) \rho(p_6) \frac{\delta^2 (\Gamma_{N,k}+\Delta \mathcal{L}_{2,k})}{\delta \Pi(p_5,p_6) \delta \Pi (p_1,p_2)} \frac{\delta^2 \mathcal{W}_{N,k} }{\delta \mathcal{J}(p_3,p_4)\delta \mathcal{J} (p_6,p_5)}=\frac{\delta(p_2-p_3)\delta (p_1-p_4)}{\rho(p_2)\rho(p_1)}\,.\label{relationinverse}
\end{equation}
In the case of a free theory, i.e., for $u=0$, we have:
\begin{equation}
\mathcal{W}_{N,k}(\varphi,\mathcal{J})=\frac{1}{2}\int \dd p_1 \dd p_2 \, \rho(p_1) \rho(p_2)\, \mathcal{J}(p_1, p_2) (m+p_1+p_2+R_k(p_1,p_2))^{-1} \mathcal{J}(p_2, p_1) \,,
\end{equation}
therefore:
\begin{equation}
\frac{\delta^2 \mathcal{W}_{N,k} }{\delta \mathcal{J}(p_3,p_4)\delta \mathcal{J} (p_6,p_5)}= \frac{1}{m+p_3+p_4+R_k(p_3,p_4)}\frac{\delta (p_3-p_5)\delta (p_4-p_6)}{\rho(p_3) \rho(p_4)}\,.
\end{equation}
In the non-perturbative regime, we will define the function $G_k(p,p^\prime)$ as follows:
\begin{equation}
\frac{\delta^2 \mathcal{W}_{N,k} }{\delta \mathcal{J}(p_3,p_4)\delta \mathcal{J} (p_6,p_5)}= G_k(p_3,p_4)\frac{\delta (p_3-p_5)\delta (p_4-p_6)}{\rho(p_3) \rho(p_4)}\,.\label{equationW2}
\end{equation}
In the symmetric phase, that is, within the basin of convergence of the power expansion of the classical field $\Pi$, one has, for $\Pi=0$:
\begin{equation}
G_k(p,p^\prime)=\frac{1}{m(k)+Z(k)(p+p^\prime)+R_k(p,p^\prime)+\mathcal{O}(p^2,(p^\prime)^2)}\,,
\end{equation}
where $m(k)$ and $Z(k)$ define the effective mass at scale $k$ and the wave function renormalization. In the same way, we will define:
\begin{equation}
 \frac{\delta^2 \Gamma_{N,k}[\varphi,\Pi]}{\delta \Pi(p_5,p_6) \delta \Pi (p_1,p_2)}=: \mathcal{K}_k(p_1,p_2)\frac{\delta (p_1-p_6) \delta (p_2-p_5)}{\rho(p_1) \rho(p_2)}\,,\label{defK}
\end{equation}
so that relation \eqref{relationinverse} can also be written as:
\begin{equation}
G_k(p,p^\prime)=\frac{1}{\mathcal{K}_k(p,p^\prime)+R_k(p,p^\prime)}\,.
\end{equation}
Note that the functions $G_k$ and $\mathcal{K}_k$ are symmetric under the exchange of their arguments. By differentiating equation \eqref{defZNk}, we obtain:
\begin{equation}
\dot{\mathcal{W}}_{N,k}=-\frac{1}{2}\, \int \dd p_1 \dd p_2\, \rho(p_1) \rho(p_2)\, \dot{R}_k(p_1,p_2) \left(\frac{\delta^2 \mathcal{W}_{N,k}}{\delta \mathcal{J}(p_1,p_2) \delta \mathcal{J}(p_2,p_1)}+\frac{\delta \mathcal{W}_{N,k}}{\delta \mathcal{J}(p_1,p_2)} \frac{\delta\mathcal{W}_{N,k}}{\delta \mathcal{J}(p_2,p_1)}\right)\,,
\end{equation}
where the dot denotes here the derivative with respect to $\ln k$ at fixed $\mathcal{J}$. It is advantageous to work not at fixed source but at fixed $\Pi$, by exploiting the relation:
\begin{equation}
k \frac{\dd \Gamma_{N,k}}{\dd k}\big\vert_{\mathcal{J}}=k \frac{\dd \Gamma_{N,k}}{\dd k}\big\vert_{\Pi}+\int \dd p_1 \dd p_2 \rho(p_1) \rho(p_2)\frac{\delta \Gamma_{N,k}}{\delta \Pi (p_1,p_2)} \dot{\Pi}(p_1,p_2)\,,
\end{equation}
we find:
\begin{equation}
\dot{\Gamma}_{N,k}=\frac{1}{2} \int \dd p_1 \dd p_2 \, \rho(p_1) \rho(p_2)\, \dot{R}_k(p_1,p_2) \, \frac{\delta^2 \mathcal{W}_{N,k}}{\delta \mathcal{J}(p_1,p_2) \delta \mathcal{J}(p_2,p_1)}\,.
\end{equation}
Since, moreover, in the large-$N$ limit:
\begin{equation}
\frac{\delta^2 \mathcal{W}_{N,k}}{\delta \mathcal{J}(p_1,p_2) \delta \mathcal{J}(p_2,p_1)}= G_k(p_1,p_2) \frac{\delta(0) \delta(0)}{\rho(p_1) \rho (p_2)} \approx N^2\, G_k(p_1,p_2)\,,
\end{equation}
we deduce from this, for $\Pi=0$:
\begin{equation}
\boxed{\dot{\Gamma}_{N,k}=\frac{N^2}{2} \int \dd p_1 \dd p_2 \, \rho(p_1) \rho(p_2)\, \dot{R}_k(p_1,p_2) \,G_k(p_1,p_2)\,.}\label{equationWett}
\end{equation}

\begin{remark}
Let us note once again that, while, morally speaking, the spectral distribution $\varphi(x)$ is not independent of the field $M$, the concentration of the measure at large $N$ freezes $\varphi$ onto its saddle-point configuration $\varphi_0(x)$. Consequently, $\varphi$ completely decouples from the quantum dynamics of the bilocal fluctuating field $\Psi(p,p^\prime)$. This spectral background field therefore acts as a rigid, classical "background geometry" on which we can construct an exact renormalization group.
\end{remark}

Equation \eqref{equationWett} cannot be solved exactly and therefore requires approximations, which, by choosing an ansatz for ${\Gamma}_{N,k}$, amount to projecting the flow onto a lower-dimensional subspace. Given the non-local nature of the theory, most standard approximations are ineffective \cite{delamotte2012introduction}. Here, we will focus on the symmetric phase at the lowest order of the derivative expansion and choose:
\begin{equation}
\boxed{\Gamma_{N,k}[\varphi,\Pi]:=\frac{1}{2}\int \dd p\, \rho(p) (m(k)+2 Z(k) p) \Pi\star\Pi(p,p)+\sum_{n=2}^\infty\frac{u_{2n}(k)}{2n N^{n-1}}\,\int \dd p \rho(p) \Pi^{\star 2n} (p,p)\,,}\label{truncationvertex}
\end{equation}
where $u_{2n}(k)$ denotes the effective coupling at scale $k$, and:
\begin{equation}
\Pi^{\star 2n}(p,p):=\int \, \left[\prod_{m=1}^{2n-1} \,\rho(p_m) \dd p_m\right]\, \Pi(p,p_1) \Pi(p_1,p_2) \cdots \Pi(p_{2n-1},p)\,
\end{equation}
While the vertex expansion is inherent to the symmetric phase (and also delineates the scope of our approximation), the validity of the derivative expansion, which justifies restricting the kinetic term to order $p$ and omitting any $p$-dependence in the interactions, is certainly a stronger condition. We will see in the final part that the Ward identities force the anomalous dimension $\eta(k):=-k\, \dd \ln Z(k)/\dd k$ to vanish in the deep IR. Given that a small anomalous dimension is expected to be a strong condition for the convergence of the derivative expansion \cite{balog2019convergence}, the aforementioned approximation is therefore expected to be quite robust.

\begin{remark}
The gauge is fixed by choosing to work in the basis where the $\sigma$ field is diagonal, but the initial $U(N)$ gauge invariance is preserved by construction. The introduction of the regulator does not break the gauge invariance any further, and one could just as well re-express the transformed action $\mathcal{L}_{2,k}$ in a manifestly invariant form. Noting that $R_{k}(p_1,p_2)$ expands in powers of the momenta:
\begin{equation}
R_k(p_1,p_2)=\sum_{n,m} R_k^{(n,m)}\, p_1^n p_2^m\,,
\end{equation}
one has $\sum_{p_1,p_2} M_{p_1p_2} R_k^{(n,m)}\, p_1^n p_2^m M_{p_2p_1}=R_k^{(n,m)}\, \operatorname{Tr}\, (\sigma^n M \sigma^m M)$. We will return to this topic in section \ref{Sec_Ward}. 
\end{remark}

\subsection{Deep IR Wilson-Fisher fixed point}\label{IRfixedpoint}

In this section, we will investigate the flow behavior by retaining only the quartic and sextic interactions (the just-renormalizable sector in the deep IR). Our complementary study in section \ref{Sec_Ward} will provide further support for the robustness of the results, while a detailed study of higher orders, the convergence of the vertex expansion, and the derivative expansion is left for future work. Note that part of this analysis has already been carried out in the complex case \cite{Lahoche:2024gal,Lahoche:2024hox}, with the model studied there corresponding to a quenched and complex version of the Hermitian model considered here (for $\beta\in [0,\beta_c)$, quenched = annealed, at least in the deep IR). We will detail the calculation only for the mass and the anomalous dimension, as the derivation of the flow equations for the other couplings follows the same procedure. An alternative derivation, up to the quartic sector, is provided in appendix \ref{App4}.

To compute the mass flow equation, we differentiate both sides of equation \eqref{equationWett} twice with respect to $\Pi(0,0)$ and use the renormalization condition:
\begin{equation}
\frac{\delta^2\Gamma_{N,k}}{\delta \Pi(0,0) \delta \Pi(0,0)}= \frac{\delta^2(0)}{\rho^2(0)} m(k)=N^2\, m(k)\,,
\end{equation}
and, graphically, the flow equation is written, defining the dot as the derivative with respect to $\ln k$:
\begin{equation}
\dot{m}\,= \,-\,\vcenter{\hbox{\includegraphics[scale=0.6]{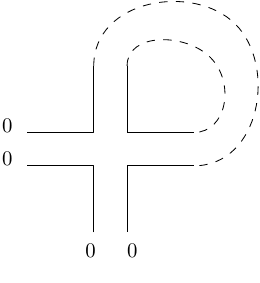}}}\,,
\end{equation}
where the loop involves the propagator $G_k$, multiplied by $\dot{R}_k\, G_k$.
The derivation of the right-hand side of the flow equation \eqref{equationWett} requires a bit more care. According to \eqref{relationinverse}, the differentiation of $\delta^2\mathcal{W}_{N,k}/\delta \mathcal{J}(p_1,p_2) \delta \mathcal{J}(p_2,p_1)$ follows the ordinary rule for differentiating the inverse of a matrix:
\begin{equation}
\delta \, A^{-1}=-A^{-1} \, \delta A\, A^{-1}\,,
\end{equation}
which reads here:
\begin{align}
\nonumber \frac{\delta}{\delta \Pi (0,0)} \frac{\delta^2 \mathcal{W}_{N,k}}{\delta \mathcal{J}(p_1,p_2) \delta \mathcal{J}(p_2,p_1)}&= - \int \, \left[\prod_{q=3}^6\dd p_q  \, \rho(p_q)\right]\,  \frac{\delta^2 \mathcal{W}_{N,k}}{\delta \mathcal{J}(p_1,p_2) \delta \mathcal{J}(p_3,p_4)}\\
&\times \frac{\delta^3 \Gamma_{N,k}}{\delta \Pi(p_3,p_4) \delta \Pi (p_5,p_6) \delta \Pi (0,0)} \frac{\delta^2 \mathcal{W}_{N,k}}{\delta \mathcal{J}(p_5,p_6) \delta \mathcal{J}(p_2,p_1)}
\end{align}
Taking the second derivative with respect to $\Pi(0,0)$, and setting $\Pi=0$ at the end of the calculation (symmetric phase), we thus find, using the definition \eqref{equationW2},
\begin{align}
\nonumber\frac{\delta^2}{\delta \Pi(0,0)^2}\, \frac{\delta^2 \mathcal{W}_{N,k}}{\delta \mathcal{J}(p_1,p_2) \delta \mathcal{J}(p_2,p_1)}\bigg\vert_{\mathcal{J}=0} = - \,G_k^2(p_1,p_2)\times \frac{\delta^4 \Gamma_{N,k}[\varphi,\Pi]}{\delta \Pi(p_2,p_1) \delta \Pi (p_1,p_2) \delta \Pi(0,0)\delta \Pi(0,0)}\,.
\end{align}
Finally, the calculation of the fourth derivative yields, with the regularization $\delta^2 (p_1)=N \rho(p_1) \delta(p_1)$,
\begin{equation}
\frac{\delta^4 \Gamma_{N,k}[\varphi,\Pi]}{\delta \Pi(p_2,p_1) \delta \Pi (p_1,p_2) \delta \Pi(0,0)\delta \Pi(0,0)}=2u_4 \frac{N^2}{\rho(0)}\, \left(\delta (p_1)+\delta (p_2)\right)\,,
\end{equation}
so that the complete flow equation reads:
\begin{equation}
\dot{m}=-2 u_4 J_2\,,
\end{equation}
having defined $J_2:=J_2(p=0)$, with:
\begin{equation}
J_n(p):=\int \, \dd q\, \rho(q) G^{n}_k(p,q) \dot{R}_k(p,q)\,.
\end{equation}
To derive the flow equation for the coupling $Z$, we differentiate the Wetterich equation \eqref{equationWett} with respect to $\Pi(p,p)$, we have:
\begin{equation}
\frac{\delta^2\Gamma_{N,k}}{\delta \Pi(p,p) \delta \Pi(p,p)}= \frac{\delta^2(0)}{\rho^2(p)} (m(k)+2Z(k)p)=N^2\, (m(k)+2 Z(k) p)\,,
\end{equation}
and:
\begin{equation}
\frac{\delta^4 \Gamma_{N,k}[\varphi,\Pi]}{\delta \Pi(p_2,p_1) \delta \Pi (p_1,p_2) \delta \Pi(p,p)\delta \Pi(p,p)}=2u_4 \frac{N^2}{\rho(p)}\, \left(\delta (p_1-p)+\delta (p_2-p)\right)\,,
\end{equation}
so that by differentiating with respect to $p$ on both sides of the equation, and setting $p=0$ at the end of the calculation, we find:
\begin{equation}
\dot{Z}=-u_4 J_2^\prime\,,
\end{equation}
where:
\begin{equation}
J_n^\prime:=\frac{\dd}{\dd p}J_n(p) \bigg\vert_{p=0}\,.
\end{equation}
The flow equation for $u_4$ is derived in the same way, and corresponds to the translation of the following diagrammatic equation (see Appendix \ref{App4}):
\begin{equation}
\dot{u}_4\,=\,-\quad \vcenter{\hbox{\includegraphics[scale=0.6]{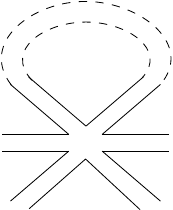}}}\quad+\quad \vcenter{\hbox{\includegraphics[scale=0.6]{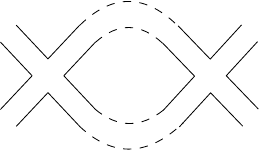}}}\,.
\end{equation}
Here again, the dotted lines correspond to the effective propagator $G_k$, one of them being assigned the additional weight $\dot{R}_k\, G_k$.

To compute the flow equations, a regulator must be chosen, and we will use the following regulator here, already considered in \cite{Lahoche:2024gal}:
\begin{equation}
R_k(p_1,p_2)=Z(k)k\left(\frac{\Lambda}{\Lambda-k}\right)\left(2-\frac{p_1+p_2}{k}\right) f\left(\frac{p_1}{k}\right)f\left(\frac{p_2}{k}\right)\label{eqR}
\end{equation}
where the function $f(x)$ is inspired by Litim's regulator:
\begin{equation}
f(x):=\theta(1-x)\,,
\end{equation}
The reason motivating the choice \eqref{eqR}, inspired by Litim's regulator \cite{litim2000optimisation}, relies on the boundary conditions at $k=\Lambda$, $R_{k\to \Lambda}(p,p^\prime) \to \infty$ and on the independence of the indices (not constrained by rotational invariance as for ordinary field theories defined on $\mathbb{R}^D$). In the deep IR, $k\ll 1$, the regulator simplifies to:
\begin{equation}
R_k(p_1,p_2)\approx Z(k)k \left(2-\frac{p_1+p_2}{k}\right) f\left(\frac{p_1}{k}\right)f\left(\frac{p_2}{k}\right)\,,\label{eqR2}
\end{equation}
Furthermore, up to a rescaling of the couplings, the asymptotic distribution reads $\rho(p)=\sqrt{p}$, so that by defining the dimensionless couplings\footnote{If $\rho(p)=\alpha \sqrt{p}$, it suffices to redefine $u_{2n} \to \alpha^{n-1} u_{2n}$.}:
\begin{equation}
\bar{m}=Z^{-1}(k)k^{-1} m\,,\qquad \bar{u}_{2n}=Z^{-n}(k) k^{-\frac{3-n}{2}} u_{2n}\,,\label{rescalingIR}
\end{equation}
the flow equations become\footnote{Note that the compensation between the anomalous dimension and the vertex correction for $u_4$ does not hold. This is a consequence of the mass regulation of the Wetterich formalism.}:
\begin{align}
\beta_{\bar{m}} &= - (1-\eta) \bar{m} - 2 \bar{u}_4 I_2 \\
\beta_{\bar{u}_4} &= - \left(\frac{1}{2} - 2\eta\right) \bar{u}_4 + 4 \bar{u}_4^2 I_3 - 4 \bar{u}_6 I_2 \\
\beta_{\bar{u}_6} &= 3\eta \bar{u}_6 + 12 \bar{u}_4 \bar{u}_6 I_3 - 6 \bar{u}_4^3 I_4 - 6 \bar{u}_8 I_2 \\
\beta_{\bar{u}_8} &= \left(\frac{1}{2} + 4\eta\right) \bar{u}_8 + 8(\bar{u}_6^2 + 2\bar{u}_4 \bar{u}_8)I_3 - 24 \bar{u}_4^2 \bar{u}_6 I_4 + 8 \bar{u}_4^4 I_5\,,
\end{align}
with:
\begin{equation}
\eta(\bar{m}, \bar{u}_4):=-k\frac{\dd}{\dd k} \, \ln Z(k) \equiv - \frac{3 \bar{u}_4}{3(\bar{m}+2)^2 - 2\bar{u}_4}
\end{equation}
and:
\begin{equation}
I_n = \frac{1}{(\bar{m}+2)^n} \left( \frac{7}{3} - \frac{14}{15}\eta \right), \quad I_n' = \frac{1}{(\bar{m}+2)^n} \left( \frac{2}{3}\eta - 1 \right)
\end{equation}
which are the continuous and dimensionless versions of the previous integrals $J_n$ and $J_n^\prime$ (see also \cite{Lahoche:2024gal} for more details).

\begin{remark}
The previous equation for $\eta$ exhibits a singularity line given by:
\begin{equation}
\bar{m}(\bar{u}_4):=\sqrt{\frac{2 \bar{u}_4}{3}}-2\,.\label{singline}
\end{equation}
Consequently, the physical region connected to the Gaussian fixed point is bounded by this singularity, where the approximation used to solve the flow equation breaks down; fixed point solutions outside this region are thus considered irrelevant. The physical region is further bounded by the singularity at $\bar{m}=-2$. 
Moreover, the condition $R_{k\to 0}=0$ imposes $R_{k}\sim k^r$ with $r>0$. In the vicinity of a fixed point with anomalous dimension $\eta_*$, $Z\sim k^{-\eta_*}$, thus $R_{k}\sim k^{1-\eta_*}$, and we therefore have the bound
\begin{equation}
\boxed{\eta_*<1\,.}\label{boundeta}
\end{equation}
\end{remark}

\begin{table}[htbp]
\centering
\resizebox{\textwidth}{!}{%
\begin{tabular}{cccccc}
\toprule
\textbf{Truncation} & \textbf{Fixed Point} & \textbf{Coordinates} & $\eta$ & \textbf{Critical Exponents ($\theta$)} & \textbf{Remarks} \\
\midrule
\textbf{Order 4} & FP 1 & $m = 8.2376$, $u_4 = -100.3460$ & $0.5844$ & $\{1.6363, -0.2715\}$ &  —  \\
                 & FP 2 & $m = -1.1668$, $u_4 = 2.1789$ & $2.8730$ & $\{-2.2491, -50.4773\}$ &  —  \\
                 & \textbf{FP 3} & $m = -0.4994$, $u_4 = 0.2581$ & $-0.1241$ & $\{0.8738, -0.9645\}$ & \textbf{WF ($\mathcal{O}(\Pi^4)$)} \\
\midrule
\textbf{Order 6} & FP 1 & $m = 17.4179$, $u_4 = -536.426$, $u_6 = -14577.6$ & $0.7302$ & $\{1.6449, -0.3014, -1.7028\}$ &  —  \\
                 & FP 2 & $m = -1.5684$, $u_4 = 0.5393$, $u_6 = 0.4225$ & $3.1135$ & $\{53.6889, -2.7084, -60.0797\}$ &  —  \\
                 & \textbf{FP 3} & $m = -0.9822$, $u_4 = 0.2521$, $u_6 = 0.0353$ & $-0.2904$ & $\{0.7804, -1.7253, -9.8371\}$ & \textbf{WF ($\mathcal{O}(\Pi^6)$)} \\
                 & FP 4 & $m = -0.7420$, $u_4 = 5.3408$, $u_6 = -32.7285$ & $2.7003$ & $\{-1.9702, -6.5804, -61.7574\}$ &  —  \\
                 & FP 5 & $m = 2.2504$, $u_4 = -6.8999$, $u_6 = 9.5458$ & $0.3044$ & $\{2.3758, 1.5670, -0.1717\}$ &  —  \\
\midrule
\textbf{Order 8} & FP 1 & $m = -1.7398$, $u_4 = 0.1884$, $u_6 = 0.1092$, $u_8 = 0.0423$ & $3.2560$ & $\{164.426, 65.552, -2.994, -80.723\}$ &  —  \\
                 & FP 2 & $m = -0.5186$, $u_4 = 7.6624$, $u_6 = -125.669$, $u_8 = 1490.85$ & $2.6297$ & $\{-1.845, -5.339, -11.388, -80.080\}$ &  —  \\
                 & FP 3 & $m = 5.8525$, $u_4 = -47.6175$, $u_6 = 83.4321$, $u_8 = 1445.09$ & $0.5098$ & $\{3.143, 1.646, -0.235, -1.728\}$ &  —  \\
                 & \textbf{FP 4} & $m = -1.2966$, $u_4 = 0.1683$, $u_6 = 0.0298$, $u_8 = 0.0034$ & $-0.4400$ & $\{0.7452, -2.7404, -13.2531, -30.503\}$ & \textbf{WF ($\mathcal{O}(\Pi^8)$)} \\
                 & FP 5 & $m = -1.1879$, $u_4 = 2.0621$, $u_6 = 0.2306$, $u_8 = -12.7366$ & $2.8833$ & $\{28.487, -2.330, -9.582, -50.969\}$ &  —  \\
                 & FP 6 & $m = -0.1084$, $u_4 = 0.0843$, $u_6 = -0.0138$, $u_8 = -0.0011$ & $-0.0239$ & $\{1.025, 0.485, -0.037, -1.363\}$ &  —  \\
                 & FP 7 & $m = 0.6082$, $u_4 = -0.8242$, $u_6 = 0.4339$, $u_8 = -0.1177$ & $0.1121$ & $\{1.338\pm 0.875\,i, 1.254, -0.075\}$ &  —  \\
\bottomrule
\end{tabular}%
}
\caption{Summary of interactive fixed points and their critical exponents according to the truncation order.}
\label{tab:fixed_points}
\end{table}

The flow equations can be solved numerically, and table \ref{tab:fixed_points} summarizes the fixed points discovered for the quartic, sextic, and octic truncations. Several fixed points are automatically disqualified. For instance, the fixed point $FP2$ of the quartic order violates condition \eqref{boundeta}. However, we discover a stable fixed point (in bold in table \ref{tab:fixed_points}) satisfying the bound \eqref{boundeta} and connected to the Gaussian point by condition \eqref{singline}. This fixed point possesses the characteristics of a Wilson-Fisher-type fixed point: one relevant direction, essentially stable (up to the octic truncation considered here). For a sextic truncation, the critical exponent\footnote{We recall that critical exponents are the opposite of the eigenvalues of the stability matrix with entries $\partial_j \beta_i$, computed at the fixed point.} associated with the unstable direction is $\theta\approx 0.78$. Note that this fixed point is the same as the one discovered in \cite{Lahoche:2024gal} for a complex matrix model coupled to a quenched matrix disorder, which comes as no surprise since we saw earlier that the condition $\beta<\beta_c$ is equivalent to working in the quenched regime, at least in the deep IR. Note, however, that the anomalous dimension at this fixed point grows in magnitude with the truncation order ($\eta_*\approx -0.12,\,-0.29,\,-0.44$ at orders 4, 6 and 8), which contrasts with the result $\eta=0$ obtained from the Ward identities in Section \ref{Sec_Ward}. This shows that the vertex expansion alone does not capture the infrared constraint imposed by the Ward identities. As we shall see in Section \ref{sec_etazero}, once the constraint $\eta=0$ is imposed, the vertex expansion converges with the truncation order and resums into a single functional equation whose relevant exponent is exactly $\theta=1/2$; the values $\theta\approx 0.75$--$0.87$ of Table \ref{tab:fixed_points} are therefore artefacts of the spurious anomalous dimension generated by the unconstrained truncation.

The interpretation of these exponents relies on the observation that in the deep IR, in the vicinity of the fixed point, the scaling behavior is essentially determined along the relevant direction by the dimensionless parameter $\bar{h}$ evaluated along the transition line. This parameter is related to the dimensionful coupling $h$ (the equivalent of the physical temperature) by $\bar{h}:=h k^{-\theta}$. Furthermore, in the deep IR, we must have $k\sim N^{-\ell}$, which leads to $\bar{h} \sim h N^{\ell \theta}$. Identifying the dimensionless parameter $\bar{h}$ with the double scaling parameter fixes:
\begin{equation}
\ell \theta = \frac{2}{2-\gamma}\,,
\end{equation}
where $\gamma$ is the entropic exponent. {Since the deep-IR flow takes place near the soft spectral edge where $\rho(p) \sim \sqrt{p}$, the number of eigenvalues below the scale $k$ scales as $N \int_0^k \rho(p)\,\dd p \sim N k^{3/2}$. The natural soft-edge scaling is therefore $k \sim N^{-2/3}$, i.e., $\ell = 2/3$. For pure gravity ($\gamma = -1/2$), this gives $\ell\theta = 4/5$ ($\theta = 6/5 = 1.2$), whereas for the confining double-well 1-cut to 2-cut transition (Painlevé II universality class, $\gamma = -1$), it gives $\ell\theta = 2/3$ ($\theta = 1$). The Ward-constrained value $\theta=1/2$ obtained in Section \ref{sec_etazero} gives instead $\ell\theta=1/3$, which would correspond formally to $\gamma=-4$, a value that does not belong to the standard list of matrix-model critical points. The identification of $\bar h$ with the double-scaling parameter must therefore be understood as qualitative: the fixed point found here is not the pure-gravity point, but rather describes a different (condensation-type) transition, as discussed in the remark below.}

{
\begin{remark}
For $\beta \in (0, \beta_c)$, the pure-gravity critical point $a_4 = -1/12$ of the original quartic matrix model maps to a negative bare coupling $u = a_4 + 2\beta^2 < 0$, whereas the reliable Wilson-Fisher-like fixed point identified in Table \ref{tab:fixed_points} (FP~3 at orders 4 and 6, FP~4 at order 8) and confirmed by the Ward identities in Section \ref{Sec_Ward} lies in the confining quadrant $\bar{u}_{4*} > 0$, $\bar{m}_* < 0$. One can alternatively expect that this fixed point governs the continuous symmetry-breaking (condensation) transition of the associated quenched matrix theory $\mathcal{F}_4^\star(\beta)$, which corresponds in the matrix-model language to the confining double-well potential ($a_4 > 0$, $m_0 < 0$) undergoing a 1-cut to 2-cut transition (the Painlevé II universality class, with $\gamma = -1$, see \cite{di19952d}). With the soft-edge scaling $k \sim N^{-2/3}$ ($\ell = 2/3$) dictated by $\rho(p) \sim \sqrt{p}$, the critical exponent $\theta = (2\nu)^{-1}$ found here once the Ward constraint $\eta=0$ is imposed, $\theta=1/2$ exactly (Section \ref{sec_etazero}), i.e.\ $\nu=1=1/(d-2)$ for $d=3$, coincides with the large-$N$ spherical value, also realized by the $p=2$ spherical spin glass, rather than with mean-field / Painlevé II scaling ($\theta = 1$). This is natural: at large $N$, the planar quartic trace interaction behaves as a large-$N$ ($O(N)$-like) interaction for the three-dimensional effective theory. Moreover, although a candidate fixed point in the non-confining sector $\bar{u}_{4*} < 0$, $\bar{m}_* > 0$ appears at orders 4 and 6 (FP1), its large couplings, and the fact that at order 8 it does not disappear but persists as FP 3 ($\bar m_*\approx 5.85$, $\bar u_{4*}\approx -47.6$, $\theta\approx 1.65$) with an additional relevant direction ($\theta\approx 3.14$), indicate that a polynomial vertex expansion around $\Pi = 0$ is ill-suited to capture a deconfinement instability. Exploring the $u < 0$ critical regime reliably therefore probably requires a non-polynomial (LPA) treatment of the effective potential, which is left for future work\footnote{The vector case treated in \cite{lahoche2026relationalbackgroundfieldformulation} lends support to this intuition. See also the next subsection.}
\end{remark}
}

{
\subsection{Critical behavior near the Gaussian fixed point}\label{secCritical}

One can support the preceding remarks by examining the theory precisely at the critical value $\beta=\beta_c$. This choice naturally sets the critical point at $u=0$, shifting this physics to an effective Gaussian fixed point. Let us show that this reproduces the exact critical exponents of the double-scaling limit.

Let us begin by studying the behavior of the asymptotic distribution in this specific case. To this end, we use the subordination relation \eqref{subordination} derived earlier:
\begin{equation}
g_{\sigma}(z)=g_{-\beta M^2}(z-R_A(g_{\sigma}(z)))\,,\label{subordination2}
\end{equation}
together with the exact relation \eqref{equationcarg}, which, taking into account $g_{-\beta M^2}(w) = -\frac{1}{\beta}\, g_{M^2}\left(-\frac{w}{\beta}\right)$, can also be written, at the point $g \equiv g_\sigma(z)$:
\begin{equation}
-(z - g)\beta^2 g^2 + \beta(1 + a_4 m_2) - a_4(z - g) + \left[-(z - g) + \frac{a_4}{\beta}(z - g)^2\right]\beta g = 0\,.
\end{equation}
Setting $a_4=-1/12$ (for which $m_2=4/3$) and $\beta=\beta_c$, one obtains:
\begin{equation}
-\frac{1}{2}\, g^3 + \left(\frac{3}{2}z + \sqrt{6}\right) g^2 - \left(z^2 + \sqrt{6}z + 1\right) g + z + \frac{8\sqrt{6}}{9} = 0\,.
\end{equation}
Let $g_*$ be the value at the lower edge, $z=z_*$. According to the subordination relation, $g_*=g_{-\beta_c M^2}(z_*-g_*)$. As explained in Section \ref{sectionfree}, this relation must be supplemented by \eqref{equationboundary}, which, from the explicit expression of $g^\prime_{-\beta_c M^2}$, gives $z_*-g_*=-8\beta_c=-2\sqrt{6}/3$. The subordination relation then fixes:
\begin{equation}
z_*=-\sqrt{6}\,,\qquad g_*=-\frac{\sqrt{6}}{3}\,.
\end{equation}
Let us expand the preceding cubic equation around these values by setting $z=z_*+\delta z$ and $g=g_*+\delta g$. We obtain:
\begin{equation}
-\frac{1}{2}\,\delta g^3 + \frac{3}{2}\,\delta z\,\delta g^2 - (\delta z)^2\,\delta g + \frac{\sqrt{6}}{3}\,(\delta z)^2 = 0\,,
\end{equation}
which yields $\delta g \sim (\delta z)^{2/3}$. In other words, at this point one has instead $\rho(p)\sim p^{2/3}$, corresponding to an effective dimension $D=10/3\approx 3.33$. In this regime, in particular, the sextic coupling becomes irrelevant. Indeed, the mass still has canonical dimension $1$, whereas the quartic coupling\footnote{The coupling of order $2n$ has canonical dimension $(5-2n)/3$.} has dimension $1/3$.

Let us now examine how the spectral edge $z_*(a_4)\equiv x_-(a_4)$ varies for $a_4=-1/12+\epsilon$ ($\epsilon>0$). Following the same strategy, the subordination relation $z_*=w_*+g_{-\beta_c M^2}(w_*)$ fixes the edge in terms of $w_*$, which is then eliminated via the condition $g^\prime_{-\beta_c M^2}(w_*)=-1$. A direct calculation yields:
\begin{equation}
z_*(a_4) = -\sqrt{6} + 4\sqrt{2}\times 3^{1/4}\,\epsilon^{3/4} - 8\sqrt{6}\,\epsilon + \mathcal{O}(\epsilon^{5/4})\,,
\end{equation}
so that the bare mass scales with $\epsilon$ as:
\begin{equation}
m(a_4) = \beta_c^{-1} - 2\vert\,z_*(a_4)\vert = 8\sqrt{2}\times 3^{1/4}\,\epsilon^{3/4} + \mathcal{O}(\epsilon)\,.
\end{equation}
In other words, $m \sim \vert a_4-a_{4}^* \vert^{3/4}$. Furthermore, since the canonical dimension of the mass is equal to $1$, we have $m=\bar{m}\,k$, and in the IR the last integrated eigenvalue lies at a scale $k$ satisfying $N \int_0^k \rho(p)\,\dd p \sim 1$, which gives here $N k^{5/3}\sim 1$. We therefore obtain the scaling relation $\vert a_4-a_{4}^* \vert^{3/4} \sim \bar{m}\, N^{-3/5}$, or equivalently $\bar{m} \sim (\vert a_4-a_{4}^* \vert N^{4/5})^{3/4}$, so that we indeed recover the double-scaling behavior by shifting the critical point to the Gaussian fixed point.
}

\section{Ward identities}\label{Sec_Ward}

The quantization of ordinary gauge field theories following the Faddeev–Popov method \cite{weinberg1995quantum,Zinn-Justin:2019jix} introduces gauge fixing as well as ghost fields. The initial gauge invariance is nevertheless encoded, at the classical level, by BRST symmetry, and at the quantum level by the Zinn-Justin equation, which generalizes the Ward–Takahashi identities in the Abelian case and the Slavnov–Taylor identities in the non-Abelian case. These identities imply non-trivial relations between observables, reflecting the underlying gauge invariance. In particular, the background field method in $f(R)$ gravity also exploits Ward identities to restore global diffeomorphism invariance, which is seemingly broken by the choice of background and the decomposition of transformations with fluctuations.

Here too, such Ward identities exist, tied to the particular gauge choice we have made, which consists in viewing the theory in the basis where $\sigma$ is diagonal (which is analogous to fixing the gauge according to the Faddeev–Popov method, see Appendix \ref{App1}). However, $U(N)$ symmetry is not truly broken, and the formal invariance of the partition function expresses this invariance in the form of relations between observables. Note that, from this perspective, gauge invariance translates into a transformation on the field indices, thereby taking on the appearance of a non-local gauge transformation from the point of view of the coordinates where $\sigma$ is diagonal: the transformation mixes different generalized momenta.

This section is devoted to studying the consequences of these identities in the deep IR: in particular, we show that, under certain approximations, these identities allow one to close the hierarchy of flow equations.

\subsection{Gauge fixing and Ward identities}

The original action $N \Tr \mathcal{V}(M)$, for the potential \eqref{Quarticini} for example, is invariant under the general transformation:
\begin{equation}
M\to U^\dagger M U\,,
\end{equation}
where $U$ is an $N \times N$ unitary matrix. In the partition function:
\begin{equation}
Z=\int \dd M\, e^{-N \Tr \mathcal{V}+N \Tr \, J M}\,,
\end{equation}
the source term explicitly breaks the $U(N)$ symmetry, and this breaking translates into constraints on the observables, which constitute the Ward identities. Indeed, since the integration measure is invariant under the change of variables $\dd M= \dd (U^\dagger M U)$, the partition function is formally invariant. Expressing this invariance for an infinitesimal transformation, $U=I+\epsilon$; $\epsilon=-\epsilon^\dagger$, we have:
\begin{equation}
\Tr \, JM \to \Tr \, J U^\dagger M U=\Tr\, JM+\Tr J(-\epsilon M+M\epsilon)\,,
\end{equation}
and therefore, to first order:
\begin{equation}
\delta Z=0=\int \dd M \,\Tr J(-\epsilon M+M\epsilon) e^{-N \Tr \mathcal{V}+N \Tr \, J M}\,,
\end{equation}
which can also be written, with $W:=\ln Z$,
\begin{equation}
0= \sum_{i,j,k}  \, J_{ij}\left(-\epsilon_{jk} \frac{\partial W}{\partial J_{ik}}+\frac{\partial W}{\partial J_{kj}}\epsilon_{ki}\right)\,.
\end{equation}
By relabeling the indices in the first sum, and taking into account the fact that the equality must hold for any $\epsilon$, we find:
\begin{equation}
\boxed{0= \sum_{j} \, \left(-J_{jk} \frac{\partial W}{\partial J_{ji}}+J_{ij}\frac{\partial W}{\partial J_{kj}}\right)\,.}
\end{equation}
Gauge fixing in the basis where the intermediate field is diagonal modifies the Ward identities, since the kinetic term then transforms as follows:
\begin{equation}
\Tr (M\sigma M) \to \Tr (M U \sigma U^\dagger M) =\Tr (M\sigma M)+\Tr (M[\epsilon,\sigma] M)+\mathcal{O}(\epsilon^2)\,,
\end{equation}
and therefore:
\begin{equation}
\sum_{p_1,p_2} M_{p_1p_2} (2 p_1) M_{p_2p_1} \to \sum_{p_1,p_2} M_{p_1p_2} (2 p_1) M_{p_2p_1} +2\sum_{p_1,p_2,p_3} M_{p_3p_1} (p_2-p_1) M_{p_2p_3} \epsilon_{p_1p_2}+\mathcal{O}(\epsilon^2)\,.
\end{equation}
The Ward identity therefore now reads
\begin{equation}
\boxed{0= \sum_{p_3}  \, \left(\frac{p_2-p_1}{N} \left(\frac{\partial^2 W}{\partial J_{p_1p_3}\partial J_{p_3p_2}}+\frac{\partial W}{\partial J_{p_1p_3}}\frac{\partial W}{\partial J_{p_3p_2}}\right)+J_{p_3p_1} \frac{\partial W}{\partial J_{p_3p_2}}-J_{p_2p_3}\frac{\partial W}{\partial J_{p_1p_3}}\right)\,.}
\end{equation}
The introduction of the regulator leads to a further modification of the Ward identity, arising from the variation of the term involving $R_k$, which reads:
\begin{equation}
\delta\sum_{p_1,p_2}\, M_{p_1p_2}R_k(p_1,p_2)M_{p_2p_1}=2\sum_{p_1,p_2,p_3}\epsilon_{p_1p_2} M_{p_3 p_1} M_{p_2p_3} (R_k(p_2,p_3)-R_k(p_1,p_3))+\mathcal{O}(\epsilon^2)\,,
\end{equation}
we obtain:
\begin{align}
\boxed{0= \sum_{p_3}  \,\bigg(\frac{\mathcal{K}_k(p_2,p_1,p_3)}{N} \left[\frac{\partial^2 W_k}{\partial J_{p_1p_3}\partial J_{p_3p_2}}+\frac{\partial W_k}{\partial J_{p_1p_3}}\frac{\partial W_k}{\partial J_{p_3p_2}}\right]+J_{p_3p_1} \frac{\partial W_k}{\partial J_{p_3p_2}}-J_{p_2p_3}\frac{\partial W_k}{\partial J_{p_1p_3}}\bigg)\,,}\label{WIF}
\end{align}
with:
\begin{equation}
\mathcal{K}_k(p_2,p_1,p_3) :=p_2-p_1+R_k(p_2,p_3)-R_k(p_1,p_3)\,.
\end{equation}

\subsection{The deep IR limit: anomalous dimension and fixed point}
The identity \eqref{WIF} encodes an infinite number of relations among the observables, which can be derived recursively by differentiating it repeatedly with respect to $J$. However, it is more convenient and useful to first rewrite relation \eqref{WIF} in the following equivalent form\footnote{For convenience, the calculations are performed here in the discrete formalism. See also Appendix \ref{App4}.}:
\begin{align}
\nonumber 0= \sum_{p_3}&  \,\bigg(\frac{\mathcal{K}_k(p_2,p_1,p_3)}{N} \left[\frac{\partial^2 W_k}{\partial J_{p_1p_3}\partial J_{p_3p_2}}+N^2\pi_{p_3p_1}\pi_{p_2p_3}\right]\\
&+\left(\frac{\partial \Gamma_k}{\partial \pi_{p_1 p_3}}+N R_k(p_1,p_3)\pi_{p_3p_1}\right) \pi_{p_2p_3}-\left(\frac{\partial \Gamma_k}{\partial \pi_{p_3 p_2}}+N R_k(p_2,p_3)\pi_{p_2p_3}\right)\pi_{p_3p_1}\bigg)\,,\label{WIF2}
\end{align}
where are defined:
\begin{equation}
\Gamma_k[\pi]+\frac{N}{2}\sum_{p,p^\prime}\, R_k(p,p^\prime) \,\pi_{pp^\prime} \pi_{p^\prime p} =N \sum_{p,p^\prime} J_{pp^\prime} \pi_{p^\prime p}- W_{k}[J]\,,
\end{equation}
with $W_{k}[J]:=\ln (\mathcal{Z}_{k}(J))$, implying in particular:
\begin{equation}
\frac{\partial }{\partial J_{pp^\prime}} W_{k}[J]=N \pi_{p^\prime p}\,,\qquad \frac{\partial \Gamma_k[\pi]}{\partial \pi_{pp^\prime}}+N R_k(p,p^\prime) \pi_{p^\prime p}= N J_{p^\prime p}\,,\label{saddlepointcond2}
\end{equation}
We will specifically consider the relation between the 4-point and 2-point functions. To this end, we will differentiate with respect to $\pi_{p_4p_5}$ and $\pi_{p_6p_7}$, and set $\pi=0$ (working in the symmetric phase) at the end of the calculation. Let us first note that the terms involving the regulator in the second part of the equation are exactly canceled by those present in the term $N\mathcal{K}_k(p_2,p_1,p_3) \pi_{p_3p_1}\pi_{p_2p_3}$. Let us briefly show this. The terms linear in the derivatives of $\Gamma_k$ read,
\begin{equation}
\frac{\partial \Gamma_k}{\partial \pi_{p_1 p_3}}=N (m(k)+Z(k)(p_1+p_3)) \pi_{p_3p_1}+\mathcal{O}(p_1^2,p_2^2,\pi^2)\,,
\end{equation}
the derivative expansion reflecting the fact that we are working in the deep IR\footnote{We will return to this assumption at the end of the section.}. We therefore find, at quadratic order:
\begin{align}
\nonumber&\left(\frac{\partial \Gamma_k}{\partial \pi_{p_1 p_3}}+N R_k(p_1,p_3)\right) \pi_{p_2p_3}-\left(\frac{\partial \Gamma_k}{\partial \pi_{p_3 p_2}}+N R_k(p_2,p_3)\right)\pi_{p_3p_1}\\
&\qquad =-N(Z(k)-1)(p_2-p_1)\pi_{p_3p_1}\pi_{p_2p_3}-N\mathcal{K}_k(p_2,p_1,p_3) \pi_{p_3p_1}\pi_{p_2p_3}+\cdots\,,
\end{align}
so that in the symmetric phase, taking the second derivative with respect to $\pi$, we find in the continuum limit:
\begin{figure}
\begin{center}
\includegraphics[scale=1]{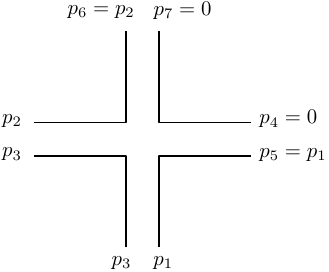}
\end{center}
\caption{Configurations of the external momenta of the inserted 1PI 4-point function.}\label{figconfig}
\end{figure}
\begin{proposition}
The relation between the $2$- and $4$-point functions, in the symmetric phase and at leading order of the derivative expansion, reads:
\begin{align}
\nonumber &(Z(k)-1)(p_2-p_1)( \delta_{p_7 p_4}\delta_{p_5p_1}\delta_{p_2p_6}+\delta_{p_6p_5}\delta_{p_1p_7}\delta_{p_2p_4})\\
&=-\sum_{p_3} \frac{\mathcal{K}_k(p_2,p_1,p_3)}{N^2}\, G_k(p_1,p_3)  \frac{\partial^4 \Gamma_k}{\partial \pi_{p_3p_1}\partial \pi_{p_2p_3}\partial \pi_{p_4p_5}\partial \pi_{p_6p_7}} G_{k}(p_2,p_3)\,,
\end{align}
where
\begin{equation}
\frac{\partial^2 W_k}{\partial J_{p_1p_2}\partial J_{p_3p_4}}=:NG_k(p_1,p_2)\delta_{p_1p_4}\delta_{p_2p_3}\,.
\end{equation}
\end{proposition}
Figure \ref{figconfig} illustrates the configuration of the external momenta we choose for the 1PI four-point function, setting $p_4=p_7=0$ and $p_1=p_5$, $p_2=p_6$, assuming $p_1\neq p_2$. Taking the limit\footnote{The typical difference between two eigenvalues is of order $\mathcal{O}(1/N)$.} $p_2\to p_1$, and then setting $p_1=0$, we obtain, in the continuum limit:
\begin{corollary}
In the vanishing external momenta limit, and at the leading order in the $1/N$ expansion, the Ward identity between the $4$- and $2$-point functions reduces to
\begin{equation}
Z(k)-1+\int \dd q \rho(q) \left(1+\frac{\partial R_k(p_1,q)}{\partial p_1}\bigg\vert_{p_1=0}\right) G_k^2(0,q) \gamma_{q,0,0,0}^{(4)}=0\,,\label{corWI}
\end{equation}
where $\gamma_{q,0,0,0}^{(4)}$ corresponds to the unique configuration of $\Gamma^{(4)}_{q0,0q,00,00}$ that creates a face, the other contributions being of order $1/N$ relative to it. 
\end{corollary}

For the next step, it is worth noting that differentiating $R_k(p,p^\prime)$ with respect to $p$ opens exactly the same momentum window as $\dot{R}_k$. Therefore, no additional approximation is introduced by using the expansion employed to construct the truncation (see also \cite{lahoche2020pedagogical,lahoche2018nonperturbative}). We thus obtain:

\begin{equation}
\int\dd q \rho(q) \frac{\partial R_k(p_1,q)}{\partial p_1}\bigg\vert_{p_1=0} G_k^2(0,q) \gamma_{q,0,0,0}^{(4)}\approx u_4 \times  \int\dd q \rho(q) \frac{\partial R_k(p_1,q)}{\partial p_1}\bigg\vert_{p_1=0} G_k^2(0,q)\,,
\end{equation}
where, within this approximation: $G_k^{-1}=(m(k)+Z(k)q+R_k(0,q))$. Since:
\begin{equation}
\frac{\partial R_k}{\partial p_1}(0,q)=-\tilde{Z}(k)\theta(k-q)\,,
\end{equation}
we have, in the deep IR:
\begin{equation}
\int\dd q \rho(q) \frac{\partial R_k(p_1,q)}{\partial p_1}\bigg\vert_{p_1=0} G_k^2(0,q)=-\frac{2C}{3}\frac{1}{k^{1/2} Z(k)}\frac{1}{(2+\bar{m})^2}\,,
\end{equation}
where we have assumed that, asymptotically, $\rho(q)\sim C\sqrt{q}$, for some constant $C$. As discussed above, this constant can be absorbed into a redefinition of the couplings, and we shall therefore set $C=1$.

\begin{figure}
\begin{center}
\includegraphics[scale=0.6]{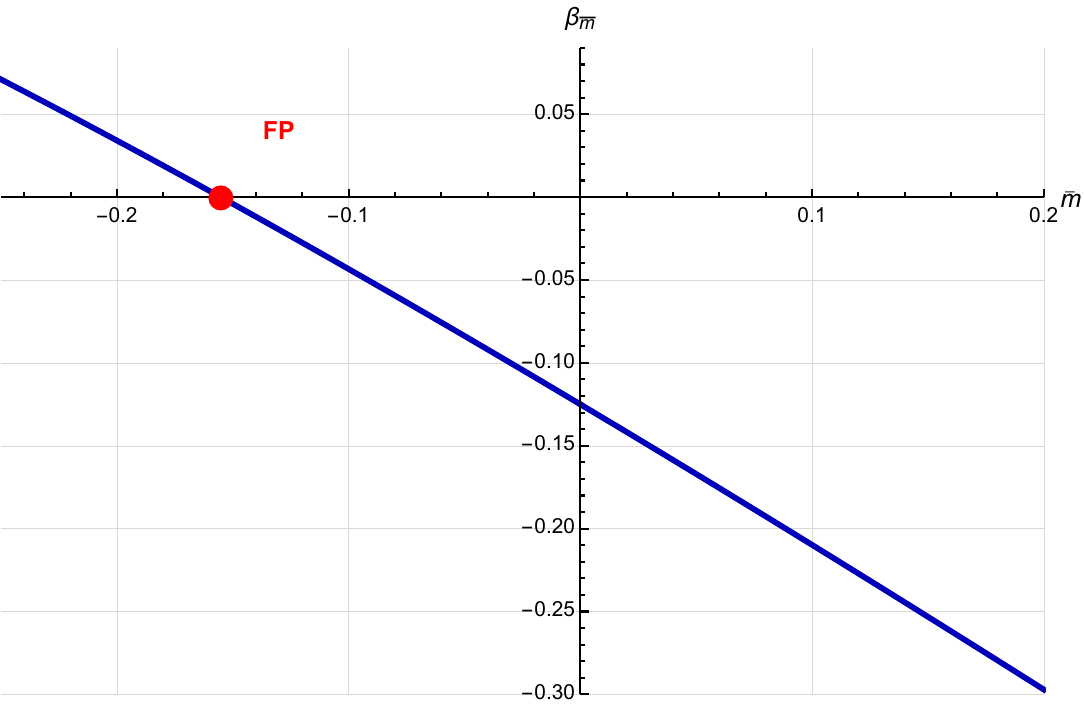}
\end{center}
\caption{Behavior of the flow equation $\beta_{\bar{m}}^{\text{eff}}(\bar{m})$ around its fixed point.}\label{flotfinal}
\end{figure}

The evaluation of the second part of the integral is more subtle, as it is not restricted to the deep IR and therefore lies outside the regime of validity of the truncation. To address this issue, and following the strategy developed in \cite{Lahoche:2024hox}, we exploit the structure of the flow equation itself, which, in the regime $k \ll 1$, behaves as follows\footnote{The inner loop selects momenta $\vert \vec{q} \vert \approx k$, so that $\Gamma^{(4)}_k(\vec{p},\vec{p},\vec{q},\vec{q})\approx \Gamma^{(4)}_k(\vec{p},\vec{p},\vec{0},\vec{0})$.}:
\begin{equation}
\dot{\Gamma}_k^{(2)}(0,q) \approx - 2\gamma_{q,0,0,0}^{(4)} J_2(0)\,,
\end{equation}
where $J_2(0) = \int \dd p \rho(p) G_k^2(0,p) \dot{R}_k(0,p)$. One can then rewrite the combination appearing in the Ward identity as follows:
\begin{align}
G_k^2(0,q) \gamma_{q,0,0,0}^{(4)} &= -\frac{1}{2J_2(0)} G_k^2(0,q) \dot{\Gamma}_k^{(2)}(0,q) \nonumber \\
&= -\frac{1}{2J_2(0)} G_k^2(0,q) \left( \dot{\Gamma}_k^{(2)}(0,q) + \dot{R}_k(0,q) \right) + \frac{1}{2J_2(0)} G_k^2(0,q) \dot{R}_k(0,q)\,.
\end{align}
Integrating over $q$ with the weight $\rho(q)$, the integral of the first term is identified exactly as a total derivative with respect to the flow time $s = \ln(k)$. The second term integrates to give exactly $1/2$, by definition of $J_2(0)$. We thus obtain:
\begin{equation}
\int \dd q \rho(q) G_k^2(0,q) \gamma_{q,0,0,0}^{(4)} = \frac{1}{2J_2(0)} \frac{\dd}{\dd s} \left( \int \dd q \rho(q) G_k(0,q) \right) + \frac{1}{2}\,.
\end{equation}
The remaining integral can be evaluated using the asymptotic behavior $\rho(q) \sim \sqrt{q}$ and the expression for the IR truncation. Given the form of the regulator $R_k(0,q) = Z(k)(2k-q)\theta(k-q)$ in the infrared, the integral is split into two contributions (for $q < k$ and $q > k$):
\begin{equation}
I_k:=\int \dd q \sqrt{q} G_k(0,q) = \frac{1}{Z(k)} \left[ \int_0^k \frac{\sqrt{q} \dd q}{k(2+\bar{m})} + \int_k^{\Lambda} \frac{\rho(q) \,\dd q}{q + \bar{m}k} \right]\,.
\end{equation}
The first integral can be computed using the approximation $\rho(p) \sim \sqrt{p}$. The second integral, on the other hand, requires an additional approximation, since the square root approximation is only valid at the edge of the spectrum. In reference \cite{Lahoche:2024hox}, the model was quenched, and the distribution of the intermediate field was fixed to the Wigner distribution:
\begin{equation}
\rho_Q(q):= \frac{\sqrt{q(4\sigma - q)}}{2\pi\sigma^2}\,.\label{disWig}
\end{equation}
Here, we only know the asymptotic form of the distribution. However, the numerical results presented in Section \ref{secscaling} suggest that the dimension depends only weakly on the precise form of the distribution. We therefore propose a method based on specifying an ansatz for evaluating the integral. As a first approximation, we choose the Wigner distribution \eqref{disWig}. In the infrared limit (for $q \to 0$), its expansion yields $\rho(q) \approx \frac{\sqrt{4\sigma}}{2\pi\sigma^2} \sqrt{q} = \frac{1}{\pi\sigma^{3/2}} \sqrt{q}$. To ensure perfect agreement with the constant $C=1$ in the asymptotic form $\rho(q) \sim C\sqrt{q}$ used above, we fix the parameter $\sigma$ such that $\frac{1}{\pi\sigma^{3/2}} = 1 \implies \sigma = \pi^{-2/3}$. We thus rewrite $I_k$ as:
\begin{equation}
I_k = \int_0^k \rho(q) \left[ \frac{1}{Zk(2+\bar{m})} - \frac{1}{Z(q+k\bar{m})} \right] \dd q + \int_0^{4\sigma} \frac{\rho_Q(q)}{Z(q+k\bar{m})} \dd q\,.
\end{equation}
The second integral is computed exactly from the Stieltjes transform of the Wigner distribution\footnote{The Stieltjes transform gives $g(z) = \frac{z - 2\sigma - \sqrt{z(z - 4\sigma)}}{2\sigma^2}$, and it is easy to see that $J(k) = \int_0^{4\sigma} \frac{\rho(q)}{(k\bar{m} + q )} \dd q = -g(-k\bar{m})$.}, yielding at leading order in $k$: $1/\sigma - \pi\sqrt{k\bar{m}} + \mathcal{O}(k)$. The first integral (over the window $[0, k]$) is evaluated at leading order using the local approximation $\rho(q) \approx \sqrt{q}$. Summing these contributions, the total integral becomes:
\begin{equation}
I_k = \frac{1}{Z(k)} \left[ \frac{1}{\sigma} + \sqrt{k} g(\bar{m}) \right]\,,
\end{equation}
with:
\begin{equation}
g(\bar{m}) = \frac{2}{3(2+\bar{m})} - 2 - \pi \sqrt{\bar{m}} + 2\sqrt{\bar{m}} \arctan\left(\frac{1}{\sqrt{\bar{m}}}\right)\,.
\end{equation}
Since the fixed point found below has $\bar m_*<0$, the function $1/(q+k\bar m)$ has a pole inside $[0,k]$, and $g(\bar m)$ must be understood as a principal value, i.e., by analytic continuation: for $-1<\bar m<0$, $g(\bar m)=\frac{2}{3(2+\bar m)}-2+\sqrt{\vert\bar m\vert}\,\ln\frac{1+\sqrt{\vert\bar m\vert}}{1-\sqrt{\vert\bar m\vert}}$. This prescription is legitimate because the pole is not physical: the regularized propagator is finite on $[0,k]$, and the pole is only introduced by adding and subtracting the term $1/(q+k\bar m)$, so that it cancels between the two integrals at leading order (both densities behave as $\sqrt{q}$ near $q=0$).
In the Ward identity, we find the term $-\frac{1}{2J_2(0)} \frac{\dd I_k}{\dd s}$, and furthermore $J_2(0) = \frac{7\sqrt{k}}{3 Z(2+\bar{m})^2}$, consistently with $I_n$ in Section \ref{IRfixedpoint} (the boundary contribution $k\partial_k\theta(1-q/k)$ does not vanish here, since $R_k(0,k)=Zk\neq 0$). Thus, applying the total flow time derivative $\frac{\dd}{\dd s}$ with the anomalous dimension defined by $\eta = -k \partial_k \ln Z = -\frac{\dot{Z}}{Z}$, we finally find:
\begin{equation}
\frac{\dd I_k}{\dd s} = -\frac{\eta}{\sigma Z} + \frac{\sqrt{k}}{Z} \left[ \frac{1}{2} g(\bar{m}) - \eta g(\bar{m}) + \beta_{\bar{m}} g'(\bar{m}) \right]\,.
\end{equation}
Multiplied by the inverse of $2J_2(0)$, the first term of the derivative immediately generates a divergent pole proportional to $k^{-1/2}$:
\begin{equation}
-\frac{1}{2J_2(0)} \frac{\dd I_k}{\dd s} = \frac{3(2+\bar{m})^2}{14\sigma} \eta k^{-1/2} + \mathcal{O}(1)\,.
\end{equation}
Since all the other terms constituting the Ward identity (such as the constant $Z-1$ and the interaction term in $\bar{u}_4$) are strictly of order $\mathcal{O}(1)$, the equation can physically only be satisfied if the divergent coefficient vanishes identically, that is to say if:
\begin{equation}
\boxed{\eta=0 \,,\quad \text{in the deep IR}\,.}
\end{equation}
\begin{remark}
Note that this result was also obtained in the quenched model of reference \cite{Lahoche:2024hox} and does not a priori depend strongly on the choice of the ansatz for $\rho(p)$.
\end{remark}
Note that $\eta=0$ only implies that $Z(k)$ tends to a constant $Z_{\text{IR}}$ in the deep IR, and not that $Z_{\text{IR}}=1$. Restoring $Z_{\text{IR}}$, the term $Z-1$ and the factor $Z$ multiplying $\bar u_4$ in the last term below remain in the identity. In what follows we make the additional assumption $Z_{\text{IR}}=1$, i.e., we neglect the finite renormalization of the kinetic term accumulated along the flow; the resulting closure must therefore be regarded as approximate. Finally, with $\eta = 0$ and $Z = 1$, the Ward identity at first order in $k$ reads:
\begin{equation}
 + \frac{1}{2J_2(0)} \frac{\dd I_k}{\dd s} + \frac{1}{2} + u_4 \int_0^k \dd q \rho(q) \frac{\partial R_k}{\partial p_1} G_k^2(0,q) = 0\,,
\end{equation}
leading to:
\begin{equation}
\frac{1}{2} + \frac{3(2+\bar{m})^2}{14} \left[ \frac{1}{2} g(\bar{m}) + \beta_{\bar{m}} g'(\bar{m}) \right] - \frac{2 \bar{u}_4}{3(2+\bar{m})^2} = 0\,.
\end{equation}
Using the flow equation for $\beta_{\bar{m}}$ obtained in the previous section, $\beta_{\bar{m}} = -\bar{m} - \frac{14\bar{u}_4}{3(2+\bar{m})^2}$ (i.e. $-2\bar u_4 I_2$ at $\eta=0$), we obtain, solving for $\bar{u}_4$:
\begin{gather}
\nonumber\\
\boxed{\bar{u}_4^{(W)}(\bar{m}) = \frac{1 + \frac{3}{14}(2+\bar{m})^2 g(\bar{m}) - \frac{3}{7}\bar{m}(2+\bar{m})^2 g'(\bar{m})}{2 g'(\bar{m}) + \frac{4}{3(2+\bar{m})^2}}\,.}
\end{gather}
Thus, the one-dimensional flow function for the mass (closing the hierarchy) reads:
\begin{equation}
\boxed{\beta_{\bar{m}}^{\text{eff}}(\bar{m}) = -\bar{m} - \frac{14}{3(2+\bar{m})^2} \bar{u}_4^{(W)}(\bar{m})\,.}
\end{equation}
The fixed-point solutions of this effective flow equation can be evaluated numerically. Figure \ref{flotfinal} shows the behavior of $\beta_{\bar{m}}^{\text{eff}}(\bar{m})$, and a fixed point is found for the values:
\begin{equation}
\bar{m}_*\approx -0.1552\,,\qquad \bar{u}_{4*}\approx 0.1132\,,
\end{equation}
with the critical exponent:
\begin{equation}
\theta \approx 0.77\,.
\end{equation}
This closure relies on three approximations: the normalization $Z_{\text{IR}}=1$, the Wigner ansatz for the UV part of $\rho$, and the relation $\dot{\Gamma}_k^{(2)}(0,q) \approx - 2\gamma_{q,0,0,0}^{(4)} J_2(0)$, which is used for all $q$, including $q\gg k$, i.e., outside the window where the truncation is reliable. The numerical proximity of this value with the unconstrained vertex expansion ($\theta\approx 0.75$--$0.78$) should therefore not be over-interpreted: the two fixed points do not lie at the same location ($\bar m_*\approx-0.16$ here, versus $\bar m_*\approx -0.5$ to $-1.3$ in Table \ref{tab:fixed_points}). The robust output of the Ward identities is the constraint $\eta=0$ itself. In the next subsection, we show that inserting this constraint into the vertex expansion leads to an exact and truncation-independent result.

{
\begin{remark}
In the preceding calculation, the results remain identical when using the distribution computed to first order in $\beta$ in Appendix \ref{App3}.
\end{remark}
}

\subsection{Ward-constrained vertex expansion: exact resummation}\label{sec_etazero}

We now combine the two previous results: the flow equations of Section \ref{IRfixedpoint} and the constraint $\eta=0$ imposed by the Ward identities. For $\eta=0$, the threshold integrals reduce to $I_n=c/(2+\bar m)^n$, with $c=7/3$, and the flow equations take a remarkably simple form. Let us introduce the generating function:
\begin{equation}
q(y):=\bar m+\sum_{n\geq 2}\bar u_{2n}\, y^{n-1}\,.
\end{equation}
One checks directly that the $\beta$-functions of Section \ref{IRfixedpoint} (at $\eta=0$) are exactly the first Taylor coefficients in $y$ of the functional equation:
\begin{equation}
\boxed{\partial_s q=-q+\frac{y}{2}\,q'-\frac{2c\,q'}{(2+q)^2}\,,}\label{resummedflow}
\end{equation}
where $q':=\partial_y q$; the first two terms encode the canonical dimensions $(n-3)/2$ of the couplings $\bar u_{2n}$, and the last one the loop contributions. We take \eqref{resummedflow} as the definition of the Ward-constrained vertex expansion at arbitrary order. Note that the rescaling $y\to c\,y$ eliminates $c$ from \eqref{resummedflow}: the fixed-point solution reads $q_*(y)=Q_*(y/c)$ with $Q_*$ independent of $c$. Consequently, $\bar m_*$ and all critical exponents are independent of the normalization $c$ of the threshold functions, while $\bar u_{2n*}\propto c^{1-n}$.

\paragraph{Fixed point.} The fixed-point equation $\left(\frac{y}{2}-\frac{2c}{(2+q)^2}\right)q'=q$ can be integrated by quadrature by regarding $y$ as a function of $q$ along the solution:
\begin{equation}
\frac{\dd y}{\dd q}=\frac{y}{2q}-\frac{2c}{q(2+q)^2}\,.
\end{equation}
The homogeneous solution $\propto\sqrt{\vert q\vert}$ is non-analytic at $q=0$, and analyticity of the fixed-point potential requires its absence. For $\bar m_*\leq q<0$, this yields:
\begin{equation}
y(q)=-2c\,\sqrt{\vert q\vert}\;\mathrm{Pf}\!\int_0^{\vert q\vert}\frac{\dd s}{s^{3/2}(2-s)^2}\,,
\end{equation}
where $\mathrm{Pf}$ denotes the Hadamard finite part, so that $y(q\to 0^-)=c$. The condition $y(\bar m_*)=0$ then fixes the mass at the fixed point: setting $\bar m_*=-2u^2$, it reduces to
\begin{gather}
3u(1-u^2)\,\mathrm{artanh}\, u=2-3u^2\,,\\
\Longrightarrow\qquad \bar m_*=-0.776693\ldots\,,\qquad \bar u_{4*}=\frac{0.58116\ldots}{c}=0.24907\ldots\,.\nonumber
\end{gather}
The fixed point lies in the confining quadrant $\bar m_*<0$, $\bar u_{4*}>0$, as the Wilson-Fisher-like fixed point of Table \ref{tab:fixed_points}.

\paragraph{Critical exponent.} Linearizing \eqref{resummedflow} around $q_*$, $q=q_*+\delta q$, one finds:
\begin{equation}
\partial_s\delta q=-\delta q+\frac{y}{2}\,\delta q'-\frac{2c\,\delta q'}{(2+q_*)^2}+\frac{4c\,q_*'\,\delta q}{(2+q_*)^3}\,.
\end{equation}
On the other hand, differentiating the fixed-point equation with respect to $y$ shows that $\delta q=q_*'$ satisfies $\partial_s q_*'=-\frac{1}{2}\,q_*'$. Since $q_*'$ is analytic at $y=0$, it is an admissible perturbation (it corresponds to a well-defined deformation of all the couplings $\bar m,\bar u_4,\bar u_6,\ldots$), and it defines the relevant direction of the fixed point, with critical exponent:
\begin{equation}
\boxed{\theta=\frac{1}{2}\quad \text{(exactly)}\,.}
\end{equation}
This result is exact within the leading-order derivative expansion supplemented by the Ward constraint $\eta=0$, and independent of $c$. Table \ref{tab:etazero} shows that the polynomial truncations of \eqref{resummedflow} converge toward it: the exponent $\theta\approx 0.79$ at quartic order decreases and converges to $1/2$ (with small oscillations of relative amplitude below $2\%$ from order 10 on), and the first irrelevant exponent converges to $-1/2$.

\begin{table}[htbp]
\centering
\begin{tabular}{ccccc}
\toprule
\textbf{Order} & $\bar m_*$ & $\bar u_{4*}$ & $\theta_1$ & $\theta_2$ \\
\midrule
4 & $-0.400000$ & $0.21943$ & $0.79057$ & $-0.7906$ \\
6 & $-0.666667$ & $0.25397$ & $0.61610$ & $-0.8576$ \\
8 & $-0.769231$ & $0.24969$ & $0.51901$ & $-0.7007$ \\
10 & $-0.785414$ & $0.24828$ & $0.49081$ & $-0.5376$ \\
12 & $-0.778293$ & $0.24893$ & $0.49675$ & $-0.4840$ \\
16 & $-0.776481$ & $0.24908$ & $0.50042$ & $-0.5018$ \\
20 & $-0.776718$ & $0.24906$ & $0.49995$ & $-0.4999$ \\
24 & $-0.776691$ & $0.24907$ & $0.50000$ & $-0.5000$ \\
\midrule
Exact & $-0.776693$ & $0.24907$ & $1/2$ & -- \\
\bottomrule
\end{tabular}
\caption{Wilson-Fisher-like fixed point of the Ward-constrained ($\eta=0$) vertex expansion, with $c=7/3$, as a function of the truncation order, compared with the exact solution of \eqref{resummedflow}.}
\label{tab:etazero}
\end{table}

\begin{remark}
The value $\theta=(2\nu)^{-1}=1/2$, i.e., $\nu=1=1/(d-2)$ for the effective dimension $d=3$, is the large-$N$ spherical value. The Ward identities, by forbidding the spurious anomalous dimension generated by the unconstrained truncation, thus select a fixed point in the spherical universality class. The approximate closure of the previous subsection ($\theta\approx 0.77$) and the unconstrained vertex expansion ($\theta\approx 0.75$--$0.87$) should therefore be regarded as less reliable approximations, the former because of the approximations listed above, the latter because it violates the Ward constraint $\eta=0$.
\end{remark}

\section{Information geometry perspective: a first look}\label{info}

We have just shown that there exists a particular splitting of the field's degrees of freedom, $M\to (M,\sigma)$, such that, in the $N\to \infty$ limit, the dynamics of $\sigma$ essentially decouples from that of $M$. Its spectrum can therefore be used as a scale to define a Wilsonian renormalization group. As we have emphasized several times, although this approach is new in the context of matrix theories, the underlying logic is not unusual: it is precisely that of background-field methods, used notably in $f(R)$ gravity, where part of the gravitational field is treated classically and defines the spectrum of the kinetic operator for the metric fluctuations. A Wilsonian RG flow is then constructed by progressively integrating out UV modes.
In more conventional field theories, such as those encountered in particle physics or condensed matter physics, the situation is essentially no different. In field theory, the Laplacian, which sets the relevant scale, can itself be viewed as being determined by a particular configuration of the gravitational field. In condensed matter physics, by contrast, the spectrum of the kinetic term is fixed by the background crystal lattice. Our position here is that this situation is universal: there is no absolute notion of scale. Rather, a scale is always defined relative to a ``cold'' background whose dynamics is effectively frozen.

The construction we propose nevertheless sheds light on the nature of the Wilsonian RG, and in particular on its connections with information geometry, as widely suggested in the literature \cite{beny2015information,berman2023bayesian,maity2015information,kar2001geometry,quinn2023information,strandkvist2020beyond,floerchinger2023information,floerchinger2023exact} \cite{amari2016information,amari2000methods}. This section is, in a sense, a side step: an open reflection on the notion of RG, which is both powerful and still not fully understood.
Information geometry provides a natural framework for probabilistic theories, of which ordinary Euclidean field theories are an example. The set of parameters, sources, and couplings (including the kinetic term) defines a manifold, which is moreover endowed with a canonical metric, the \textit{Fisher metric}. According to Chentsov's theorem, this is the unique metric invariant under sufficient statistics. Denoting by $\theta:={\theta_i}$ the set of parameters, a field theory associated with a given random field $\phi$ is described by a conditional probability $p(\phi\vert \theta)$. Field theories usually take the form of exponential distributions, linear in the couplings, $p(\phi\vert \theta)=\exp \left(-W[\theta]-\sum_i \theta_i \mathcal{O}_i(\phi)\right)$, $W=\ln Z$ (otherwise \eqref{Fishermetric} holds with $-\partial^2W$), where the $\mathcal{O}_i$ define observables of the theory\footnote{Trace invariants in the case of a random matrix model, local interactions in ordinary Euclidean field theories, etc.}, while $W$, the free energy, ensures normalization. The Fisher metric is defined by:
\begin{equation}
g_{ij}:= \int \dd \phi\, p(\phi\vert \theta)\, \partial_i \ln p(\phi\vert \theta) \partial_j \ln p(\phi\vert \theta) \equiv \frac{\partial^2 W}{\partial \theta_i \partial \theta_j}\,,\label{Fishermetric}
\end{equation}
and measures local distinguishability. In this perspective, the RG flow is a trajectory in parameter space, $\theta(t)$, along which the free energy does not change\footnote{More generally, the free energy can change by a constant depending only on the Gaussian couplings and the regulator. This dependence can be absorbed into a redefinition of $W$, ensuring that its derivative is rigorously zero. In the Wegner-Morris formalism which we will discuss below, this condition is imposed from the start.}, which is also expressed by the condition\footnote{The invariance of the free energy is written as $W^\prime(t^\prime)=W(t)$, with $t^\prime=t+\delta t$, which implies $g_{ij}^\prime(\theta,t^\prime)=g_{ij}(\theta,t)$, which is written as \eqref{equationgeo} noting that $g_{ij}^\prime(\theta,t)-g_{ij}(\theta,t)=\delta t\mathcal{L}_{-\dot{\theta}}\, g_{ij}+\mathcal{O}(\delta t^2).$}:
\begin{equation}
\frac{\partial g_{ij}}{\partial t}=\mathcal{L}_{\dot{\theta}}\,g_{ij}\,,\label{equationgeo}
\end{equation}
where $\mathcal{L}_{\dot{\theta}}\,g_{ij}:=\dot{\theta}^k \partial_k g_{ij}+\partial_i \dot{\theta}^k g_{k j}+\partial_j \dot{\theta}^k g_{k i}$ is the ordinary Lie derivative along the flow. In \cite{strandkvist2020beyond}, the authors propose broadening the notion of RG to that of a parametric flow, based on equation \eqref{equationgeo}. Such a broader notion of RG could potentially provide a way to address the problems encountered by the standard approaches discussed in Section \ref{sec1}. However, this generalization immediately raises a central question: what should be retained from the Wilsonian RG framework, and what can be relinquished?
The RG is intimately linked to the idea of simplification. As the authors point out, however, it can sometimes be difficult to determine a priori when such simpler effective descriptions are possible. The Wilsonian RG has the advantage of providing a systematic prescription, provided that (1) one has a scale, i.e., a canonical notion of UV and IR, and (2) the fluctuations associated with the degrees of freedom can be factorized and treated independently, so that they can be partially integrated out. Let us therefore examine in a little more detail what makes the ordinary RG distinctive.

This exercise is already complicated by the fact that the RG admits a multitude of incarnations, linked by a common underlying idea but not by a common formalism. In fact, there is no axiomatic definition of the RG from which all of its incarnations could be recovered depending on the context. The purpose of this section is not to provide a rigorous formulation of such axioms, but rather to initiate a discussion that may pave the way toward such a formulation, which we intend to develop in future work.

Polchinski's equation \cite{Zinn-Justin:2019jix,polchinski1984renormalization} is probably the purest realization of Wilson's idea, and we will therefore take it as our reference framework. Let us first clarify the notation. In the Hamiltonian $H:=\sum_i \theta_i \mathcal{O}_i(\phi)$, we distinguish the Gaussian part $H_{\text{kin}}=:\sum_{i,j}\frac{1}{2} \phi_i C^{-1}_{ij} \phi_j$ from the ``interaction'' part $H_{\text{int}}$. The Gaussian couplings are encoded in the symmetric, invertible matrix $C_{ij}(t)$, which we assume to depend on a single parameter $t\in \mathbb{R}$. The following theorem comes from the properties of Gaussian integrals \cite{zinn2021quantum}:

\begin{theorem}
Let an infinitesimal variation be $C_{ij}(t)=C_{ij}(t^\prime)+\Delta_{ij} \delta t+\mathcal{O}(\delta t^2)$ ($\Delta$ being assumed symmetric and positive definite), and the starting Hamiltonian $H(t):=\sum_{i,j}\frac{1}{2} \phi_i C^{-1}_{ij}(t) \phi_j+H_{\text{int}}$. Moreover, consider the Hamiltonian $H_2$ defined by:
\begin{equation}
H_2[\phi,\varphi]:=\frac{1}{2}\sum_{i,j} \phi_i (C)^{-1}_{ij}(t^\prime)\phi_j+\frac{1}{2}\sum_{i,j} \varphi_i \Delta^{-1}_{ij} \varphi_j+ H_{\text{int}}[\phi+\varphi]\,.
\end{equation}
Then, the following equation holds (up to irrelevant numerical factors):
\begin{equation}
\int \dd \phi \, \dd \varphi \, e^{-H_2[\phi,\varphi]}\propto \left(\frac{\det \Delta \det C^\prime}{\det C}\right)^{1/2}\, \int \dd \phi\, e^{-H[\phi]}\,.\label{statementGaussian}
\end{equation}
\end{theorem}
Applying this theorem, the variation of the interaction Hamiltonian reads \cite{zinn2021quantum}:
\begin{equation}
\boxed{\frac{\dd}{\dd t}H_{\text{int}}[M,\phi]=- \frac{1}{2}\sum_{i,j} \frac{\dd C_{ij}(t)}{\dd t} \left(\frac{\partial^2 H_{\text{int}}}{\partial \phi_i \partial \phi_j}-\frac{\partial H_{\text{int}}}{\partial \phi_i}\frac{\partial H_{\text{int}}}{\partial \phi_j}\right)\,.}\label{Polchinski}
\end{equation}
In the usual interpretation of coarse-graining, the variation results from a partial integration over "high" modes. For a standard scalar Euclidean field theory on $\mathbb{R}^D$, the propagator typically reads $C(p)=f(p^2/\Lambda^2)/(p^2+m^2)$ in momentum space, with the role of the parameter $t$ being played here by $\ln \Lambda$. The function $f$ cuts off high momenta relative to the scale $\Lambda$, so that a variation $\delta \Lambda/\Lambda$ of the scale parameter physically amounts to partially integrating over modes in the interval $\sim [\Lambda-\delta \Lambda, \Lambda]$. The interpretation of equation \eqref{Polchinski} is however more general, since it only assumes the given data of a certain one-parameter curve $C_{\mu\nu}(t)$ in coupling space.
The interpretation of this curve and of Polchinski's equation are brought to light by rewriting it in terms of probabilities:
\begin{equation}
\boxed{\frac{\dd}{\dd t}\, p(\phi\vert \theta(t))=-\sum_i \frac{\partial}{\partial \phi_i}\, \left(\Psi_i\, p(\phi\vert \theta(t))\right)\,,}\label{equationWM}
\end{equation}
where:
\begin{equation}
\Psi_i:= - \frac{1}{2} \sum_j \dot{C}_{ij} \frac{\partial}{\partial \phi_j} (H-H_0)\,,\qquad H_0(t):=\sum_{i,j} \phi_i C^{-1}_{ij}(t)\phi_j\label{equationPsi}
\end{equation}
This rewriting shows an important property of the Wilsonian RG: the existence of a reference distribution $p(\phi\vert \theta_0(t))\equiv\exp (-W_0(t)-H_0(\phi) )$, such that $\dd_t p(\phi\vert \theta_0(t))=0$. Equation \eqref{equationWM} describes the most general form of the RG, developed by Wegner in the 1970s \cite{wegner1974some}, and then by Morris \cite{morris1993exact}. These authors notably emphasized the fact that equation \eqref{equationWM} is the general form of an infinitesimal one-parameter change of variable: $\phi_i \to \phi^\prime_i=\phi_i+ \Psi_i\, \delta t$. The function $\Psi_i(t)$ induces a certain flow $X_t(\phi)=\Psi$, and Wegner's observation can be summarized by saying that this flow $X_t$ induces a flow $\tilde{X}_t$ in parameter space, such that\footnote{Note that from this point of view, the invariance of the Fisher metric generally results from the translation invariance of the functional measure $\dd \phi$: 

$$\frac{\dd g_{ij}}{\dd t}=\int \dd \phi\, p(X_t(\phi)\vert \theta)\, \partial_i \ln p(X_t(\phi)\vert \theta) \partial_j \ln p(X_t(\phi)\vert \theta)\,,$$

 because $\dd X_t(\phi)=\lim_{\delta t\to 0}\frac{1}{\delta t} (\dd (\phi+\Psi \delta t)-\dd \phi)$.}:
\begin{equation}
p(X_t(\phi)\vert \theta)=:p(\phi\vert \tilde{X}_t(\theta)) \equiv \frac{\dd}{\dd t}p(\phi\vert \theta(t))\,,
\end{equation}
equation \eqref{equationWM} explicating the definition of $\tilde{X}_t$. However, any one-parameter change of variable does not necessarily correspond to an RG transformation, and equation \eqref{equationPsi} specifies its properties in the case where the transformation follows a trajectory $C_{\mu\nu}(t)$ in coupling space; without it specifically being a coarse-graining yet. This transformation, however, already possesses several properties, among which\footnote{The preceding list does not a priori exhaust all the properties of the RG.}:
\begin{enumerate}
\item $\tilde{X}_t(p(\phi\vert \theta_0(t)))=0$\,,
\item $\tilde{X}_t (D_{KL}(\theta(t)\vert{}\vert{} \theta_0(t)))=\int \, \dd \phi \,p(\phi\vert \theta ) \sum_i \Psi_i^2 \geq 0$\,,
\item $C_{\mu\nu}(t)$ is not a geodesic\,,
\item $\dd_t p(\phi\vert \theta_0(t))$ is irreversible\,,
\end{enumerate}
where $D_{KL}$ is the Kullback-Leibler (KL) divergence\footnote{From the perspective of information theory, the KL divergence $D_{KL}(P\vert{}\vert{}Q)$ quantifies the average number of bits of information used by an optimal code to simulate with $Q$ a signal generated by $P$.}:
\begin{equation}
D_{KL}(\theta(t)\vert{}\vert{} \theta_0(t)) = \int p(\phi\vert \theta(t))\ln\left(\frac{p(\phi\vert \theta(t))}{p(\phi\vert \theta_0(t))}\right) \dd \phi\,.
\end{equation}
Point 1) was discussed above. Point 2) shows that along the flow, the relative entropy always increases. This condition is linked to the existence of a "monotone" of the flow, a fundamental condition connected to Zamolodchikov's famous $c$-theorem and its generalizations, as pointed out by the authors of the recent paper \cite{cotler2023renormalization}. Here, this condition shows that along the flow, (relative) information is progressively degraded. Condition 3), on the other hand, refers back to the central idea of "simplification".

To understand the origin of point 3), let us begin by noting that in the case of an ordinary Euclidean scalar field theory, the reference distribution is, up to a normalizing factor:
\begin{equation}
p(\phi\vert\theta_0(t))\propto \exp\left(-\int \dd p \phi(p)f^{-1}(p^2/\Lambda^2)(p^2+m^2)\phi(-p)\right)\,,\label{parametrisation0}
\end{equation}
In this specific case, the construction of this reference family follows a clear intuition underlying the idea of coarse-graining: all distributions are assumed to share the same IR limit, while differing progressively in their UV predictions. More precisely, all distributions yield the same results for any IR observable, but progressively deviate from one another in the UV as an increasing portion of their variance is removed by the cutoff function $f(p^2/\Lambda^2)$.

Let us examine more precisely the geodesic condition for a Gaussian distribution described by a curve $C_{ij}(t)$ in coupling space. The Fisher metric can also be understood as the infinitesimal form of the KL divergence,
\begin{equation}
D_{KL}(\theta_0(t)\vert{}\vert{} \theta_0(t+\delta t)) = \frac{1}{2}\, g_{tt}\, \delta t^2+\mathcal{O}(\delta t^3)\,.\label{infinDKL}
\end{equation}
Along the curve $C_{\mu\nu}(t)$, the variation \eqref{infinDKL} is also written as:
\begin{equation}
D_{KL}(\theta_0(t)\vert{}\vert{} \theta_0(t+\delta t)) = \frac{1}{2}\, \sum_{ij,kl}\, g_{ij,kl} \frac{\dd C^{-1}_{ij}}{\dd t} \frac{\dd C^{-1}_{kl}}{\dd t}\, \delta t^2\,,
\end{equation}
where $g_{ij,kl}$ denotes the components of the Fisher metric which can be explicitly calculated from the definition \eqref{Fishermetric},
\begin{equation}
g_{ij,kl}=\frac{1}{4}\, (C_{ik}C_{jl}+C_{il}C_{jk})\,.
\end{equation}
Geodesics can be calculated by looking for trajectories minimizing the action: 
\begin{equation}
S=\int\sqrt{ \sum_{ij,kl}\, g_{ij,kl} \dot{C}_{ij}^{-1} \dot{C}_{kl}^{-1}}\, \dd t\,.
\end{equation}
Following the standard method in general relativity and introducing the Lagrangian $\mathcal{L}:=\sum_{ij,kl}\, g_{ij,kl} \dot{C}_{ij}^{-1} \dot{C}_{kl}^{-1}$, one finds that the geodesic curves $\delta S/\delta C^{-1}=0$ satisfy:
\begin{equation}
\frac{\dd}{\dd t} \frac{\partial \mathcal{L}}{\partial \dot{C}_{ij}^{-1}}-\frac{\partial \mathcal{L}}{\partial C^{-1}_{ij}}=\frac{1}{2} \frac{1}{\mathcal{L}} \, \frac{\dd \mathcal{L}}{\dd t}\, \frac{\partial \mathcal{L}}{\partial \dot{C}_{ij}^{-1}}\,,
\end{equation}
which gives here:
\begin{equation}
\dot{H}_t=\dot{F}(t)\, H_t\,,\label{geodesic}
\end{equation}
where $F(t):=\ln \, \sqrt{\mathcal{L}}$ and $H_t:=-\dot{C} C^{-1}$. Equation \eqref{geodesic} has the general solution $H_t=H_0 \exp(F(t)-F(0))$, corresponding to a global dilatation. The equation \eqref{geodesic} ensures that $C(t)$ remains symmetric for all time. However, to obtain a transformation that can be equivalent to a redefinition for the field, $H_0$ (and also $H(t)$) must be symmetric. Requiring that $H_t=H_t^T$ for all $t$ is equivalent to imposing the commutation relation $[C,\dot{C}]=0$ along the flow, which means that $C$ and $\dot{C}$ must be simultaneously diagonalizable (or diagonalized in the same basis). If this condition holds, it follows that $[H_0,C(0)]=0$, and hence:
\begin{equation}
[H(t),C(0)]=[H(t),C(t)]=0\,.
\end{equation}
If these conditions hold, we then deduce:
\begin{equation}
C(t)= \exp \left(-H_0\int_0^t e^{(F(t^\prime)-F(0))}\dd t^\prime\right)\,C(0)\,,
\end{equation}
which corresponds to the field redefinition:
\begin{equation}
\phi_i \to \phi^{\prime}_i=\sum_j\,\left(\exp \left(-\frac{1}{2}H_0\int_0^t e^{(F(t^\prime)-F(0))}\dd t^\prime\right)\right)_{ij}\,\phi_j\,,
\end{equation}
corresponding to infinitesimal transformations of the type $\Psi_i=-\sum_j\frac{(H_0)_{ij}}{2} e^{(F(t)-F(0))} \phi_j$. It is easy to see that the transformation \eqref{parametrisation0} satisfies $[C,\dot{C}]=0$, but not the geodesic equation \eqref{geodesic}. In contrast, a pure dilation $C(t)=\alpha(t) C(0)$, where $\alpha(t)$ is a scalar, satisfies the geodesic equation. 

According to the classification proposed by Wegner \cite{wegner1974some}, dilations correspond to redundant transformations. In other words, the flow is defined only up to a global reparameterization, including for fixed points, as Wegner writes "A large class of equivalent fixed points exists for each fixed point [...] the fixed point describing a certain critical behaviour is not uniquely defined." Wegner further shows that if these redundant operators modify the Hamiltonian, they do not change physical quantities (the critical exponents): the Hamiltonian remains in the same universality class. In this very special case (without interactions), we see that this equivalence takes on a kinematic meaning regarding scheme independence: the redundant directions corresponding to geodesics, Wegner's equivalence amounts to saying that the choice of the reference curve $\theta_0(t)$ is defined "up to an inertial transformation" or else that the flow is essentially independent of the inertial frame of reference, defined as the flow along a Gaussian geodesic line. Presented from this angle, and reflecting scheme independence (the choice of the physical regulator function) the flow presents the analogue of a Galilean invariance in the space of Gaussian models; provided that these reference distributions are viewed as the analogue of a reference frame. 

For the Hermitian matrix model considered in this article, the observables $\mathcal{O}_n(M)$ are identified with traces $\mathcal{O}_n(M):=\Tr \, M^{2n}$. The only trajectories $\theta_0(t)$ compatible with strict $U(N)$ invariance are of the form $P(M\vert \theta_0(t)) \propto \exp \left(-f^{-1}(t) \Tr M^2\right)$, and therefore correspond to pure gauge transformations: there is no coarse-graining, and the reference frames are only geodesics. Note that condition 3) is insufficient by itself to define the irreversibility of the flow, which is guaranteed by condition 4.
In the latter case, the kinetic term corresponding to $\theta_0(t)$ can be eliminated by a simple global dilatation $\phi\to f^{1/2}(t) \phi$, such that for this equivalent theory, the corresponding Polchinski equations vanish identically. In fact, in that case, the apparent flow reduces to a purely parametric trajectory that is directly integrated and reversible, with each coupling constant being scaled by a power of $f^{1/2}(t)$ without escaping far from the initial finite dimensional subspace of the full theory space.

This kinematic interpretation paves the way toward another reading of the RG construction proposed in this article. We have just seen that strict rotation invariance of the model is incompatible with coarse-graining, since the kinetic operator is strictly proportional to the identity. The introduction of the auxiliary field $\sigma$ offers a relational resolution to this impasse. From a probabilistic point of view, our approach replaces the initial marginal distribution $P(M)=\int\dd\sigma\, P(M,\sigma)$ of the Hermitian matrix by a conditional distribution $P(M \vert \varphi_0)$, where $\varphi_0$ corresponds to the eigenvalue distribution $\varphi$ evaluated at its saddle point. In the limit where this Hubbard-Stratonovich intermediate field is ``frozen'' by measure concentration at large $N$, the basis in which this field is diagonal becomes singular and defines a privileged reference frame. Relative to this frame, one can define a family of non-auto-parallel (non-geodesic) curves associated with coarse-graining over the spectrum, which becomes non-trivial in this limit, of the propagator, thereby providing a canonical notion of scale. Mathematically, choosing a reference subsystem to define a notion of scale translates into conditioning and, therefore, into an entropy loss:
\begin{equation}
S[M,\varphi]=S[\varphi]+S[M\vert \varphi]\,.
\end{equation}

In \cite{strandkvist2020beyond}, the authors propose extending the notion of RG to that of metric flow and introduce the concept of a ``question of interest''. For ordinary field theories, such questions make it possible to define a reference distribution $p(\phi\vert \theta_0(t))$, which in turn determines the curve $C_{\mu\nu}(t)$ in parameter space, thereby defining a family of models that all exhibit the same long-distance behaviour for IR observables. However, it may happen that no question of interest exists in the sense proposed by the authors. In the case of the matrix model, for example, the only question that can be addressed takes the form of a global dilation, which is perfectly reversible.

We defend here the idea that the definition of an RG necessarily requires the definition of a notion of scale, the existence of which is not an \textit{a priori} property of physical theories, but rather an emergent feature in certain limiting cases, as explicitly demonstrated by the large $N$ construction proposed in this article. Usually, the field relative to which this notion of scale is defined is the gravitational field, which provides a common denominator for all others, since everything gravitates. In many contexts, however, other emergent structures take over this role. In condensed matter physics, for example, the crystal lattice is a condensed structure formed from fundamental fields whose dynamics is essentially decoupled from the fast internal dynamics, such as that of the electrons. While the positions of the lattice atoms are ultimately determined by local gravity, it is the lattice itself that acts as the intermediate structure defining the scale of the effective theory. In our model, the effective spectrum of the intermediate field $\sigma$ plays precisely the role of such a background ``lattice''. Relative to this cold background, it becomes possible to define unambiguously the notions of UV and IR, and hence to construct a Wilsonian flow relative to a family of reference distributions that is no longer geodesic. Several recent works give substance to this intuition by defining exotic RG flows. In \cite{Achitouv:2025ghy,Achitouv:2024kqh,Lahoche:2024huc,Natta:2024nke,Lahoche:2024qqq,Lahoche:2024puq,Lahoche:2024hox,Lahoche:2024gal,Lahoche_2024PFP}, the authors consider spin glasses and define a flow relative to the spectrum of the disorder matrix, while in \cite{finotello2026datafieldtheorytheory,finotello2026fieldtheorydataanomaly2,Finotello_2026} (see also the references therein), the same authors consider the flow induced by the spectrum of the data correlation matrix to construct effective models describing these correlations. These flows are abstract and are not associated with an ordinary physical scale, but they share the property of being realized in a self-averaging regime, which is precisely the analogue of the statistical conditioning defined above. It is only after this conditioning has been performed that questions in the sense of \cite{strandkvist2020beyond} can be meaningfully formulated and a physical, simplifying flow constructed.

More fundamentally, we can extend this relational paradigm to the physical world as a whole. The universe in its entirety can be formally regarded as an immense joint probability distribution\footnote{The square of the modulus of the Universe's wave function in a Euclidean signature.} $P(\phi_1, \phi_2, \dots, \phi_n)$, encompassing all physical fields. Within this global distribution, there is, strictly speaking, no notion of absolute scale or absolute hierarchy: the properties of the fields are purely relational, with each degree of freedom existing and defining itself only through its correlations with the rest of the system.

The very notion of ‘‘scale'', and, by extension, that of a structured spacetime serving as an arena for physics and allowing the construction of a coarse-graining, appears only in specific limiting cases. Such a notion of scale, and therefore the very possibility of coarse-graining, emerges when the dynamics of certain fields, or of a subset of their degrees of freedom, undergoes a condensation phenomenon and ‘‘freezes''. The passage from a fundamental theory without scale to an effective theory structured by an RG flow thus corresponds, in probabilistic terms, to the transition from the joint probability $P(\phi_1, \phi_2, \dots)$ to a conditional probability $P(\phi_{\text{fluctuating}} \vert \phi_{\text{frozen}})$. This kinematic freezing, whether induced by spontaneous symmetry breaking, quantum decoherence, or a measure-concentration phenomenon at large $N$, as in our matrix model, splits reality into two asymmetrical strata: a cold and structuring background, which acts as the relational frame of reference, and residual thermal or quantum fluctuations. From this perspective, a scale is therefore always the result of a particular conditioning.

\section{Conclusion and outlook}\label{Sec_Conclu}

In this paper, we have presented a new RG approach for matrix models based on a ``relational'' background-field method. By exploiting the self-averaging property of large random matrices in the $N \to \infty$ limit, we define a non-trivial notion of scale through the effective spectral density of the intermediate Hubbard-Stratonovich field. Unlike standard approaches, this method preserves the exact $U(N)$ gauge symmetry of the model by avoiding the artificial symmetry breaking typically induced by kinetic-term regulators. Our analysis shows that the theory flows in the deep infrared toward a non-local 3D Euclidean model. We identify a non-trivial Wilson-Fisher-type fixed point with a single relevant direction in the confining sector, Furthermore, we demonstrate that the Ward identities derived from the gauge-fixing procedure force the anomalous dimension to vanish in the deep IR. Once this constraint is imposed, the vertex expansion converges and resums into a single functional equation, whose fixed point ($\bar m_*\approx-0.7767$) has exactly one relevant direction with critical exponent $\theta=1/2$, independently of the normalization of the threshold functions: the fixed point belongs to the large-$N$ spherical universality class. The unconstrained truncations ($\theta\approx0.75$--$0.87$) and an approximate closure of the Ward hierarchy ($\theta\approx0.77$) give compatible orders of magnitude, but violate or only approximately implement this constraint. The comparison with the double-scaling exponents remains qualitative, and the critical point of the original model at $\beta=\beta_c$ is shown to reproduce the double-scaling behavior exactly by shifting it to the Gaussian fixed point.

This work opens several prospects:
\begin{enumerate}
\item \textbf{Extension to tensors and higher dimensions:} A natural next step is to generalize this relational construction to tensor models and higher-rank GFTs, where gauge symmetries are more intricate. Unlike matrix models, tensor models offer greater flexibility, as the quartic sector is richer (for $d>3$), allowing for a broader variety of background fields. This approach could also help resolve a recent open problem concerning Ward identities in TGFTs \cite{Wahabou_Kpera_2024}, which typically exhibit anomalies due to the way gauge symmetry is broken by the propagator.

\item \textbf{Geometric interpretation:} As discussed in Section 5, the connection between this ``background-driven'' RG and information geometry suggests that the choice of background field can be interpreted as a specific conditioning. This opens a new perspective on renormalization and on the physical nature of scale, which we intend to investigate in future work.

\item \textbf{Higher-order truncations and non-polynomial potentials:} Although our current results are qualitatively robust in the confining sector, further investigation is needed to systematically assess the convergence of the vertex and derivative expansions at higher orders, as well as to implement a non-polynomial treatment using LPA techniques of the effective potential in order to probe the non-confining $u < 0$ sector reliably. This issue, already highlighted in \cite{Lahoche:2024gal}, remains particularly challenging for non-local theories, where most standard techniques fail. A complementary result for random vectors \cite{lahoche2026relationalbackgroundfieldformulation} nevertheless demonstrates the relevance of this intuition, and in that case, the double scaling limit corresponds to a ``shock''-type singularity.

\item \textbf{Refining the ansatz in the Ward identities:} Solving the Ward identities required approximating the exact distribution by a test distribution. This choice is clearly not optimal and deserves further investigation. It therefore represents another limitation of the method proposed in this paper.

\item \textbf{Influence of the regulator:} In our analysis, we considered only a single regulator. However, it is well known that the approximations employed to solve the flow equations introduce a dependence on this choice, which should be systematically investigated.
\end{enumerate}

By providing a gauge-invariant renormalization scheme that bridges quantum gravity and the standard Wilsonian RG, we hope that this framework will contribute to a deeper understanding of how continuous geometry may emerge from microscopic spacetime quanta.

\pagebreak
\appendix

\section{The eigenvalue problem for finite polynomial potentials}\label{App1}

We recall here some basic results concerning the spectrum of large-size random matrices characterized by $U(N)$ gauge invariance. For further details, the reader may consult the recent reference \cite{potters2020first}, as well as \cite{di19952d}. Let us first consider a generic model, defined by the partition function:
\begin{equation}
Z_N:= \int \dd M \, e^{-N \Tr \, \mathcal{V}(M)}\,,
\end{equation}
where $\mathcal{V}(M)$ is a polynomial of degree $2K$:
\begin{equation}
\mathcal{V}(M):=\sum_{l=1}^K \, \frac{a_{2 l}}{2l} \, M^{2 l}\,.
\end{equation}
The matrix $M$ is an $N \times N$ Hermitian matrix, $M=M^\dagger$, and $dM$ denotes the Lebesgue measure on this set. Hermitian matrices can be diagonalized by a unitary transformation $\mathbf{U}$, such that $M=\mathbf{U} \Lambda \mathbf{U}^\dagger$, where $\Lambda=\mathrm{diag}\{\lambda_1, \lambda_2, \dots, \lambda_N\}$ is a diagonal matrix. One can explicitly construct the Haar measure in terms of so-called "angular" and "radial" variables:
\begin{equation}
\dd M = \dd \mathbf{U} [\Delta(\Lambda)]^2 \prod_{i=1}^N \dd \lambda_i \,,
\end{equation}
where $d\mathbf{U}$ is the invariant Haar measure on $U(N)$ and $\Delta(\Lambda) := \prod_{k < l} |\lambda_k - \lambda_l|$ is the Vandermonde determinant, which is the Jacobian of the transformation. In this representation, the integral is written as:
\begin{equation}
Z_N=\int \prod_{i=1}^N \dd \lambda_i e^{-N \sum_{i=1}^N \mathcal{V}(\lambda_i)+\sum_{i\neq j} \ln \vert \lambda_i-\lambda_j \vert}\,,\label{Eigendis}
\end{equation}
where by convention $\int d\mathbf{U} = 1$. Note that the square of the Vandermonde determinant can be justified by a "Faddeev-Popov"-like trick (see \cite{di19952d}, page 14). Let us consider a particular realization of the matrix $M$, and a unitary matrix $\mathbf{U}_0$ that diagonalizes $M$: $M = \mathbf{U}_0 \Lambda \mathbf{U}_0^\dagger$. We first define $[\Delta(\Lambda)]^2$ in a way that introduces a suitable form of the identity into the path integral:
\begin{equation}
1=\int \prod_{i=1}^N \dd \lambda_i^\prime\, \dd \mathbf{U} \, \delta^{N^2}(\mathbf{U}^\dagger M \mathbf{U}-\Lambda^\prime) [\Delta(\Lambda^\prime)]^2\,.\label{GF}
\end{equation}
By substituting this expression into the initial path integral, the integration over $M$ is immediate, and the integration over $\mathbf{U}$ decouples due to the unitary invariance of the traces. We thus directly recover the form \eqref{Eigendis} determining the joint distribution $P(\Lambda)$ for a particular realization of $\Lambda$. The determination of $\Delta(\Lambda)$ can be carried out as follows. By noting that in the integral of \eqref{GF}, only the matrices in the vicinity of $\mathbf{U}_0$ contribute, let $\mathbf{U} = \mathbf{U}_0 (1 + i T)$, where $T$ is an infinitesimal Hermitian matrix. We therefore have:
\begin{equation}
\mathbf{U}^\dagger M \mathbf{U}-\Lambda^\prime = \Lambda-\Lambda^\prime+i[\Lambda,T]\,.
\end{equation}
It is easy to see that the matrix $[\Lambda, T]$ is not diagonal: $[\Lambda, T]_{ij} = (\lambda_i - \lambda_j) T_{ij}$. Therefore:
\begin{equation}
\delta^{N^2}(\mathbf{U}^\dagger M \mathbf{U}-\Lambda^\prime)=\delta^{N}(\Lambda-\Lambda^\prime)\times \delta^{N(N-1)}(i[\Lambda,T])\,,
\end{equation}
and \eqref{GF} becomes, after integration over $\Lambda^\prime$:
\begin{equation}
1=i^{N(N-1)}\int \, \dd T\, \delta^{N(N-1)}(i[\Lambda,T]) [\Delta(\Lambda)]^2\,.\label{GF2}
\end{equation}
By expressing the Dirac delta functions $\delta((\lambda_i - \lambda_j) T_{ij}) = |\lambda_i - \lambda_j|^{-1} \delta(T_{ij})$, we find, up to a numerical factor, $[\Delta(\Lambda)]^2 := \prod_{i < j} (\lambda_i - \lambda_j)^2$.

One approach to determining the eigenvalues consists in introducing the density $$\mu(x) := N^{-1} \sum_{\mu} \delta(\lambda_\mu - x)$$ and rewriting the integral as an integral over densities, before determining the most probable density using the saddle-point method in the limit $N \to \infty$. The change of variables $\prod_{i=1}^N d\lambda_i \to d\mu(x)$ is accompanied by a Jacobian, which can be evaluated by counting the number of configurations ${\lambda_i}$ corresponding to the same distribution $\mu(x)$. To this end, let us partition the $N$ eigenvalues into $k$ subsets of sizes $N_l$ ($\sum_l N_l = N$). Assuming that each $N_l$ corresponds to an interval $\delta x$ containing eigenvalues around $x_l$, such that $N_l \approx \mu(x_l) N \delta_x$, the number $\Omega$ of configurations is:
\begin{equation}
\Omega= \frac{N!}{\prod_l N_l!}\,,\qquad \log\Omega \approx - N \int \dd x\, \mu(x) \log \mu(x)+\mathrm{const.}\,,
\end{equation}
Taking this additional entropy term into account, the probability $P[\mu(x)]$ of a realization $\mu(x)$ is therefore:
\begin{equation}
P[\mu(x)] \propto \int  \dd \mu(x) e^{-N^2 \left( \int \dd x \mu(x)  \mathcal{V}(x) -\fint  \dd x\dd y \mu(x)\mu(y) \log \vert x-y \vert +\frac{1}{N} \int \dd x \mu(x) \log \mu(x) \right)} \delta \left(\int \mu(x) \dd x-1\right)\,.\label{Eigendis3}
\end{equation}
At $N \to \infty$, the entropy term does not contribute to the saddle point, which is written as:
\begin{equation}
\mathcal{V}(x)=2\fint \dd y \mu(y) \log \vert x-y \vert +\mathrm{cte}\,,
\end{equation}
where the constant '$\mathrm{cte}$' is a Lagrange multiplier imposing the constraint $\int d x \, \mu(x) = 1$. By taking the first derivative,
\begin{equation}
\mathcal{V}^\prime(x)=2\fint \dd y \mu(y)\, \frac{1}{x-y}\,.
\end{equation}
This is an equation whose solution is given by the Tricomi formula \cite{tricomi1985integral},
\begin{equation}
\mu(x)=-\frac{1}{\pi^2 \sqrt{(x-a)(b-x)}} \left(\frac{1}{2}\fint_a^b \dd y \sqrt{(y-a)(b-y)} \frac{\mathcal{V}^\prime(y)}{x-y}+C\right)\,,\label{Tricomi}
\end{equation}
where it is assumed that the density is confined within the interval $[a, b]$, and $C$ is a constant.

\begin{figure}
\begin{center}
\includegraphics[scale=0.5]{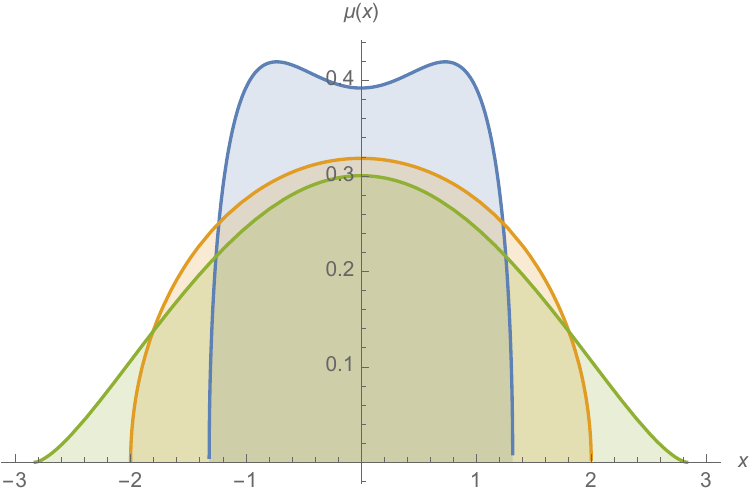}\qquad \includegraphics[scale=0.5]{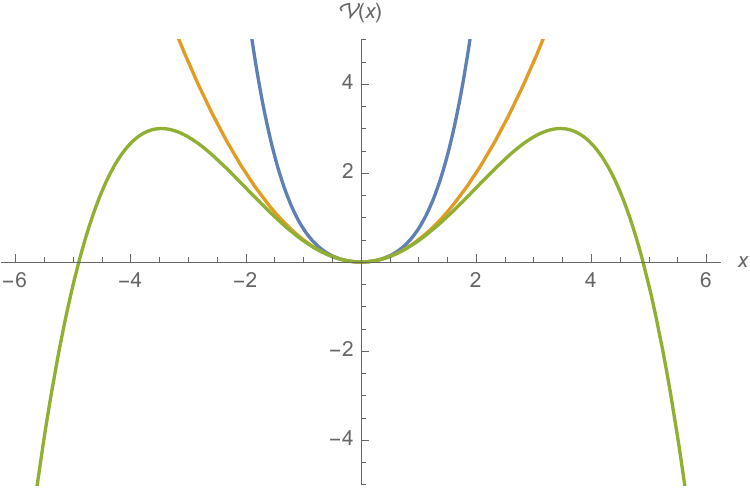}
\end{center}
\caption{On the left: distribution $\mu(x)$ for $a_4=1$ (blue curve), $a_4=0$ (yellow curve), and $a_4=-1/12$ (green curve). On the right, the corresponding potentials $\mathcal{V}(x):=x^2/2+a_4 x^4/4$.}\label{plotmuM}
\end{figure}

Let us show how this works for the potential $\mathcal{V}(x) := x^2/2 + a_4 x^4/4$. Since the potential is symmetric, $\mathcal{V}(x) = \mathcal{V}(-x)$, we have $a = -b$, and by calculating the Cauchy integrals:
\begin{equation}
\mu(x)=-\frac{1}{2}\frac{2C-\frac{\pi  b^2}{2}-\frac{a_4}{8} \pi  \left(b^4+4 b^2 x^2-8 x^4\right)+\pi  x^2}{\pi^2 \sqrt{(x+b)(b-x)}}\,.
\end{equation}
As $\mu(b) = 0$, we have $2C = -\frac{3 a_4 \pi b^4}{8} - \frac{\pi b^2}{2}$, so we have:
\begin{equation}
\mu(x)=\frac{\sqrt{(b-x) (b+x)} \left(a_4 \left(b^2+2 x^2\right)+2\right)}{4\pi }\,.\label{densityM4}
\end{equation}
Finally, we determine $b$ by imposing the normalization condition on the distribution, and we find:
\begin{equation}
b^2=\frac{2}{3} \frac{\sqrt{12 a_4+1}-1}{a_4}\,.
\end{equation}
For $a_4>-1/12$ (the critical point, see \ref{sec1}), the density \eqref{densityM4} vanishes as a square root $\sim \sqrt{(b-x)}$ at the right edge of the spectrum (Fig \ref{plotmuM}). At the critical point, $b=\sqrt{8}$, and the polynomial factoring the square roots becomes $a_4 \left(b^2+2 x^2\right)+2=(8-x^2)/6$, so that the behavior at the edge is $(b-x)^{3/2}$ at the critical point. For a general symmetric potential, but one that is sufficiently confining to capture all the eigenvalues in a single well, the density will be:

\begin{equation}
\mu(x)=\frac{Q(x)\sqrt{(b-x) (b+x)}}{4\pi}\,,
\end{equation}
where $Q(x)$ is a polynomial depending on the potential. This polynomial can have $k$ roots equal to $b$, so that the asymptotic behavior of the distribution follows the law $\sim (b-x)^\theta$, with $\theta=(2k+1)/2$.

\section{Adding a Wigner random matrix}\label{App2}

Let us provide a quick review of the result regarding the free sum of two random matrices, one of which is a Wigner matrix. Free random matrix theory generalizes probability theory for independent variables to the case of non-commutative variables; the reader will find an introduction in \cite{potters2020first}. Here, we consider an elementary approach based on Dyson’s Brownian motion, which is more accessible to a reader unfamiliar with free probability theory.

Let $A$ be an $N \times N$ matrix, and $B$ a Gaussian Wigner matrix, i.e., an $N \times N$ Hermitian matrix with variance $\sigma^2/N$ (Theorem \ref{th0}). Dyson’s Brownian motion consists in considering the properties of the sum $C = A + B$, and in particular the distribution of the eigenvalues of $C$ (in the limit $N \to \infty$), by breaking the problem down into a multitude of steps in which a Gaussian matrix of infinitesimal variance is added. Let $C(t) = A + B(t)$, where $B(t)$ is a Wigner matrix with variance $\sigma^2 t/N$. In an infinitesimal step $\dd t$ of the variance, $\dd C = \sqrt{\dd t}\, B$. The variation of the eigenvalues of $C(t)$ falls under standard perturbation theory in quantum mechanics. If $\lambda_\mu(t)$ denotes the eigenvalues of $C(t)$, we have, to first order in $\dd t$:
\begin{equation}
\lambda_\mu(t+\dd t)=\lambda_\mu(t)+\sqrt{\dd t} \tilde{B}_{\mu\mu}+ \dd t \sum_{\mu\neq \nu}\, \frac{\vert \tilde{B}_{\mu\nu} \vert^2}{\lambda_\mu(t)-\lambda_\nu(t)}+\mathcal{O}(\dd t^{3/2})\,,
\end{equation}
where $\tilde{B}$ is the projection of $B$ onto the eigenvector (assumed to be non-degenerate) of $\lambda_\mu(t)$. It is easy to see that $\tilde{B}_{\mu\mu}$ is a centered Gaussian variable with variance $\sigma^2/N$. Furthermore, the mean of $|\tilde{B}_{\mu\nu}|^2$ is $\sigma^2/N$, with a standard deviation of the same order, such that this term can be viewed as a deterministic quantity. Thus:
\begin{equation}
\dd \lambda_\mu= \sqrt{\frac{\sigma^2}{N}} \dd \eta_\mu(t)+\frac{\sigma^2}{N}\sum_{\mu\neq \nu} \frac{\dd t}{\lambda_\mu-\lambda_\nu}\,,\label{eqflowDBM}
\end{equation}
where $\dd \eta_\mu(t)$ is a Gaussian noise with variance $\dd t$. Many properties of the matrix, including the distribution of its eigenvalues, can be obtained from the (empirical) Stieltjes transform:
\begin{equation}
g_N(t,z):=\frac{1}{N}\sum_{\mu=1}^N \, \frac{1}{z-\lambda_\mu(t)}\,.
\end{equation}
It is easy to see that the expansion in $1/z$ generates the moments of the matrix $C(t)$. At large $N$, the moments converge to deterministic functions, as does $g_N(t,z) \to g(t,z)$. It is easy to derive the differential equation governing the evolution of the asymptotic function $g(t,z)$ by using \eqref{eqflowDBM}. We use the Itô convention, which stipulates that $C(t)$ does not depend on $\dd\eta_\mu(t^\prime)$ for $t^\prime \geq t$, i.e., $C(t)$ depends only on previous steps\footnote{The Itô convention is equivalent to assuming $\Theta(t=0)=0$, where $\Theta$ is the standard step function.} $\tau < t$. With this convention, it is easy to see that $g(t,z)$ follows a Burgers-type equation \cite{potters2020first}:
\begin{equation}
\frac{\partial g(t,z)}{\partial t}=-\sigma^2 g(t,z) \frac{\partial g(t,z)}{\partial z}\,.
\end{equation}
The equation can be solved formally by the method of characteristics, and yields the closed-form equation:
\begin{equation}
g(t,z)=g(0,z-\sigma^2 t g(t,z))\,.
\end{equation}
where $g(0, z)$ denotes the Stieltjes transform of the initial matrix $A$. By defining the inverse function $\mathfrak{z}(t, g)$ such that $g(t, \mathfrak{z}(t, g)) = g$, the preceding relations yield:
\begin{equation}
\mathfrak{z}(t,g)=\mathfrak{z}(0,g)+\sigma^2 t g\,,
\end{equation}
and in particular, for $t=1$, which corresponds to the sum $C=A+B$,
\begin{equation}
\mathfrak{z}_{A+B}(g)=\mathfrak{z}_{A}(g)+\sigma^2 g\,.
\end{equation}
For a pure Wigner matrix with variance $\sigma^2/N$, $\mathfrak{z}(g) = \sigma^2 g + g^{-1}$, and therefore:
\begin{equation}
\boxed{\mathfrak{z}_{A+B}(g)=\mathfrak{z}_{A}(g)+\mathfrak{z}_{B}(g)-g^{-1}\,.}
\end{equation}
This relation is often written in terms of the cumulant generating function $R(g) := \mathfrak{z}(g) - g^{-1}$:
\begin{equation}
R_{A+B}(g)=R_A(g)+R_B(g)\,,
\end{equation}
where the $R$-function admits the following expansion in terms of the free cumulants $\kappa_n$ of the distribution:
\begin{equation}
R_A(g)=\sum_{n=1}^\infty \kappa_n g^{n-1}\,.
\end{equation}

\section{Complement to section \ref{sectionfree}}\label{App3}

In this section, we will provide a supplement to the results presented in Section \ref{sectionfree}. We will begin with a perturbative calculation, evaluating the corrections in powers of $\beta$ of the spectrum of the intermediate field $\sigma$. Let us start by recalling that since $\sigma$ is the free sum of a Wigner matrix $A$ of variance $1/N$ and the matrix $-\beta M^2$, we have:
\begin{equation}
R_\sigma(g)=R_A(g)+R_{-\beta M^2}(g)=R_A(g)-\beta R_{M^2}(-\beta g)\,.
\end{equation}
By expanding the second term on the right-hand side in powers of $\beta$,
\begin{equation}
R_\sigma(g)=g-\beta \kappa_2+\beta^2 g (\kappa_4+\kappa_2^2)+\mathcal{O}(\beta^3)\,,\label{perturbation}
\end{equation}
where we used $R_A(g)=g$ for a Wigner matrix, and the fact that the $R$-transform is the generating functional of the cumulants $\kappa_n$: $R_M(g):=\sum_{n=1}^\infty \kappa_n g^{n-1}$. The cumulants mentioned in \eqref{perturbation} are those of the matrix $M$:
\begin{align}
\kappa_2&=\tau(M^2)\,,\\
\kappa_4&=\tau(M^4)-2\tau(M^2)^2\,.
\end{align}
In the limit $N\to \infty$ (which is the limit in which the matrices are free), $\tau(M^{2n})=\int_{-b}^{b} \mu(x) x^{2n} \dd x$, where $\mu(x)$ is given by formula \eqref{densityM4} of Appendix \ref{App1}. Explicitly, we find:
\begin{align}
\kappa_2&=\frac{\sqrt{1+12 a_4}+6 a_4\left(2 \sqrt{1+12 a_4}-3\right)-1}{54 a_4^2}\\
\kappa_4&=\\
\kappa_4&=\frac{\left(2+33a_4+108a_4^2\right)\sqrt{1+12a_4}-2-45a_4-270a_4^2-270a_4^3}{1458\, a_4^4}\quad\Big(=\frac{1-\kappa_2}{a_4}-2\kappa_2^2\Big)
\end{align}
The definition of the inverse function of the Stieltjes transform, $\mathfrak{z}(g):=R(g)+g^{-1}$, thus gives us:
\begin{equation}
\mathfrak{z}_\sigma(g)=g(1+\beta^2 (\kappa_4+\kappa_2^2))+g^{-1}-\beta \kappa_2+\mathcal{O}(\beta^3)\,.
\end{equation}
We deduce the Stieltjes transform $g_\sigma(z)$, such that $g_\sigma(\mathfrak{z}(g))=g$ (up to order $\mathcal{O}(\beta^3)$):
\begin{align}
g_\sigma(z)&=\frac{\beta \kappa_2+z}{2(1+\beta^2 (\kappa_4+\kappa_2^2))}- \mathrm{Sign}(\Re (z+\beta \kappa_2))\frac{\sqrt{(z+\beta \kappa_2)^2-4(1+\beta^2 (\kappa_4+\kappa_2^2))}}{2(1+\beta^2 (\kappa_4+\kappa_2^2))}\\
g_\sigma(z)&=\frac{\beta \kappa_2+z}{2(1+\beta^2 (\kappa_4+\kappa_2^2))}- \frac{\sqrt{(z-\lambda_+)(z-\lambda_-)}}{2(1+\beta^2 (\kappa_4+\kappa_2^2))}\,,
\end{align}
with:
\begin{equation}
\lambda_\pm=\pm 2 \sqrt{1+\beta^2 (\kappa_4+\kappa_2^2)}-\beta \kappa_2\,.
\end{equation}
The eigenvalue distribution is then deduced from the Sokhotski-Plemelj formula:
\begin{equation}
\mu(\lambda,\beta)=\frac{\sqrt{(\lambda_+-\lambda)(\lambda-\lambda_-)}}{2\pi(1+\beta^2 (\kappa_4+\kappa_2^2))}
\end{equation}

\begin{figure}
\begin{center}
\includegraphics[scale=0.6]{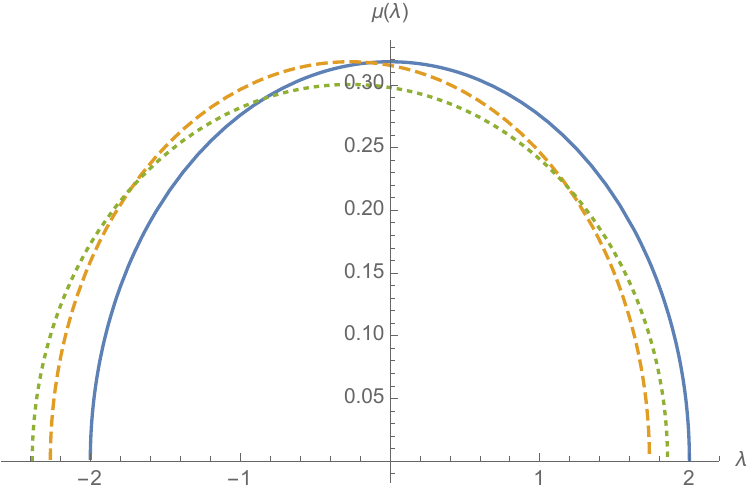}
\end{center}
\caption{Behavior of the distribution $\mu(\lambda,\beta)$ at orders $\beta^0$ (blue curve), $\beta^1$ (dashed yellow curve), and $\beta^2$ (dotted green curve). The distribution is computed for $a_4=-1/12$.}\label{figbetaexp}
\end{figure}

Figure \ref{figbetaexp} shows the behavior of the distribution $\mu(\lambda,\beta)$, at orders $\beta^0$, $\beta^1$, and $\beta^2$ respectively. Note that at order $\beta^0$, we recover an ordinary Wigner distribution. Furthermore, the distribution spreads out and shifts further and further to the left, as expected.

As a further check, let us compute the exact function $\mathfrak{z}_\sigma(g)$ of the model. To do so, we first compute $\mathfrak{z}_{-\beta M^2}(g)=-\beta \mathfrak{z}_{M^2}(-\beta g)$. One method consists of computing the Stieltjes transform of $M$, then deducing that of $M^2$, and working back to the function $\mathfrak{z}$. The saddle-point equation of \eqref{Eigendis} reads:
\begin{equation}
\mathcal{V}^\prime(\lambda_i)=\frac{2}{N}\, \sum_{j\neq i} \frac{1}{\lambda_i-\lambda_j}\,.
\end{equation}
Multiplying both sides by $1/(z-\lambda_i)$, noting that the indices $i$ and $j$ are dummy indices on the left-hand side, and symmetrizing the sums, one shows that the saddle-point equation becomes:
\begin{equation}
g_M^2(z)-\mathcal{V}^\prime(z) g_M(z)+\Pi(z)=0\,,
\end{equation}
where $g_M(z)$ is the empirical Stieltjes transform\footnote{The empirical Stieltjes transform converges almost surely to the mean value for large $N$ due to the concentration of measure.} of $M$:
\begin{equation}
g_M(z):=\frac{1}{N}\sum_{i=1}^N \frac{1}{z-\lambda_i}\,,
\end{equation}
and $\Pi(z)$ is a polynomial of degree $k-2$ if $\mathcal{V}$ is of degree $k$, explicitly:
\begin{equation}
\Pi(z)=\frac{1}{N}\sum_{i=1}^N \, \frac{\mathcal{V}^\prime(z)-\mathcal{V}^\prime(\lambda_i)}{z-\lambda_i}\,.
\end{equation}
Here, we focus on a quartic model, $\mathcal{V}(z):=z^2/2+a_4 z^4/4$, and $\Pi(z)$ will thus be of degree $2$: $\Pi(z)=a+b z+ c z^2$. The coefficients $a$, $b$, and $c$ can be determined from the asymptotic expansion of the Stieltjes transform \eqref{gAdef}. We find the conditions:
\begin{align}
a&=1+a_4 \tau(M^2)\,,\\
b&=0 \,,\quad (\text{in agreement with}\,\, \tau(M)=0)\\
c&=a_4\,,\\
1&=\tau(M^2+a_4 M^4)\,.
\end{align}
Note that the last condition is also a consequence of the Schwinger-Dyson equations\footnote{By expanding the condition$$\int \dd M\, \sum_{i,j}\frac{\partial}{\partial M_{ij}} M_{ij}\, e^{-N \Tr\, \mathcal{V}(M)}=0\,.$$}. We therefore find,
\begin{equation}
g_M^2(z)-\mathcal{V}^\prime(z) g_M(z)+(1+a_4 m_2)+a_4 z^2=0\,.\label{constraintg}
\end{equation}
Before moving on to the equation for $g_{M^2}(z)$, we need to fix $m_2$. At large $N$, assuming this quantity is self-averaging:
\begin{equation}
g_M(z)=\frac{1}{2} \left(\mathcal{V}^\prime(z) - \sqrt{D(z)} \right)\,,
\end{equation}
with:
\begin{equation}
D(z):=(\mathcal{V}^\prime(z))^2-4 \Pi(z)=a_4^2 \left(z^6+\frac{2}{a_4} z^4+\frac{1-4 a_4}{a_4^2}z^2-4\frac{1+a_4 m_2}{a_4^2}\right)\,.
\end{equation}
Setting $X=z^2$, we look for a solution of the form $(X-b^2)(X+\beta)^2$. Identifying term by term, we find:
\begin{align}
2\beta -b^2&= 2(a_4)^{-1}\\
\beta (\beta-2 b^2)&=\frac{1-4 a_4}{a_4^2}\\
b^2 \beta^2&= 4\frac{1+a_4 m_2}{a_4^2}\,.\label{eqconstraintapp3}
\end{align}
From the first two equations, we find $3 a_4 b^4+4 b^2 -16=0$, whose solution was given in Appendix \ref{App1}. As a consistency check, we indeed recover the eigenvalue distribution given in this appendix using the Sokhotski-Plemelj formula. Note also that the third constraint \eqref{eqconstraintapp3} fixes the second moment of the distribution.

Let us now express the saddle-point equation \eqref{constraintg} for $g_{M^2}(z)$. From the definition:
\begin{align}
g_{M^2}(z^2)=\frac{1}{N}\sum_{i=1}^N \frac{1}{z^2-\lambda_i^2}=\frac{1}{2 z N} \sum_i \left(\frac{1}{z-\lambda_i}-\frac{1}{-z-\lambda_i}\right)\,,
\end{align}
and since, for a symmetric potential, $g_{M}(z)=-g_{M}(-z)$, we deduce:
\begin{equation}
g_{M^2}(z)=\frac{1}{\sqrt{z}}\, g_{M}(\sqrt{z})\,,
\end{equation}
so that the saddle-point equation becomes, evaluating it for $z\to \sqrt{z}$ and using the previous relation:
\begin{equation}
\boxed{z g_{M^2}^2(z)+(1+ a_4 m_2)  + a_4 z-(z+ a_4 z^2) g_{M^2}(z)=0\,.}\label{equationcarg}
\end{equation}
Inverting this relation and choosing the branch that behaves as $1/g$ as $g\to 0$, we find:
\begin{equation}
\mathfrak{z}_{M^2}(g) = \frac{(g^2 - g + a_4) \pm  \sqrt{(g^2 - g + a_4)^2 + 4 a_4 g (1 + a_4 m_2)}}{2 a_4 g}\,,
\end{equation}
from which we finally deduce:
\begin{equation}
\boxed{\mathfrak{z}_\sigma(g) = g + \frac{(\beta^2 g^2 + \beta g + a_4) \pm \sqrt{(\beta^2 g^2 + \beta g + a_4)^2 - 4 a_4 \beta g (1 + a_4 m_2)}}{2 a_4 g}}\,.
\end{equation}
The choice of the sign is quite subtle and must satisfy a twofold requirement: 1) $\mathfrak{z}_{M^2}(g)$ must tend to $1/g$ in the limit $g\to 0$, and 2) $\mathfrak{z}_{M^2}(g)$ must tend to the solution $1/(g-g^2)$ in the limit $a_4\to 0$. More precisely, we require that the following expansions hold simultaneously:
\begin{align}
\mathfrak{z}_{M^2}(g)&= \frac{1}{g}+  \kappa_2 + \mathcal{O}(g)\,\label{limit1App}\\
\mathfrak{z}_{M^2}(g)&= \frac{1}{g-g^2}+\mathcal{O}(a_4)\,,\label{limit2App}
\end{align}
where, in the first expression, $\kappa_2$ is again the second cumulant of $M^2$. The signs will depend on the order in which the limits are taken. Let us first examine the second condition. A Taylor expansion of the square root yields:
\begin{align}
\mathfrak{z}_{M^2}(g) = \frac{g^2-g}{2ga_4}+\frac{1}{2g}\pm \left(\frac{\vert g^2-g \vert}{2 g a_4}+\frac{g^2+g}{\vert g^2-g \vert}\frac{1}{2g}\right)\,.
\end{align}
The sign in front of the parenthesis will therefore depend on that of $g^2-g$:
\begin{enumerate}
    \item If $g^2-g>0$, the minus sign is required, and after simplification, we recover the limit \eqref{limit2App}.
    \item If $g^2-g<0$, the plus sign is required, which gives the correct limit, since $\vert g^2-g \vert=g-g^2$ in this case. 
\end{enumerate}
Let us now examine the first limit \eqref{limit1App}. Let $\epsilon= \mathrm{Sign}(a_4)$; setting $a=1+\epsilon \vert a_4 \vert \kappa_2$, we have:
\begin{align}
\mathfrak{z}_{M^2}(g) &= \frac{\vert a_4 \vert-\epsilon g}{2 g \vert a_4 \vert}\pm \epsilon \frac{1}{2 g \vert a_4 \vert} \left( \vert a_4 \vert^2-2 \epsilon \vert a_4 \vert g+ 4 \epsilon \vert a_4 \vert g a  \right)^{1/2}+\mathcal{O}(g)\\
&= \frac{\vert a_4 \vert-\epsilon g}{2 g \vert a_4 \vert}\pm \epsilon \frac{1}{2 g} \left( 1- \epsilon \vert a_4 \vert^{-1} g(1-2a)  \right)+\mathcal{O}(g)\\
&= \frac{1\pm \epsilon}{2 g } -\frac{\epsilon}{2\vert a_4 \vert} (1\pm \epsilon (1-2 a)+\mathcal{O}(g)\,.
\end{align}
The last equality requires $\pm \epsilon=1$, which implies that the last term reduces to $\kappa_2$. These two limits, considered simultaneously, impose specific matching conditions on the construction of the solution for $\mathfrak{z}_\sigma(g)$, the matching points being fixed by the zeros of the polynomial:
\begin{equation}
\Delta := (\beta^2 g^2 + \beta g + a_4)^2 - 4 a_4 \beta g (1 + a_4 m_2)\,.
\end{equation}

\begin{figure}
\begin{center}
$\underset{a}{\vcenter{\hbox{\includegraphics[scale=0.35]{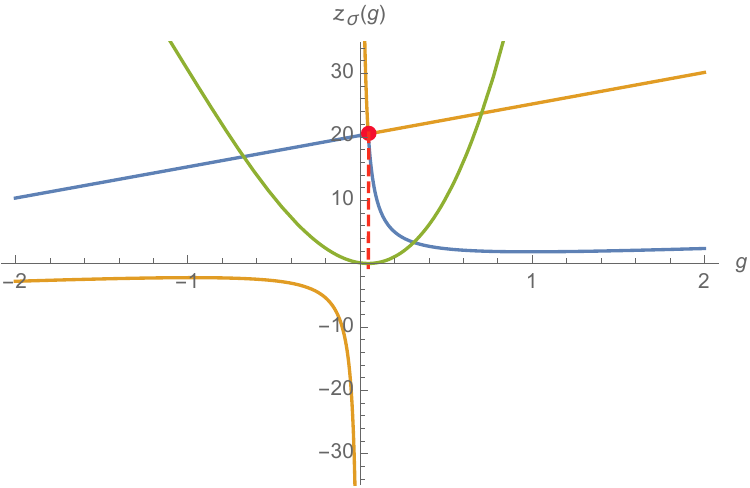}}}}$\quad $\underset{b}{\vcenter{\hbox{\includegraphics[scale=0.35]{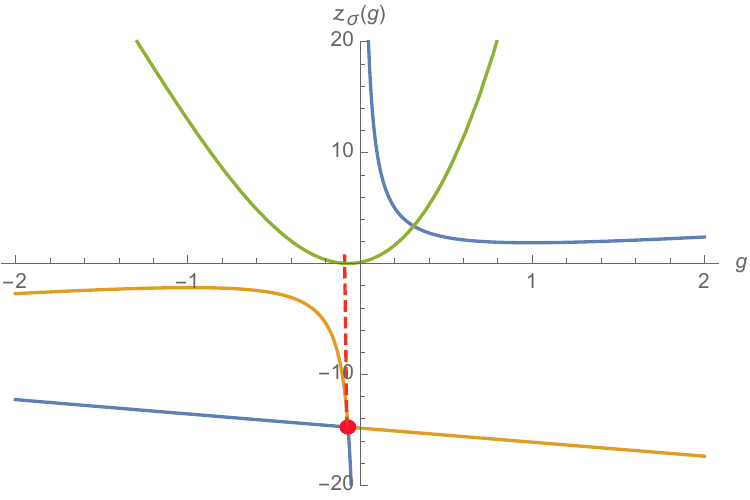}}}}$\quad $\underset{c}{\vcenter{\hbox{\includegraphics[scale=0.35]{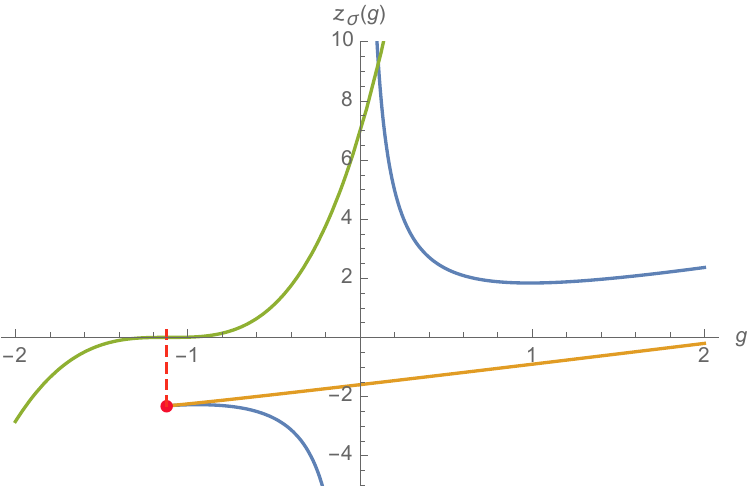}}}}$
\end{center}
\caption{From left to right, the two branches of $\mathfrak{z}_\sigma(g)$ (blue curve for the minus sign and yellow curve for the plus sign). In all three figures, $\beta=0.15$. In the left figure (a), $a_4=0.01$, in the middle figure (b), $a_4=-0.01$, and in the right figure (c), $a_4=-1/12$. In each figure, the green curve corresponds to the discriminant $\Delta$ (times 1000) and the red dot represents the branch point of the solutions.}\label{figcut}
\end{figure}

Figure \ref{figcut} illustrates the construction of the "physical" solution of $\mathfrak{z}_\sigma(g)$ for three examples that cover the essential cases. On the left (Figure \ref{figcut}a), $a_4>0$, and the branch point (the red dot at the value $g_*$ where $\Delta$ vanishes) lies in the positive half-plane $g>0$. In the negative half-plane $g<0$, the physical solution corresponds to the yellow curve (plus sign), which simultaneously satisfies both limits \eqref{limit1App} and \eqref{limit2App}. For $g>0$, however, two cases must be distinguished. When $g<g_*$, the plus sign always corresponds to the physical solution: it behaves like $1/g$ at the origin, and the expansion in powers of $a_4$ shows that the zeroth-order term is indeed $1/(g-g^2)$. On the other hand, for $g>g_*$, the solution compatible with the value $1/(g-g^2)$ at zeroth order in $a_4$ corresponds to the solution with a minus sign. By matching the solutions at the branch point, we thus obtain a function $\mathfrak{z}_\sigma(g)$ that is continuous and compatible with both limits. Figure \ref{figcut}b illustrates the same mechanism when $a_4<0$. Finally, the last panel (c) illustrates the case of the critical regime ($a_4=-1/12$). In this latter case, the cusp at the branch point is pushed to the edge of the blue curve, which determines the physical function in both the $g>0$ and $g<0$ regions.

\begin{figure}
\begin{center}
$\underset{a'}{\vcenter{\hbox{\includegraphics[scale=0.5]{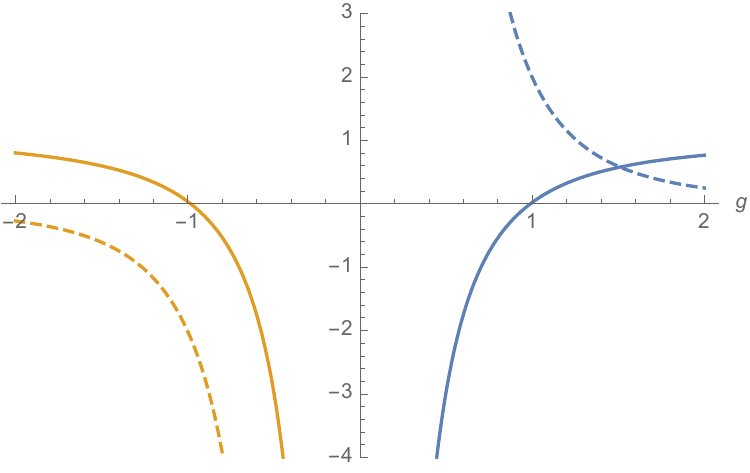}}}}$\quad $\underset{b'}{\vcenter{\hbox{\includegraphics[scale=0.5]{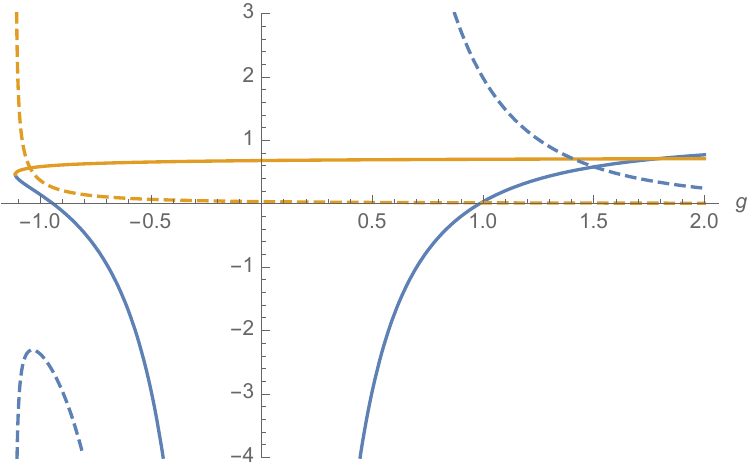}}}}$ \\ $\underset{c'}{\vcenter{\hbox{\includegraphics[scale=0.35]{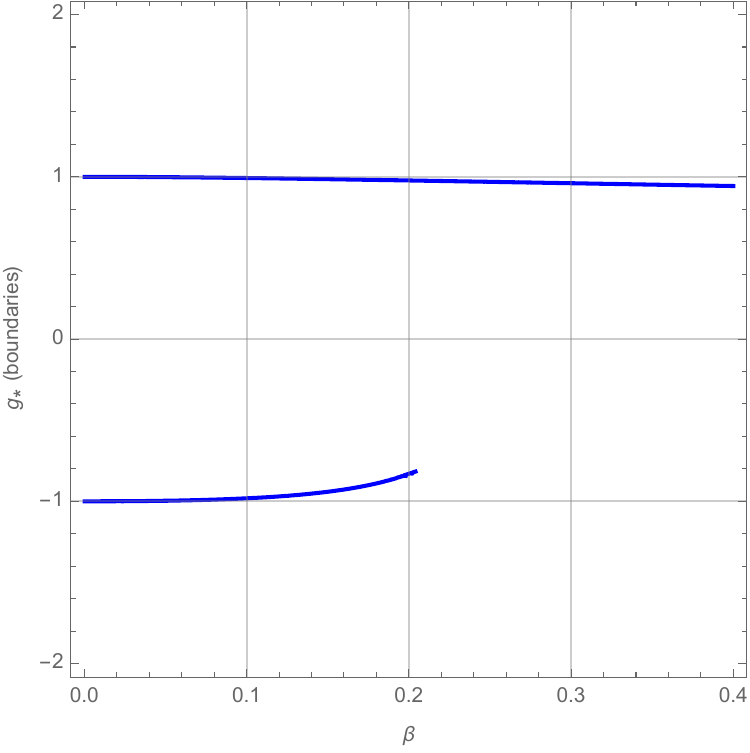}}}}$ \qquad \quad $\underset{d'}{\vcenter{\hbox{\includegraphics[scale=0.35]{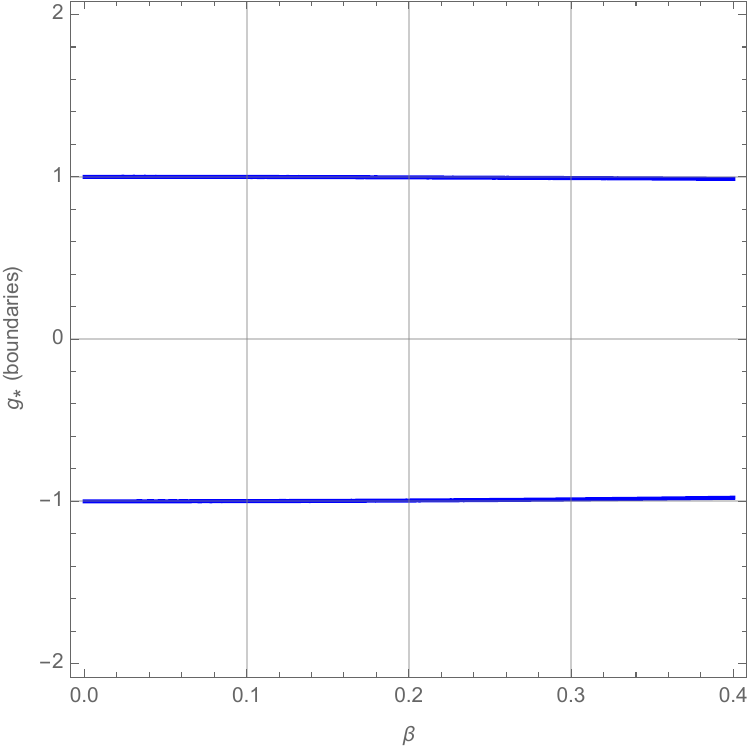}}}}$
\end{center}
\caption{Behavior of the first derivative $\mathfrak{z}_\sigma^\prime(g)$ (solid curves) and the second derivative $\mathfrak{z}_\sigma^{\prime\prime}(g)$ (dashed curves) corresponding to the physical solutions presented in Figures \ref{figcut}a ($a^\prime$) and \ref{figcut}c ($b^\prime$), respectively. Finally, panels $c^\prime$ and $d^\prime$ represent the set of solutions for $\mathfrak{z}_\sigma^\prime(g)=0$ at the critical point $a_4=-1/12$ (c) and for $a_4=1$ (d).}\label{figcut2}
\end{figure}

As explained in Section \ref{sectionfree}, the edges of the spectrum are determined by the condition $\mathfrak{z}_\sigma^\prime(g)=0$. Furthermore, the second derivative must not vanish in order to obtain a behavior $\mu(x)\sim \sqrt{\vert x_{\pm}-x\vert}$ at the spectral edge for the eigenvalue distribution (see Equation \eqref{equationexpansion}). Figure \ref{figcut2} summarizes the results for the examples chosen in Figure \ref{figcut}. We find that below $\beta_c=1/\sqrt{24}$, there exist simple zeros of the function $\mathfrak{z}_\sigma^\prime(g)$ (roots of multiplicity 1), ensuring a square-root behavior for the spectrum. Note that for each of the panels (a), (b), and (c) of Figure \ref{figcut}, the zeros are located at $x_- \approx -2.172$, $x_-\approx -2.181$, and $x_-= -2.277$, far from the value $-\beta b^2$ of the edge of the matrix $-\beta M^2$, which lies at the respective values $\lambda_-(a_4=0.01)\approx-0.583$, $\lambda_-(a_4=-0.01)\approx -0.62$, and $\lambda_-(a_4=-1/12)=-1.2$.

\section{Alternative proof of the flow equation}\label{App4}

We give here an alternative proof of the flow equation \eqref{equationWett}, starting from the discrete theory and taking the continuum limit at the end of the calculations. This justifies in particular condition \eqref{chainrule}.

We therefore start from the discrete version of the partition function $\mathcal{Z}_{N,k}(\varphi,J)$:
\begin{equation}
{\mathcal{Z}}_{N,k} (\{p\},J) := \int \dd M \, e^{-N\sum_{p}\left(\frac{1}{2} (m+2 p)\, (M^2)_{pp} +\frac{u}{4}\,(M^4)_{pp} +\frac{1}{2}\sum_{p^\prime} R_k(p,p^\prime) M_{pp^\prime} M_{p^\prime p}- \sum_{p^\prime}J_{pp^\prime} M_{p^\prime p}\right)}\,.\label{equationzdisc}
\end{equation}
We will also define the discrete version of \eqref{Gammafunc}:
\begin{equation}
\Gamma_k[\{p\},\pi]+\frac{N}{2}\sum_{p,p^\prime}\, R_k(p,p^\prime) \,\pi_{pp^\prime} \pi_{p^\prime p} =N \sum_{p,p^\prime} J_{pp^\prime} \pi_{p^\prime p}- W_{N,k}[\{p\},J]\,,
\end{equation}
where $W_{N,k}[\{p\},J]:=\ln (\mathcal{Z}_{N,k}(\{p\},J))$, implying in particular:
\begin{equation}
\frac{\partial }{\partial J_{pp^\prime}} W_{N,k}[\{p\},J]=N \pi_{p^\prime p}\,,\qquad \frac{\partial \Gamma_k[\{p\},\pi]}{\partial \pi_{pp^\prime}}+N R_k(p,p^\prime) \pi_{p^\prime p}= N J_{p^\prime p}\,.\label{saddlepointcond}
\end{equation}
At this stage, compatibility with the continuous definition \eqref{Gammafunc} implies in particular:
\begin{equation}
\Pi(p,p^\prime):=N^{3/2} \pi_{pp^\prime}\,,\qquad \mathcal{J}(p,p^\prime):=N^{3/2} J_{pp^\prime}\,.
\end{equation}
By differentiating equation \eqref{equationzdisc} with respect to $k$, we obtain:
\begin{equation}
\frac{\dd }{\dd k} W_{N,k}[\{p\},J]= -\frac{1}{2N} \sum_{p,p^\prime} \, \frac{\dd R_k(p,p^\prime)}{\dd k}\left( \frac{\partial^2 W_{N,k}}{\partial J_{pp^\prime} J_{p^\prime p}} + \frac{\partial W_{N,k}}{\partial J_{p p^\prime}} \frac{\partial W_{N,k}}{\partial J_{p^\prime p}}\right)\,.
\end{equation}
The same manipulation that, in section \ref{flowequationsection}, allows passing from the derivative with respect to $k$ at fixed source to the derivative with respect to $k$ at fixed classical field gives here
\begin{equation}
\dot{\Gamma}_{N,k}[\{p\},J]=\frac{1}{2N} \sum_{p,p^\prime} \,\dot{R}_k(p,p^\prime)\, \frac{\partial^2 W_{N,k}}{\partial J_{pp^\prime} J_{p^\prime p}} \,.\label{flotdiscret}
\end{equation}
By differentiating the left-hand relation of equation \eqref{saddlepointcond}, we have:
\begin{equation}
\sum_{q,q^\prime}\, \frac{\partial^2 W_{N,k}}{\partial J_{qq^\prime}\partial J_{pp^\prime}}\frac{\partial J_{qq^\prime}}{\partial \pi_{p^{\prime\prime} p^{\prime\prime\prime}}}=N\delta_{p^{\prime\prime}  p^\prime} \delta_{p^{\prime\prime\prime} p}\,,
\end{equation}
and using the right-hand equation of \eqref{saddlepointcond},
\begin{equation}
\sum_{q,q^\prime}\, \frac{\partial^2 W_{N,k}}{\partial J_{qq^\prime}\partial J_{pp^\prime}} \left(\frac{\partial^2 \Gamma_{N,k}}{\partial \pi_{p^{\prime\prime} p^{\prime\prime\prime}} \partial \pi_{q^\prime q}}+N R_k(p^{\prime\prime},p^{\prime\prime\prime})\delta_{q^\prime  p^{\prime\prime\prime}} \delta_{q p^{\prime\prime}}\right)=N^2\delta_{p^{\prime\prime}  p^\prime} \delta_{p^{\prime\prime\prime} p}\,,
\end{equation}
and we define:
\begin{equation}
\frac{\partial^2 \Gamma_{N,k}}{\partial \pi_{p^{\prime\prime} p^{\prime\prime\prime}} \partial \pi_{q^\prime q}}\bigg\vert_{\pi=0}= N K_k(p^{\prime\prime},p^{\prime\prime\prime}) \, \delta_{q^\prime  p^{\prime\prime\prime}} \delta_{q p^{\prime\prime}}\,,
\end{equation}
implying:
\begin{equation}
\dot{\Gamma}_{N,k}[\{p\}]=\frac{1}{2} \sum_{p,p^\prime} \,\dot{R}_k(p,p^\prime)\, \frac{1}{K_k(p,p^\prime)+R_k(p,p^\prime)}\,.
\end{equation}
Taking the continuum limit,
\begin{equation}
\dot{\Gamma}_{N,k}[\{p\}] \to \frac{N^2}{2}\int \, \dd p \dd p^\prime\, \rho(p)\rho(p^\prime)\, \frac{\dot{R}_k(p,p^\prime)}{K_k(p,p^\prime)+R_k(p,p^\prime)}\,.
\end{equation}
The analogue of truncation \eqref{truncationvertex} is written here as:
\begin{equation}
{\Gamma}_{N,k}[\{p\},\pi]:=N \left( \sum_p \frac{1}{2} (m(k)+2 Z(k) p)\, (\pi^2)_{pp}+\sum_{n>1}\, \frac{u_{2n}(k)}{2n} \Tr \pi^{2n}\right)\,.\label{troncation_discrete}
\end{equation}
By substituting the truncation into the flow equation, it is easy to see that one recovers the beta functions derived in section \ref{IRfixedpoint}.

Let us show this by explicitly calculating the beta functions for $m$, $\eta$, and $u_4$. To do this, we will project the flow equation \eqref{flotdiscret} onto a diagonal classical field: $\pi_{ij}=\chi_i \delta_{ij}$, and use the truncation \eqref{troncation_discrete}, which becomes:
\begin{equation}
{\Gamma}_{N,k}[\{p\},\chi]:=N \sum_p\, \left(  \frac{1}{2} (m(k)+2 Z(k) p)\, \chi_p^2+\sum_{n>1}\, \frac{u_{2n}(k)}{2n}  \chi^{2n}_p\right)\,.\label{troncation_discrete2}
\end{equation}
Consider the derivative of a typical monomial $\Tr \, \pi^{2n}$; we find terms of this kind:
\begin{equation}
\frac{\partial^2}{\partial \pi_{pp^\prime} \partial \pi_{p^\prime p}}\, \Tr \, \pi^{2n}= 2n \sum_{k=0}^{2n-2} \chi_p^k \chi_{p^\prime}^{2n-2-k}\,.
\end{equation}
Let $F_{pp^\prime}$ be the contribution of the second-derivative matrix containing the interactions, such that the flow equation becomes:
\begin{equation}
\dot{\Gamma}_{N,k}[\{p\},\chi] = \frac{1}{2} \sum_{p,p^\prime} \, \frac{\dot{R}_k(p,p^\prime)}{K_k(p,p^\prime)+R_k(p,p^\prime)+F_{pp^\prime}}\,.
\end{equation}
Explicitly,
\begin{equation}
F_{pp\prime}=u_4 (\chi^2_p+\chi_p\chi_{p^\prime}+\chi_{p^\prime}^2)+u_6(\chi^4_p+\chi_p^3\chi_{p^\prime}+\chi_p\chi_{p^\prime}^3+\chi_p^2\chi_{p^\prime}^2+\chi_{p^\prime}^4)+\cdots\,.
\end{equation}
Denoting $G_k^{-1}(p,p^\prime):=K_k(p,p^\prime)+R_k(p,p^\prime)$, we have, expanding in powers of $F$ (dropping here the zeroth-order term that corrects the vacuum energy):
\begin{equation}
\dot{\Gamma}_{N,k}[\{p\},\chi] =\frac{1}{2} \sum_{p,p^\prime} \, \dot{R}_k(p,p^\prime) \left( -G^2_k(p,p^\prime) F_{pp^\prime}+  G^3_k(p,p^\prime) F_{pp^\prime}^2+\cdots\right)\,.
\end{equation}
To read off the flow equations, it now suffices to isolate on both sides of the equation the single-trace contributions of the same power. Consider for example the flow of $m$. We have:
\begin{equation}
-\sum_{p,p^\prime} \, \dot{R}_k(p,p^\prime)G^2_k(p,p^\prime) F_{pp^\prime}=-2 u_4 \sum_{p,p^\prime} \, \dot{R}_k(p,p^\prime) G^2_k(p,p^\prime) \chi_p^2+\cdots\,,
\end{equation}
and thus we directly obtain the identification
\begin{equation}
\sum_p \chi_p^2(\dot{m}+2\dot{Z}p)+\cdots=-\frac{2 u_4}{N} \sum_{p,p^\prime} \, \dot{R}_k(p,p^\prime) G^2_k(p,p^\prime) \chi_p^2\,,
\end{equation}
from which we deduce, in the continuum limit:
\begin{align}
\dot{m}&=-2 u_4 \int \rho(p)\, \dd p\, G^{2}_k(p,0) \dot{R}_k(p,0)\\
\dot{Z}&=-u_4 \frac{\dd}{\dd p^\prime}\, \int \rho(p)\, \dd p\, G^{2}_k(p,p^\prime) \dot{R}_k(p,p^\prime)\bigg\vert_{p\prime=0}\,.
\end{align}
To derive the flow equations for $u_4$, one must consider the terms of order $F^2$. Those contributing to the flow of $u_4$ involve $\chi_p^4$ and $\chi_{p^\prime}^4$, so there are two terms. Furthermore, there are two identical terms, at linear order, coming from the sextic interaction, thus we have:
\begin{equation}
\frac{N}{4}\sum_p\dot{u}_4 \chi_p^4+\cdots=-u_6\sum_{p,p^\prime} \, \dot{R}_k(p,p^\prime) G^2_k(p,p^\prime) \chi_p^4+u_4^2\sum_{p,p^\prime} \, \dot{R}_k(p,p^\prime) G^3_k(p,p^\prime)\chi_p^4\,,
\end{equation}
from which we now deduce:
\begin{equation}\dot{u}_4=-4u_6\int \rho(p)\, \dd p\, \dot{R}_k(p,0) G^2_k(p,0)+4 u_4^2 \int \rho(p)\, \dd p\,  \dot{R}_k(p,0) G^3_k(p,0)\,.
\end{equation}

\section*{Acknowledgments}

V.L., for his part, extends his warm thanks to the little crab and the swan of destiny. 

\printbibliography[heading=bibintoc]

@article{wigner1967random,
  title={Random matrices in physics},
  author={Wigner, Eugene P},
  journal={SIAM review},
  DOI={10.1137/100900},
  volume={9},
  number={1},
  pages={1--23},
  year={1967},
}

@article{penner1988perturbative,
  title={Perturbative series and the moduli space of Riemann surfaces},
  author={Penner, Robert C},
  journal={Journal of Differential Geometry},
  DOI={https://doi.org/10.4310/JDG/1214441648},
  volume={27},
  number={1},
  pages={35--53},
  year={1988},
}

@article{wetterich1993exact,
  title={Exact evolution equation for the effective potential},
  author={Wetterich, Christof},
  journal={Physics Letters B},
  DOI={https://doi.org/10.1016/0370-2693(93)90726-X},
  volume={301},
  number={1},
  pages={90--94},
  year={1993},
}

@book{voiculescu1992free,
  title={Free random variables},
  author={Voiculescu, Dan V and Dykema, Ken J and Nica, Alexandru},
  volume={1},
  year={1992},
  publisher={American Mathematical Soc.}
}

@article{Rivasseau_2015,
   title={Why are tensor field theories asymptotically free?},
   volume={111},
   ISSN={1286-4854},
   url={http://dx.doi.org/10.1209/0295-5075/111/60011},
   DOI={10.1209/0295-5075/111/60011},
   number={6},
   journal={EPL (Europhysics Letters)},
   author={Rivasseau, V.},
   year={2015},
   month=sep, pages={60011} }

@article{reuter1998nonperturbative,
  title={Nonperturbative evolution equation for quantum gravity},
  author={Reuter, Martin},
  journal={Physical Review D},
  DOI={10.1103/PhysRevD.57.971},
  volume={57},
  number={2},
  pages={971},
  year={1998},
}

@book{mehta2004random,
  title={Random matrices},
  author={Mehta, Madan Lal},
  DOI={9780124880511/random-matrices},
  volume={142},
  year={2004},
  publisher={Elsevier}
}

@article{rovelli1993statistical,
  title={Statistical mechanics of gravity and the thermodynamical origin of time},
  author={Rovelli, Carlo},
  journal={Classical and Quantum Gravity},
  DOI={10.1088/0264-9381/10/8/015},
  volume={10},
  number={8},
  pages={1549--1566},
  year={1993}
}

@book{tricomi1985integral,
  title={Integral equations},
  author={Tricomi, Francesco Giacomo},
  volume={5},
  year={1985},
  publisher={Courier corporation}
}

@book{potters2020first,
  title={A first course in random matrix theory: for physicists, engineers and data scientists},
  author={Potters, Marc and Bouchaud, Jean-Philippe},
  year={2020},
  DOI={10.1017/9781108768900},
  publisher={Cambridge University Press}
}

@article{di19952d,
  title={2D gravity and random matrices},
  author={Di Francesco, Philippe and Ginsparg, Paul and Zinn-Justin, Jean},
  journal={Physics Reports},
  DOI={https://doi.org/10.1016/0370-1573(94)00084-G},
  volume={254},
  number={1-2},
  pages={1--133},
  year={1995},
}

@article{eynard2015random,
  title={Random matrices},
  author={Eynard, Bertrand and Kimura, Taro and Ribault, Sylvain},
  DOI={https://doi.org/10.48550/arXiv.1510.04430},
  journal={ arXiv:1510.04430},
  year={2015}
}

@article{barrett2018tullio,
  title={Tullio Regge's legacy: Regge calculus and discrete gravity},
  author={Barrett, John W and Oriti, Daniele and Williams, Ruth M},
  journal={ arXiv:1812.06193},
  DOI={https://doi.org/10.48550/arXiv.1812.06193},
  year={2018}
}

@article{brezin1990exactly,
  title={Exactly solvable field theories of closed strings},
  author={Br{\'e}zin, Edouard and Kazakov, Vladimir A},
  journal={Physics Letters B},
  DOI={10.1016/0370-2693(90)90818-Q},
  volume={236},
  number={2},
  pages={144--150},
  year={1990},
}

@article{duplantier2011liouville,
  title={Liouville quantum gravity and KPZ},
  author={Duplantier, Bertrand and Sheffield, Scott},
  DOI={10.1007/s00222-010-0308-1},
  journal={Invent. math. },
  volume={185},
  number={2},
  pages={333--393},
  year={2011},
}

@article{seiberg1990notes,
  title={Notes on quantum Liouville theory and quantum gravity},
  author={Seiberg, Nathan},
  journal={Progress of Theoretical Physics Supplement},
  DOI={https://doi.org/10.1143/PTP.102.319},
  volume={102},
  pages={319--349},
  year={1990},
}

@article{gurau2024quantum,
  title={Quantum gravity and random tensors},
  author={Gurau, Razvan and Rivasseau, Vincent},
  journal={arXiv:2401.13510},
  DOI={https://doi.org/10.48550/arXiv.2401.13510},
  year={2024}
}

@article{freidel2005group,
  title={Group field theory: an overview},
  author={Freidel, Laurent},
  journal={International Journal of Theoretical Physics},
  DOI={10.1007/s10773-005-8894-1},
  volume={44},
  number={10},
  pages={1769--1783},
  year={2005},
}

@book{carrozza2014tensorial,
  title={Tensorial methods and renormalization in group field theories},
  author={Carrozza, Sylvain},
  DOI={10.1007/978-3-319-05867-2},
  year={2014},
  publisher={Springer Science \& Business Media}
}

@article{boulatov1992model,
  title={A model of three-dimensional lattice gravity},
  author={Boulatov, Dimitri V},
  journal={Modern Physics Letters A},
  DOI={https://doi.org/10.1142/S0217732392001324},
  volume={7},
  number={18},
  pages={1629--1646},
  year={1992},
}

@article{Lahoche_2024PFP,
   title={An intriguing connection between Pisarski's fixed point and (2+3)-spin glasses},
   volume={525},
   ISSN={0375-9601},
   url={http://dx.doi.org/10.1016/j.physleta.2024.129906},
   DOI={10.1016/j.physleta.2024.129906},
   journal={Physics Letters A},
   author={Lahoche, Vincent and Ousmane Samary, Dine},
   year={2024},
   month=Nov, pages={129906} }

@book{rovelli2004quantum,
  title={Quantum Gravity},
  author={Rovelli, Carlo},
  year={2004},
  DOI={10.1017/CBO9780511755804},
  publisher={Cambridge University Press},
  address={Cambridge, UK}
}

@article{loll2020quantum,
  title={Quantum gravity from causal dynamical triangulations: a review},
  author={Loll, Renate},
  journal={Classical and Quantum Gravity},
  DOI={10.1088/1361-6382/ab57c7},
  volume={37},
  number={1},
  pages={013002},
  year={2020},
}

@book{gurau2017random,
  title={Random tensors},
  author={Gurau, Razvan},
  DOI={https://doi.org/10.1093/acprof:oso/9780198787938.001.0001},
  year={2017},
  publisher={Oxford University Press}
}

@article{garban2012quantum,
  title={Quantum gravity and the KPZ formula},
  author={Garban, Christophe},
  DOI={https://www.numdam.org/book-part/AST_2013__352__315_0/},
  journal={arXiv preprint arXiv:1206.0212},
  year={2012}
}

@article{brezin1992renormalization,
  title={Renormalization group approach to matrix models},
  author={Brezin, Edouard and Zinn-Justin, Jean},
  journal={Physics Letters B},
  DOI={10.1016/0370-2693(92)91953-7},
  volume={288},
  number={1-2},
  pages={54--58},
  year={1992},
}

@article{duplantier2010liouvillequantumgravitykpz,
      title={Liouville Quantum Gravity and KPZ}, 
      author={Bertrand Duplantier and Scott Sheffield},
      journal={Invent. Math.},
      volume={185},
      pages={333--393},
      DOI={10.1007/s00222-010-0308-1},
      year={2011},
      eprint={0808.1560},
      archivePrefix={arXiv},
      primaryClass={math.PR},
      url={https://arxiv.org/abs/0808.1560}, 
}

@article{duplantier2010rigorous,
  title={A rigorous perspective on Liouville quantum gravity and the KPZ relation},
  author={Duplantier, B},
  journal={Exact methods in low-dimensional statistical physics and quantum computing},
  pages={529--561},
  year={2010},
}

@inproceedings{miller2018liouville,
  title={Liouville quantum gravity as a metric space and a scaling limit},
  author={Miller, Jason},
  DOI={https://doi.org/10.1142/9789813272880_0167},
  booktitle={Proceedings of the International Congress of Mathematicians: Rio de Janeiro 2018},
  pages={2945--2971},
  year={2018},
  organization={World Scientific}
}

@incollection{delamotte2012introduction,
  title={An introduction to the nonperturbative renormalization group},
  author={Delamotte, Bertrand},
  booktitle={Renormalization group and effective field theory approaches to many-body systems},
  editor={Schwenk, Achim and Polonyi, Janos},
  series={Lecture Notes in Physics},
  volume={852},
  DOI={10.1007/978-3-642-27320-9_2},
  pages={49--132},
  year={2012},
  publisher={Springer}
}

@misc{castro2026quantitativecharacterizationgravitationaluniversality,
      title={Towards a quantitative characterization of gravitational universality classes for order-4 random tensor models}, 
      author={Alicia Castro and Astrid Eichhorn and Razvan Gurau},
      DOI={10.1007/JHEP05(2026)117},
      year={2026},
      eprint={2602.09257},
      archivePrefix={arXiv},
      primaryClass={gr-qc},
      url={https://arxiv.org/abs/2602.09257}, 
}

@article{eichhorn2013continuum,
  title={Continuum limit in matrix models for quantum gravity from the functional renormalization group},
  author={Eichhorn, Astrid and Koslowski, Tim},
  journal={Physical Review D--Particles, Fields, Gravitation, and Cosmology},
  DOI={10.1103/PhysRevD.88.084016},
  volume={88},
  number={8},
  pages={084016},
  year={2013},
}

@article{geloun2018functional,
  title={Functional renormalization group analysis of rank-3 tensorial group field theory: The full quartic invariant truncation},
  author={Geloun, Joseph Ben and Koslowski, Tim A and Oriti, Daniele and Pereira, Antonio D},
  journal={Physical Review D},
  DOI={10.1103/PhysRevD.97.126018},
  volume={97},
  number={12},
  pages={126018},
  year={2018},
}

@article{eichhorn2018flowing,
  title={Flowing to the continuum limit in tensor models for quantum gravity},
  author={Eichhorn, Astrid and Koslowski, Tim},
  journal={Annales de l'Institut Henri Poincar{\'e} D},
  DOI={10.4171/AIHPD/52},
  volume={5},
  number={2},
  pages={173--210},
  year={2018}
}

@article{eichhorn2019towards,
  title={Towards background independent quantum gravity with tensor models},
  author={Eichhorn, Astrid and Lumma, Johannes and Koslowski, Tim and Pereira, Antonio D},
  journal={Classical and Quantum Gravity},
  DOI={10.1088/1361-6382/ab2545},
  volume={36},
  number={15},
  pages={155007},
  year={2019},
}

@article{eichhorn2014towards,
  title={Towards phase transitions between discrete and continuum quantum spacetime from the Renormalization Group},
  author={Eichhorn, Astrid and Koslowski, Tim},
  journal={Physical Review D},
  DOI={10.1103/PhysRevD.90.104039},
  volume={90},
  number={10},
  pages={104039},
  year={2014},
}

@article{eichhorn2019status,
  title={Status of background-independent coarse graining in tensor models for quantum gravity},
  author={Eichhorn, Astrid and Koslowski, Tim and Pereira, Antonio D},
  journal={Universe},
  DOI={10.3390/universe5020053},
  volume={5},
  number={2},
  pages={53},
  year={2019},
}

@article{benedetti2015functional,
  title={Functional renormalisation group approach for tensorial group field theory: a rank-3 model},
  author={Benedetti, Dario and Geloun, Joseph Ben and Oriti, Daniele},
  journal={Journal of High Energy Physics},
  DOI={10.1007/JHEP03%282015%29084},
  volume={2015},
  number={3},
  pages={1--40},
  year={2015},
}

@article{carrozza2014renormalization,
  title={Renormalization of tensorial group field theories: Abelian U (1) models in four dimensions},
  author={Carrozza, Sylvain and Oriti, Daniele and Rivasseau, Vincent},
  journal={Communications in Mathematical Physics},
  DOI={10.1007/s00220-014-1954-8},
  volume={327},
  number={2},
  pages={603--641},
  year={2014},
}

@article{carrozza2016flowing,
  title={Flowing in group field theory space: a review},
  author={Carrozza, Sylvain},
  journal={SIGMA. Symmetry, Integrability and Geometry: Methods and Applications},
  DOI={10.3842/SIGMA.2016.070},
  volume={12},
  pages={070},
  year={2016},
}

@article{geloun2015functional,
  title={Functional renormalization group analysis of a tensorial group field theory on R3},
  author={Geloun, Joseph Ben and Martini, Riccardo and Oriti, Daniele},
  DOI={10.1209/0295-5075/112/31001},
  journal={EPL},
  volume={112},
  number={3},
  year={2015},
}

@article{geloun2016functional,
  title={Functional renormalization group analysis of tensorial group field theories on $R^d$},
  author={Geloun, Joseph Ben and Martini, Riccardo and Oriti, Daniele},
  journal={Physical Review D},
  DOI={10.1103/PhysRevD.94.024017},
  volume={94},
  number={2},
  pages={024017},
  year={2016},
}

@article{benedetti2016functional,
  title={Functional renormalization group approach for tensorial group field theory: a rank-6 model with closure constraint},
  author={Benedetti, Dario and Lahoche, Vincent},
  journal={Classical and Quantum Gravity},
  DOI={10.1088/0264-9381/33/9/095003},
  volume={33},
  number={9},
  pages={095003},
  year={2016},
}

@article{lahoche2015renormalization,
  title={Renormalization of an Abelian tensor group field theory: solution at leading order},
  author={Lahoche, Vincent and Oriti, Daniele and Rivasseau, Vincent},
  journal={Journal of High Energy Physics},
  DOI={10.1007/JHEP04(2015)095},
  volume={2015},
  number={4},
  pages={1--41},
  year={2015},
}

@article{marchetti2023mean,
  title={Mean-field phase transitions in tensorial group field theory quantum gravity},
  author={Marchetti, Luca and Oriti, Daniele and Pithis, Andreas GA and Th{\"u}rigen, Johannes},
  journal={Physical Review Letters},
  DOI={10.1103/PhysRevLett.130.141501},
  volume={130},
  number={14},
  pages={141501},
  year={2023},
}

@article{lahoche2017functional,
  title={Functional renormalization group for the U (1)-T 5 6 tensorial group field theory with closure constraint},
  author={Lahoche, Vincent and Ousmane Samary, Dine},
  journal={Physical Review D},
  DOI={10.1103/PhysRevD.95.045013 },
  volume={95},
  number={4},
  pages={045013},
  year={2017},
}

@article{lahoche2020renormalization,
  title={Renormalization group flow of coupled tensorial group field theories: Towards the Ising model on random lattices},
  author={Lahoche, Vincent and Ousmane Samary, Dine and Pereira, Antonio D},
  journal={Physical Review D},
  DOI={10.1103/PhysRevD.101.064014 },
  volume={101},
  number={6},
  pages={064014},
  year={2020},
}

@article{carrozza2017renormalizable,
  title={Renormalizable Group Field Theory beyond melonic diagrams: an example in rank four},
  author={Carrozza, Sylvain and Lahoche, Vincent and Oriti, Daniele},
  journal={Physical Review D},
  DOI={10.1103/PhysRevD.96.066007 },
  volume={96},
  number={6},
  pages={066007},
  year={2017},
}

@article{lahoche2019ward,
  title={Ward-constrained melonic renormalization group flow for the rank-four $\phi$ 6 tensorial group field theory},
  author={Lahoche, Vincent and Samary, Dine Ousmane},
  journal={Physical Review D},
  DOI={10.1103/PhysRevD.100.086009},
  volume={100},
  number={8},
  pages={086009},
  year={2019},
}

@article{carrozza2017asymptotic,
  title={Asymptotic safety in three-dimensional SU (2) Group Field Theory: evidence in the local potential approximation},
  author={Carrozza, Sylvain and Lahoche, Vincent},
  journal={Classical and Quantum Gravity},
  DOI={10.1088/1361-6382/aa6d90},
  volume={34},
  number={11},
  pages={115004},
  year={2017},
}

@article{lahoche2017renormalization,
  title={Renormalization of a tensorial field theory on the homogeneous space SU(2)/U(1)},
  author={Lahoche, Vincent and Oriti, Daniele},
  journal={Journal of Physics A: Mathematical and Theoretical},
  DOI={10.1088/1751-8113/50/2/025201},
  volume={50},
  number={2},
  pages={025201},
  year={2017},
}

@article{lahoche2021no,
  title={No Ward-Takahashi identity violation for Abelian tensorial group field theories with a closure constraint},
  author={Lahoche, Vincent and Natta, B{\^e}m-Bi{\'e}ri Barth{\'e}lemy and Ousmane Samary, Dine},
  journal={Physical Review D},
  DOI={doi
10.1103/PhysRevD.104.106013},
  volume={104},
  number={10},
  pages={106013},
  year={2021},
}

@article{lahoche2018nonperturbative,
  title={Nonperturbative renormalization group beyond the melonic sector: The effective vertex expansion method for group fields theories},
  author={Lahoche, Vincent and Ousmane Samary, Dine},
  journal={Physical Review D},
  DOI={10.1103/PhysRevD.98.126010},
  volume={98},
  number={12},
  pages={126010},
  year={2018},
}

@article{pithis2020phase,
  title={Phase transitions in TGFT: functional renormalization group in the cyclic-melonic potential approximation and equivalence to O (N) models},
  author={Pithis, Andreas GA and Th{\"u}rigen, Johannes},
  journal={Journal of High Energy Physics},
  DOI={https://doi.org/10.1007/JHEP12(2020)159},
  volume={2020},
  number={12},
  pages={159},
  year={2020},
}

@article{lahoche2020pedagogical,
  title={Pedagogical comments about nonperturbative Ward-constrained melonic renormalization group flow},
  author={Lahoche, Vincent and Ousmane Samary, Dine},
  journal={Physical Review D},
  DOI={10.1103/PhysRevD.101.024001},
  volume={101},
  number={2},
  pages={024001},
  year={2020},
}

@article{Lahoche:2020pjo,
    author = "Lahoche, Vincent and Samary, Dine Ousmane",
    title = "{Reliability of the local truncations for the random tensor models renormalization group flow}",
    eprint = "2005.11846",
    archivePrefix = "arXiv",
    primaryClass = "hep-th",
    doi = "10.1103/PhysRevD.102.056002",
    journal = "Phys. Rev. D",
    volume = "102",
    number = "5",
    pages = "056002",
    year = "2020"
}

@article{Lahoche:2019ocf,
    author = "Lahoche, Vincent and Ousmane Samary, Dine",
    title = "{Revisited functional renormalization group approach for random matrices in the large-$N$ limit}",
    eprint = "1909.03327",
    archivePrefix = "arXiv",
    primaryClass = "hep-th",
    doi = "10.1103/PhysRevD.101.106015",
    journal = "Phys. Rev. D",
    volume = "101",
    number = "10",
    pages = "106015",
    year = "2020"
}

@article{Hubbard:1959ub,
    author = "Hubbard, J.",
    title = "{Calculation of partition functions}",
    doi = "10.1103/PhysRevLett.3.77",
    journal = "Phys. Rev. Lett.",
    volume = "3",
    pages = "77--80",
    year = "1959"
}

@article{pawlowski2017physics,
  title={Physics and the choice of regulators in functional renormalisation group flows},
  author={Pawlowski, Jan M and Scherer, Michael M and Schmidt, Richard and Wetzel, Sebastian J},
  journal={Annals of Physics},
  DOI={10.1016/j.aop.2017.06.017},
  volume={384},
  pages={165--197},
  year={2017},
}

@article{balog2019convergence,
  title={Convergence of nonperturbative approximations to the renormalization group},
  author={Balog, Ivan and Chat{\'e}, Hugues and Delamotte, Bertrand and Marohni{\'c}, Maroje and Wschebor, Nicol{\'a}s},
  journal={Physical review letters},
  DOI={10.1103/PhysRevLett.123.240604},
  volume={123},
  number={24},
  pages={240604},
  year={2019},
}

@article{Bojowald:2025ocr,
    author = "Bojowald, Martin and Diaz, Manuel and Duque, Erick I.",
    title = "{Singularities in loop quantum cosmology}",
    eprint = "2507.08116",
    archivePrefix = "arXiv",
    primaryClass = "gr-qc",
    doi = "10.1103/7g3z-2pgf",
    journal = "Phys. Rev. D",
    volume = "113",
    number = "8",
    pages = "084060",
    year = "2026"
}

@article{Oriti:2024elx,
    author = "Oriti, Daniele",
    title = "{Hydrodynamics on~(Mini)superspace or~a~Non-linear Extension of~Quantum Cosmology: An Effective Timeless Framework for Cosmology from Quantum Gravity}",
    doi = "10.1007/978-3-031-61860-4_11",
    journal = "Fundam. Theor. Phys.",
    volume = "216",
    pages = "221--252",
    year = "2024"
}

@article{Wegner:1972ih,
    author = "Wegner, Franz J. and Houghton, Anthony",
    title = "{Renormalization group equation for critical phenomena}",
    doi = "10.1103/PhysRevA.8.401",
    journal = "Phys. Rev. A",
    volume = "8",
    pages = "401--412",
    year = "1973"
}

@article{Wilson:1972zzb,
    author = "Wilson, K. G.",
    editor = "Jackson, J. D. and Roberts, A.",
    title = "{Field theory in less than four dimensions - the renormalization group}",
    journal = "eConf",
    number = "C720906V2",
    pages = "169--173",
    year = "1972"
}

@book{Zinn-Justin:2019jix,
    author = "Zinn-Justin, Jean",
    title = "{From Random Walks to Random Matrices}",
    isbn = "978-0-19-878775-4",
    publisher = "Oxford University Press",
    series = "Oxford Graduate Texts",
    month = "6",
    year = "2019"
}

@article{Lahoche:2025bmp,
    author = "Lahoche, Vincent and Samary, Dine Ousmane",
    title = "{Stochastic dynamics for group field theories II: Methods for nonequilibrium renormalization group}",
    journal = "Phys. Rev. D 114, 046028 (2026)",
    DOI={10.1103/kmgr-pc3w},
    eprint = "2509.05507",
    archivePrefix = "arXiv",
    primaryClass = "hep-th",
    month = "9",
    year = "2025"
}

@article{Achitouv:2025ghy,
    author = "Achitouv, Ixandra and Lahoche, Vincent and Samary, Dine Ousmane and Radpay, Parham",
    title = "{Constructing the low-temperature phase diagram for the $2+p$-quantum spin glass using the nonperturbative renormalization group}",
    journal = "Physica A 701 (2026) 132019",
    DOI={10.1016/j.physa.2026.132019 },
    eprint = "2503.12247",
    archivePrefix = "arXiv",
    primaryClass = "cond-mat.dis-nn",
    month = "3",
    year = "2025"
}

@article{Achitouv:2024kqh,
    author = "Achitouv, Ixandra and Lahoche, Vincent and Samary, Dine Ousmane and Radpay, Parham",
    title = "{Time-translation invariance symmetry breaking hidden by finite-scale singularities}",
    eprint = "2412.11619",
    archivePrefix = "arXiv",
    primaryClass = "cond-mat.dis-nn",
    doi = "10.1088/1402-4896/ae4cd1",
    journal = "Phys. Scripta",
    volume = "101",
    number = "11",
    pages = "115201",
    year = "2026"
}

@article{Lahoche:2024huc,
    author = "Lahoche, Vincent and Samary, Dine Ousmane and Radpay, Parham",
    title = "{Large time effective kinetics {\ensuremath{\beta}}-functions for quantum (2+p)-spin glass}",
    eprint = "2408.02602",
    archivePrefix = "arXiv",
    primaryClass = "cond-mat.dis-nn",
    doi = "10.1016/j.aop.2025.170102",
    journal = "Annals Phys.",
    volume = "480",
    pages = "170102",
    year = "2025"
}

@misc{Natta:2024nke,
    author = "Natta, B{\^e}m-Bi{\'e}ri Barth{\'e}l{\'e}my and Lahoche, Vincent and Samary, Dine Ousmane and Radpay, Parham",
    title = "{Large time effective kinetics {\ensuremath{\beta}}-functions for quantum (2 + p)-spin glass II: Effective vertex expansion, local potential approximation and symmetries}",
    eprint = "2411.11089",
    archivePrefix = "arXiv",
    primaryClass = "cond-mat.dis-nn",
    month = "11",
    year = "2024"
}

@article{Lahoche:2024qqq,
    author = "Lahoche, Vincent and Samary, Dine Ousmane",
    title = "{Functional renormalization group for {\textquotedblleft}p=2{\textquotedblright} like glassy matrices in the planar approximation III. Equilibrium dynamics and beyond}",
    eprint = "2404.11915",
    archivePrefix = "arXiv",
    primaryClass = "hep-th",
    doi = "10.1016/j.nuclphysb.2024.116656",
    journal = "Nucl. Phys. B",
    volume = "1006",
    pages = "116656",
    year = "2024"
}

@article{Lahoche:2024puq,
    author = "Lahoche, Vincent and Samary, Dine Ousmane",
    title = "{An intriguing connection between Pisarski's fixed point and (2+3)-spin glasses}",
    eprint = "2404.05436",
    archivePrefix = "arXiv",
    primaryClass = "cond-mat.dis-nn",
    doi = "10.1016/j.physleta.2024.129906",
    journal = "Phys. Lett. A",
    volume = "525",
    pages = "129906",
    year = "2024"
}

@article{Lahoche:2024hox,
    author = "Lahoche, Vincent and Samary, Dine Ousmane",
    title = "{Functional renormalization group for {\textquotedblleft}p=2{\textquotedblright} like glassy matrices in the planar approximation II. Ward identities method in the deep IR}",
    eprint = "2403.12217",
    archivePrefix = "arXiv",
    primaryClass = "hep-th",
    doi = "10.1016/j.nuclphysb.2024.116627",
    journal = "Nucl. Phys. B",
    volume = "1006",
    pages = "116627",
    year = "2024"
}

@article{Lahoche:2024gal,
    author = "Lahoche, Vincent and Samary, Dine Ousmane",
    title = "{Functional renormalization group for {\textquotedblleft}p=2{\textquotedblright} like glassy matrices in the planar approximation I. Vertex expansion at equilibrium}",
    eprint = "2403.07577",
    archivePrefix = "arXiv",
    primaryClass = "hep-th",
    doi = "10.1016/j.nuclphysb.2024.116582",
    journal = "Nucl. Phys. B",
    volume = "1005",
    pages = "116582",
    year = "2024"
}

@misc{finotello2026datafieldtheorytheory,
      title={Data Field Theory: Theory and Applications of the Functional Renormalization Group for Signal Detection}, 
      author={Riccardo Finotello and Vincent Lahoche and Dine Ousmane Samary and Parham Radpay},
      year={2026},
      eprint={2607.27236},
      archivePrefix={arXiv},
      primaryClass={physics.data-an},
      url={https://arxiv.org/abs/2607.27236}, 
}

@misc{finotello2026fieldtheorydataanomaly2,
      title={Field Theory of Data: Anomaly Detection via the Functional Renormalization Group. The 2D Ising Model as a Benchmark}, 
      author={Riccardo Finotello and Vincent Lahoche and Parham Radpay and Dine Ousmane Samary},
      year={2026},
      eprint={2605.11138},
      archivePrefix={arXiv},
      primaryClass={cond-mat.stat-mech},
      url={https://arxiv.org/abs/2605.11138}, 
}

@article{Finotello_2026,
   title={Functional renormalisation for signal detection: dimensional analysis and dimensional phase transition for nearly continuous spectra effective field theory},
   volume={2026},
   ISSN={1742-5468},
   url={http://dx.doi.org/10.1088/1742-5468/ae5a21},
   DOI={10.1088/1742-5468/ae5a21},
   number={4},
   journal={Journal of Statistical Mechanics: Theory and Experiment},
   author={Finotello, Riccardo and Lahoche, Vincent and Ousmane Samary, Dine},
   year={2026},
   month=Apr, pages={043403} }

@book{amari2016information,
  title={Information geometry and its applications},
  author={Amari, Shun-ichi},
  year={2016},
  publisher={Springer}
}

@book{amari2000methods,
  title={Methods of information geometry},
  author={Amari, Shun-ichi and Nagaoka, Hiroshi},
  DOI={10.1090/mmono/191 },
  volume={191},
  year={2000},
  publisher={American Mathematical Soc.}
}

@article{beny2015information,
  title={Information-geometric approach to the renormalization group},
  author={B{\'e}ny, C{\'e}dric and Osborne, Tobias J},
  journal={Physical Review A},
  DOI={10.1103/PhysRevA.92.022330},
  volume={92},
  number={2},
  pages={022330},
  year={2015},
}

@article{berman2023bayesian,
  title={Bayesian renormalization},
  author={Berman, David S and Klinger, Marc S and Stapleton, Alexander G},
  journal={Machine Learning: Science and Technology},
  DOI={10.1088/2632-2153/ad0102},
  volume={4},
  number={4},
  pages={045011},
  year={2023},
}

@article{maity2015information,
  title={Information geometry and the renormalization group},
  author={Maity, Reevu and Mahapatra, Subhash and Sarkar, Tapobrata},
  journal={Physical Review E},
  DOI={10.1103/PhysRevE.92.052101},
  volume={92},
  number={5},
  pages={052101},
  year={2015},
}

@article{kar2001geometry,
  title={Geometry of renormalization group flows in theory space},
  author={Kar, Sayan},
  journal={Physical Review D},
  DOI={10.1103/PhysRevD.64.105017},
  volume={64},
  number={10},
  pages={105017},
  year={2001},
}

@article{quinn2023information,
  title={Information geometry for multiparameter models: New perspectives on the origin of simplicity},
  author={Quinn, Katherine N and Abbott, Michael C and Transtrum, Mark K and Machta, Benjamin B and Sethna, James P},
  journal={Reports on Progress in Physics},
  DOI={10.1088/1361-6633/aca6f8},
  volume={86},
  number={3},
  pages={035901},
  year={2023},
}

@article{strandkvist2020beyond,
  title={Beyond RG: from parameter flow to metric flow},
  author={Strandkvist, Charlotte and Chvykov, Pavel and Tikhonov, Mikhail},
  DOI={https://doi.org/10.48550/arXiv.2011.12420},
  journal={arXiv:2011.12420},
  year={2011}
}

@article{floerchinger2023information,
  title={Information geometry of Euclidean quantum fields},
  author={Floerchinger, Stefan},
  DOI={10.1103/PhysRevD.110.125027},
  journal={Phys. Rev. D 110, 125027},
  year={2024}
}

@article{floerchinger2023exact,
  title={Exact flow equation for the divergence functional},
  author={Floerchinger, Stefan},
  journal={Physics Letters B},
  DOI={10.1016/j.physletb.2023.138244},
  volume={846},
  pages={138244},
  year={2023},
}

@article{polchinski1984renormalization,
  title={Renormalization and effective Lagrangians},
  author={Polchinski, Joseph},
  journal={Nuclear Physics B},
  DOI={10.1016/0550-3213(84)90287-6},
  volume={231},
  number={2},
  pages={269--295},
  year={1984},
}

@book{zinn2021quantum,
  title={Quantum field theory and critical phenomena},
  author={Zinn-Justin, Jean},
  DOI={https://doi.org/10.1093/acprof:oso/9780198509233.001.0001},
  volume={171},
  year={2021},
  publisher={Oxford university press}
}

@article{wegner1974some,
  title={Some invariance properties of the renormalization group},
  author={Wegner, FJ},
  journal={Journal of Physics C: Solid State Physics},
  DOI={10.1088/0022-3719/7/12/004},
  volume={7},
  number={12},
  pages={2098--2108},
  year={1974}
}

@article{morris1993exact,
  title={The exact renormalisation group and approximate solutions},
  author={Morris, Tim R},
  DOI={10.1142/S0217751X94000972
},
  journal={arXiv preprint hep-ph/9308265},
  year={1993}
}

@article{cotler2023renormalization,
  title={Renormalization group flow as optimal transport},
  author={Cotler, Jordan and Rezchikov, Semon},
  journal={Physical Review D},
  DOI={10.1103/PhysRevD.108.025003},
  volume={108},
  number={2},
  pages={025003},
  year={2023},
}

@article{Wahabou_Kpera_2024,
   title={Anomalous higher order Ward identities in tensorial group field theories without closure constraint},
   volume={41},
   ISSN={1361-6382},
   url={http://dx.doi.org/10.1088/1361-6382/ad7c13},
   DOI={10.1088/1361-6382/ad7c13},
   number={22},
   journal={Classical and Quantum Gravity},
   author={Wahabou Kpera, Bio and Lahoche, Vincent and Ousmane Samary, Dine and Fawaaz Zime Yerima, Seke},
   year={2024},
   month=Oct, pages={225015} }

@article{litim2000optimisation,
  title={Optimisation of the exact renormalisation group},
  author={Litim, Daniel F},
  journal={Physics Letters B},
doi="10.1016/S0370-2693(00)00748-6",
  volume={486},
  number={1-2},
  pages={92--99},
  year={2000},
}

@book{weinberg1995quantum,
  title={The quantum theory of fields},
  author={Weinberg, Steven},
  DOI={https://doi.org/10.1017/CBO9781139644174},
  volume={2},
  year={1995},
  publisher={Cambridge university press}
}

@misc{lahoche2026relationalbackgroundfieldformulation,
      title={Relational background field formulation of discrete quantum gravity models: The case of $O(N)$ vector models in zero dimensions}, 
      author={Vincent Lahoche and Dine Ousmane Samary and Parham Radpay},
      year={2026},
      eprint={2609.28032},
      archivePrefix={arXiv},
      primaryClass={hep-th},
      url={https://arxiv.org/abs/2609.28032}, 
}


\end{document}